\pdfoutput=1

\documentclass{vldb}
\usepackage{graphicx}
\usepackage[noend]{algpseudocode}
\usepackage{algorithm}
\usepackage{caption}
\usepackage[labelfont=bf]{caption}
\usepackage{subcaption}
\usepackage{url}
\usepackage{balance}  
\usepackage[normalem]{ulem}
\usepackage{xcolor}
\usepackage{float}
\definecolor{keyword-color}{rgb}{0.2, 0.2, 0.6}
\usepackage{listings}

\lstdefinelanguage{SQLPP}{
  alsoletter={-"'_:\$[]()0123456789\\},
  morekeywords={
    SELECT,
    FROM,
    WHERE,
    GROUP,
    BY,
    AS,
    by,
    with,
    order,
    desc,
    asc,
    limit,
    replace,
    delete,
    insert,
    into,
    some,
    every,
    satisfies,
    and,
    or,
    use,
    dataverse,
    set,
    simfunction,
    simthreshold,
    create,
    index,
    type,
    fuzzy,
    keyword,
    NGram,
    rtree,
    VALUE,
    WITH,
    ORDER,
    LIMIT,
    SOME,
    EVERY,
    SATISFIES,
    IN,
    DESC,
    ASC,
    AND,
    UNNEST,
    LET,
    DISTINCT
  },
  basicstyle=\sf,
  keywordstyle={\sf\textbf},
  identifierstyle=\texttt,
  commentstyle=\textit,
  literate={<=}{{\litleq}}1 {>=}{{\litgeq}}1,
  literate={~=}{$\sim{}=$}1 
}[keywords,comments,strings]

\lstdefinelanguage{AQLSchema}{
  alsoletter={-"'_:(\$\\},
  morekeywords={
    use,
    declare,
    type,
    as,
    open,
    closed,
    dataset,
    dataverse,
    nodegroup,
    on,
    partitioned,
    by,
    primary,
    key,
    create,
    index,
    bigint,
    double,
    date,
    polygon,
    with,
    filter,
    CREATE,
    TYPE,
    string,
    int,
    AS,
    DATASET,
    PRIMARY,
    KEY,
    CLOSED,
    OPEN,
    WITH,
    true
  },
  basicstyle=\sf,
  keywordstyle={{\sf\textbf}},
  identifierstyle=\texttt,
  commentstyle=\textit,
  literate={<=}{{\litleq}}1 {>=}{{\litgeq}}1
}[keywords,comments,strings]

\lstdefinelanguage{AQLSchema2}{
  alsoletter={-"'_:(\$\\},
  morekeywords={
    use,
    declare,
    type,
    as,
    open,
    closed,
    dataset,
    dataverse,
    nodegroup,
    on,
    partitioned,
    by,
    primary,
    key,
    create,
    index,
    int32,
    int64,
    string,
    datetime,
    double,
    date,
    point,
    with,
    filter,
    TwitterUserType
  },
  basicstyle=\sf,
  keywordstyle=\textbf,
  identifierstyle=\texttt,
  commentstyle=\textit,
  literate={<=}{{\litleq}}1 {>=}{{\litgeq}}1
}[keywords,comments,strings]

\lstnewenvironment{aql}
   {\lstset{language=AQL,basicstyle=\small,basewidth={.5em,.5em}}}
   {}
   
\lstnewenvironment{sqlpp}
   {\lstset{language=SQLPP,basicstyle=\hspace{4cm}\small,basewidth={.6em,.7em}}}
   {}

\lstnewenvironment{aqlscriptsize}
   {\lstset{language=AQL,basicstyle=\scriptsize,basewidth={.5em,.5em},numbers=left}}
   {}

\lstnewenvironment{aqltiny}
   {\lstset{language=AQL,space-flexible,basicstyle=\tiny,basewidth={.5em,.5em},numbers=left}}
   {}

\lstnewenvironment{aqlschema}
   {\lstset{language=AQLSchema,basicstyle=\small,basewidth={.5em,.5em}}}
   {}
   
\lstnewenvironment{inlineaqlschema}
   {\lstset{language=AQLSchema,basicstyle=\noindent\small,basewidth={.5em,.5em}}}
   {}
   
\lstnewenvironment{inlineaqlschema2}
   {\lstinline{language=AQLSchema,basicstyle=\noindent\small,basewidth={.5em,.5em}}\lstMakeShortInline[columns=fixed]|}
   {}
   
\lstnewenvironment{aqlschema2}
   {\lstset{language=AQLSchema2,basicstyle=\small,basewidth={.5em,.5em}}}
   {}
   
\lstnewenvironment{scala}
   {\lstset{language=SCALA, showstringspaces=false,tabsize=3, columns=space-flexible,basicstyle=\linespread{1}\small\ttfamily,basewidth={.5em,.5em}}}
   {}

\floatstyle{plaintop}
\restylefloat{table}

\vldbTitle{An LSM-based Tuple Compaction Framework in Apache AsterixDB}
\vldbAuthors{Wail Y. Alkowaileet, Sattam Alsubaiee and Michael J. Carey}
\vldbDOI{https://doi.org/10.14778/3397230.3397236}
\vldbVolume{13}
\vldbNumber{9}
\vldbYear{2020}
\newcommand{\rev}[1]{\textcolor{black}{#1}}
\begin{document}


\title{An LSM-based Tuple Compaction Framework \\for Apache AsterixDB (Extended Version)}



%
%
%
%

\numberofauthors{3} 

\author{
%
%
\alignauthor
Wail Y. Alkowaileet\\
       \affaddr{University of California, Irvine}\\
       \affaddr{Irvine, CA}\\
       \email{w.alkowaileet@uci.edu}
\alignauthor
Sattam Alsubaiee\\
       \affaddr{King Abdulaziz City for Science and Technology}\\
       \affaddr{Riyadh, Saudi Arabia}\\
       \email{ssubaiee@kacst.edu.sa}
\alignauthor 
Michael J. Carey\\
       \affaddr{University of California, Irvine}\\
       \affaddr{Irvine, CA}\\
       \email{mjcarey@ics.uci.edu}
}

\maketitle

\begin{abstract}

Document database systems store self-describing \sloppy{\rev{semi-structured}} records, such as JSON, ``as-is" without requiring the users to pre-define a schema. This provides users with the flexibility to change the structure of incoming records without worrying about taking the system offline or hindering the performance of currently running queries. However, the flexibility of such systems does not free. The large amount of redundancy in the records can introduce an unnecessary storage overhead and impact query performance.

Our focus in this paper is to address the storage overhead issue by introducing a tuple compactor framework that infers and extracts the schema from self-describing \rev{semi-structured} records during the data ingestion. As many prominent document stores, such as MongoDB and Couchbase, adopt Log Structured Merge (LSM) trees in their storage engines, our framework exploits LSM lifecycle events to piggyback the schema inference and extraction operations. We have implemented and empirically evaluated our approach to measure its impact on storage, data ingestion, and query performance in the context of Apache AsterixDB.
\end{abstract}
\section{Introduction}
\label{intro}

Self-describing semi-structured data formats like JSON have become the de facto format for storing and sharing information as developers moving away from the rigidity of schemas in the relational model. Consequently, NoSQL Database Management Systems (DBMSs) have emerged as popular solutions for storing, indexing, and querying self-describing semi-structured data. In document store systems such as MongoDB~\cite{mongodb} and Couchbase~\cite{couchbase}, users are not required to define a schema before loading or ingesting their data since each data instance is self-describing (i.e., each record embeds metadata that describes its structure and values). The flexibility of the self-describing data model provided by NoSQL systems attracts applications where the schema can change in the future by adding, removing, or even changing the type of one or more values without taking the system offline or slowing down the running queries. 

The flexibility provided in document store systems over the rigidity of the schemas in Relational Database Management Systems (RDBMSs) does not come without a cost. For instance, storing a boolean value for a field named \textit{hasChildren}, which takes roughly one byte to store in an RDBMS, can take a NoSQL DBMS an order of magnitude more bytes to store. Defining a schema prior to ingesting the data can alleviate the storage overhead, as the schema is then stored in the system's catalog and not in each record. However, defining a schema defies the purpose of schema-less DBMSs, which allow adding, removing or changing the types of the fields without manually altering the schema~\cite{asterix-overview-vldb2015, asterixdb-midflight}. From a user perspective, declaring a schema requires thorough a priori understanding of the dataset's fields and their types.


Let us consider a scenario where a data scientist wants to ingest and analyze a large volume of semi-structured data from a new external data source without prior knowledge of its structure. Our data scientist starts by acquiring a few instances from the data source and tries to analyze their structures; she then builds a schema according to the acquired sample. After ingesting a few data instances, our data scientist discovers that some fields can have more than one type, which was not captured in her initial sample. As a result, she stops the ingestion process, alters the schema to accommodate the irregularities in the types of those fields, and then reinitiates the data ingestion process. In this case, our data scientist has to continuously monitor the system and alter the schema if necessary, which may result in taking the system offline or stopping the ingestion of new records. Having an automated mechanism to infer and consolidate the schema information for the ingested records without losing the flexibility and the experience of schema-less stores would clearly be desirable.

In this work, we address the problem of the storage overhead in document stores by introducing a framework that infers and compacts the schema information for semi-structured data during the ingestion process. Our design utilizes the lifecycle events of Log Structured Merge (LSM) tree \cite{lsm} based storage engines, which are used in many prominent document store systems \cite{couchbase, mongodb} including Apache AsterixDB \cite{storage}. In LSM-backed engines, records are first accumulated in memory (\textit{LSM in-memory component}) and then subsequently written sequentially to disk (\textit{flush operation}) in a single batch (\textit{LSM on-disk component}).~Our framework takes the opportunity provided by LSM \textit{flush} operations to extract and strip the metadata from each record and construct a schema for each flushed LSM component. We have implemented and empirically evaluated our framework to measure its impact on the storage overhead, data ingestion rate and query performance in the context of AsterixDB. Our main contributions could be summarized as follows:

\begin{itemize}
    \item We propose a mechanism that utilizes the LSM workflow to infer and compact the schema for NoSQL systems records during flush operations. Moreover, we detail the steps required for distributed query processing using the inferred schema. 
    \vspace{-.2cm}
    \item We introduce a non-recursive physical data layout that separates data values from their metadata, which allows us to infer and compact the schema efficiently for nested data.
    \vspace{-.2cm}
	\item We introduce page-level compression in AsterixDB. This is a similar solution to these adopted by other NoSQL DBMSs to reduce the storage overhead of self-describing records.
    \vspace{-.2cm}
    \item We evaluate the feasibility of our design, prototyped using AsterixDB, to ingest and query a variety of large semi-structured datasets. We compare our ``semantic'' approach of reducing the storage overhead to the ``syntactic'' approach of compression.
\end{itemize}

The remainder of this paper is structured as follows: Section \ref{sec:asterix-arch} provides a preliminary review of the AsterixDB architecture and our implementation for page-level compression. Section \ref{sec:tuple-compactor} details the design and implementation of our tuple compaction framework in AsterixDB. Section \ref{sec:exper} presents an experimental evaluation of the proposed framework. Section \ref{sec:related-work} discusses related work on utilizing the LSM lifecycle and on schema inference for semi-structured data. Finally, Section \ref{sec:conclusion} presents our conclusions and discusses potential future directions for our work.
\section{Apache AsterixDB Overview}
In this paper, we use Apache AsterixDB to prototype our tuple compactor framework. AsterixDB is a parallel semi-structured Big Data Management System (BDMS) which runs on large, shared-nothing, commodity computing clusters. To prepare the reader, we give a brief overview of AsterixDB~\cite{asterix-overview-vldb2015, asterixdb-midflight} and its query execution engine Hyracks~\cite{hyracks}. Finally, we present our design and implementation of page-level compression in AsterixDB.
\subsection{User Model}
\label{sec:asterix-arch} 
The AsterixDB Data Model (ADM) extends the JSON data model to include types such as temporal and spatial types as well as data modeling constructs (e.g., bag or multiset). Defining an ADM \textit{datatype} (akin to a schema in an RDBMS) that describes at least the primary key(s) is required to create a \textit{dataset} (akin to a table in an RDBMS).

There are two options when defining a datatype in AsterixDB: \textit{open} and \textit{closed}. Figure~\ref{fig:empddl} shows an example of defining a dataset of employee information. In this example, we first define \textit{DependentType}, which declares two fields \textit{name} and \textit{age} of types \textit{string} and \textit{int}, respectively. Then, we define \textit{EmployeeType}, which declares \textit{id}, \textit{name} and \textit{dependents} of types \textit{int}, \textit{string} and a multiset of \textit{DependentType}, respectively. 
The symbol ``$?$'' indicates that a field is optional. Note that we defined the type \textit{EmployeeType} as \textit{open}, where data instances of this type can have additional undeclared fields.
On the other hand, we define the \textit{DependentType} as \textit{closed}, where data instances can only have declared fields. In both the open and closed datatypes, AsterixDB does not permit data instances that do not have values for the specified non-optional fields. Finally, in this example, we create a dataset \textit{Employee} of the type \textit{EmployeeType} and specify its \textit{id} field as the primary key.

\begin{figure}[h]
\begin{subfigure}{0.49\linewidth}
  \vspace{-0.25cm}
\begin{aqlschema}
CREATE TYPE DependentType 
AS CLOSED {
   name: string,
   age: int
};
\end{aqlschema}
\end{subfigure}
\begin{subfigure}{0.5\linewidth}
\begin{aqlschema}
CREATE TYPE EmployeeType 
AS OPEN {
   id: int,
   name: string,
   dependents:{{DependentType}}?
};
\end{aqlschema}
\end{subfigure}
\begin{subfigure}{1.0\linewidth}
\centering\begin{aqlschema}
CREATE DATASET Employee(EmployeeType) PRIMARY KEY id;
\end{aqlschema}
\end{subfigure}
\vspace{-0.5cm}
\caption{Defining Employee type and dataset in ADM}
\label{fig:empddl}
\end{figure}


To query the data stored in AsterixDB, users can submit their queries written in SQL++ \cite{sql++_book, sql++}, a SQL-inspired declarative query language for semi-structured data. Figure~\ref{fig:sqlpp_example} shows an example of a SQL++ aggregate query posed against the dataset declared in Figure~\ref{fig:empddl}.
\begin{figure}[h]
\centering
\begin{sqlpp}
SELECT VALUE nameGroup FROM Employee AS emp 
GROUP BY emp.name GROUP AS nameGroup
\end{sqlpp}
\vspace{-0.5cm}
\caption{An example of a SQL++ query}
\label{fig:sqlpp_example}
\end{figure}

\subsection{Storage and Data Ingestion}
\label{sec:asterix_storage_ingestion}
\begin{figure}[h]
    \centering
    \includegraphics[width=0.48\textwidth]{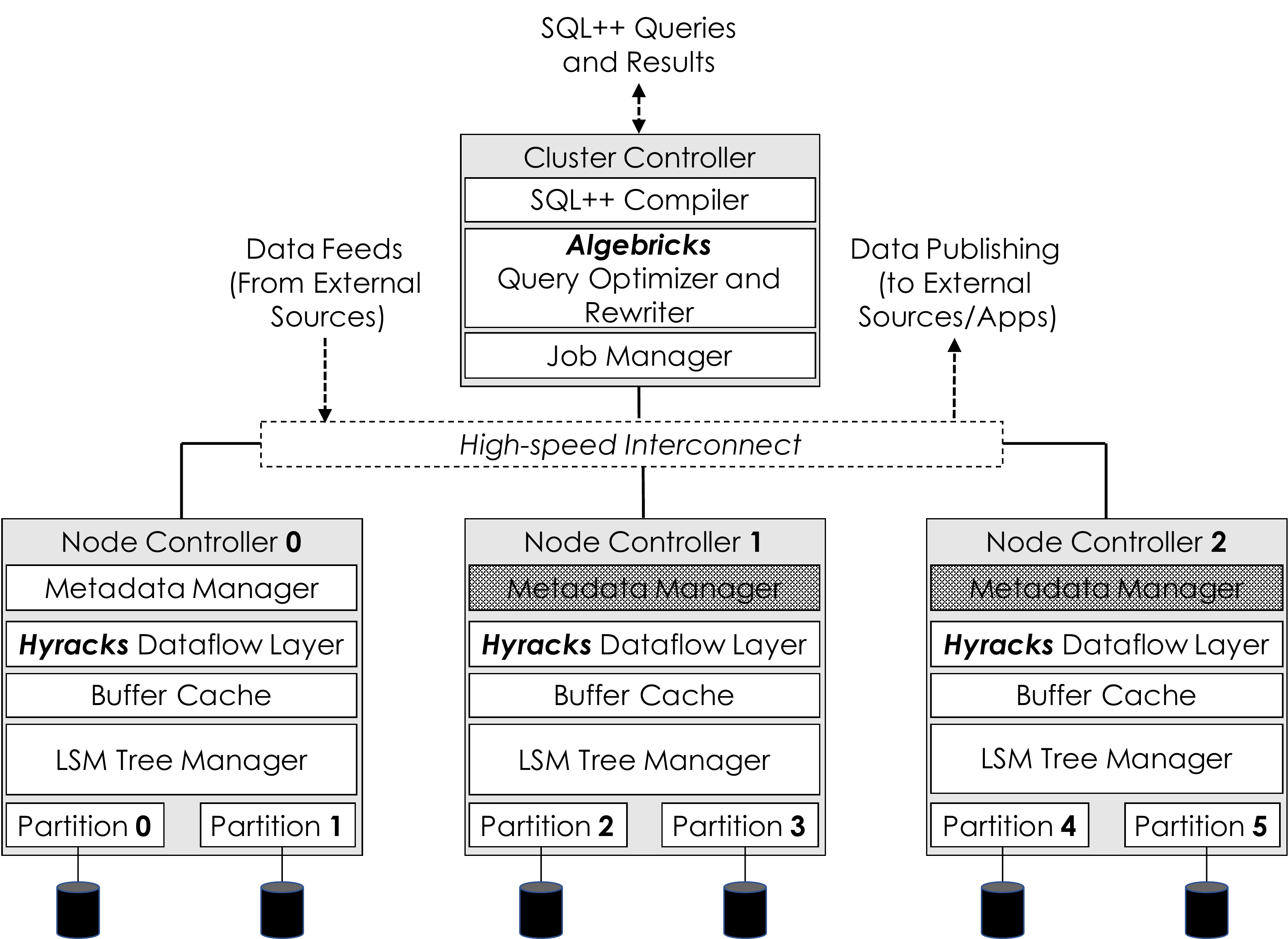}
    \caption{Apache AsterixDB cluster configured with two partitions in each of the three NCs}
    \label{fig:asterix_arch}
    \vspace{-0.4cm}
\end{figure}

In an AsterixDB cluster, each worker node (Node Controller, or NC for short) is controlled by a Cluster Controller (CC) that manages the cluster's topology and performs routine checks on the NCs. Figure~\ref{fig:asterix_arch} shows an AsterixDB cluster of three NCs, each of which has two data partitions that hold data on two separate storage devices. Data partitions in the same NC (e.g., Partition 0 and Partition 1 in NC0) share the same buffer cache and memory budget for LSM in-memory components; however, each partition manages the data stored in its storage device independently. In this example, NC0 also acts as a metadata node, which stores and provides access to AsterixDB metadata such as the defined datatypes and datasets.

\begin{figure}[h]
    \begin{subfigure}[b]{0.5\textwidth}
        \includegraphics[width=\textwidth]{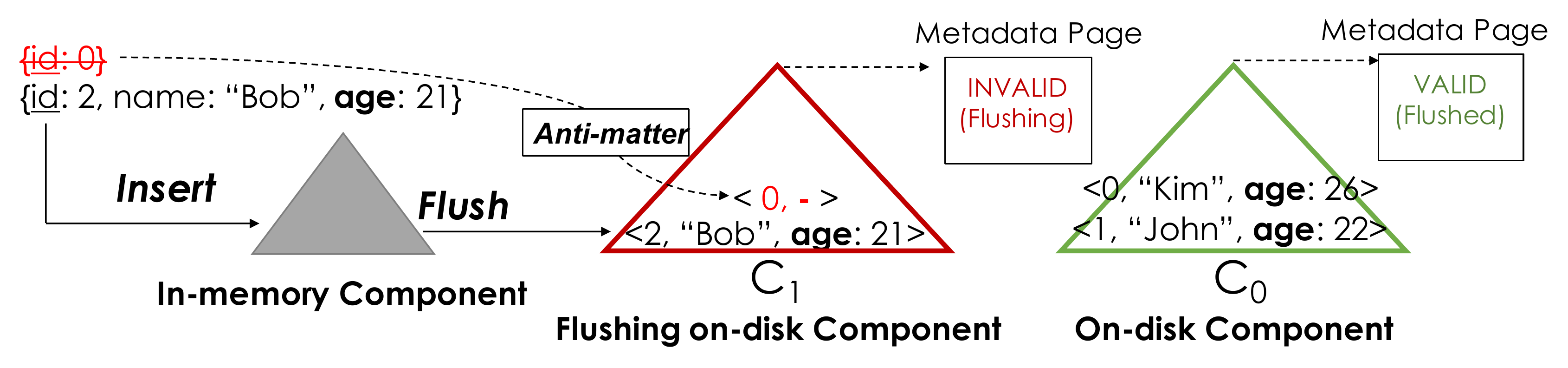}
        
        \caption{}
        \label{fig:lsm-flush}
    \end{subfigure}
    \begin{subfigure}[b]{0.5\textwidth}
        \includegraphics[width=\textwidth]{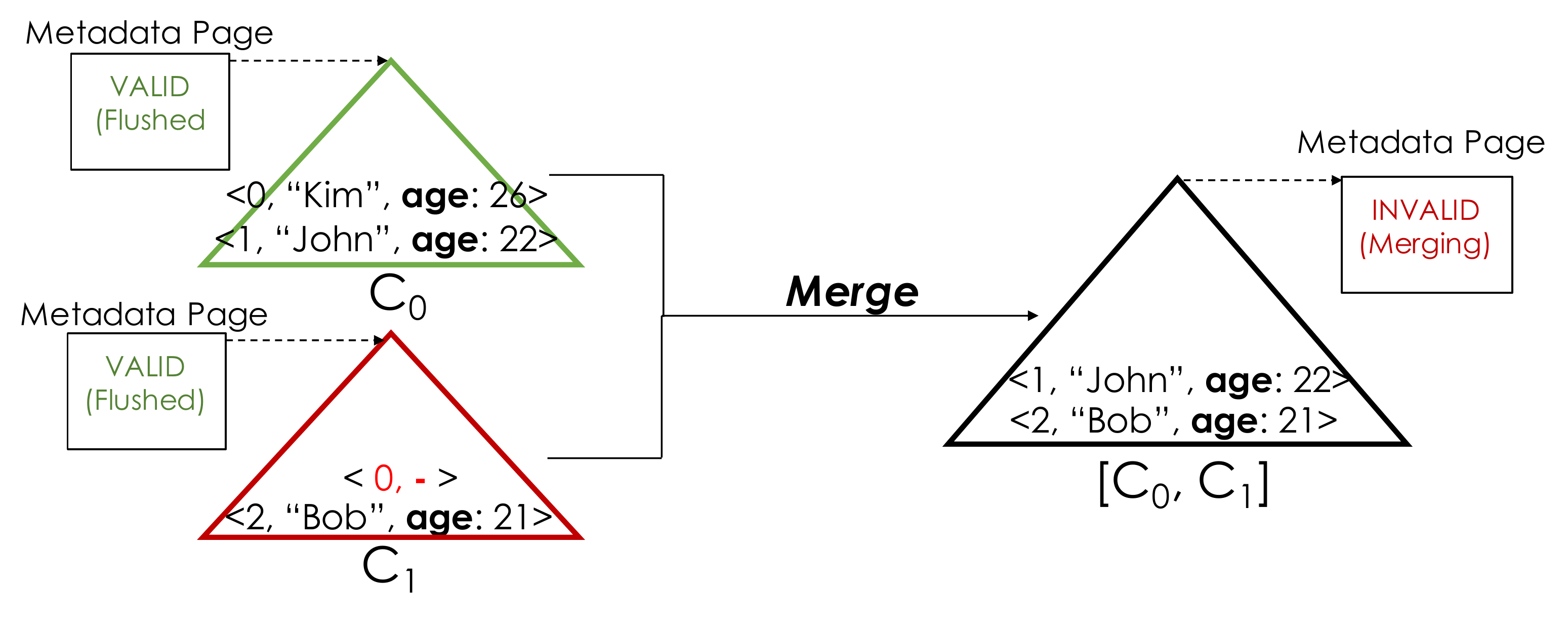}
        \vspace{-.5cm}
                \vspace{-0.2cm}
        \caption{}
        \label{fig:lsm-merge}
    \end{subfigure}
    \vspace{-.4cm}
    \caption{\textbf{(a)} Flushing component $C_1$ \textbf{(b)} Merging the two components $C_0$ and $C_1$ into a new component $[C_0, C_1]$}
    \label{fig:lsm-lifecycle}
    \vspace{-.3cm}
\end{figure}

AsterixDB stores the records of its datasets, spread across the data partitions in all NCs, in primary LSM B$^+$-tree indexes. During data ingestion, each new record is hash-partitioned using the primary key(s) into one of the configured partitions (Partition~0 to Partition~5 in Figure~\ref{fig:asterix_arch}) and inserted into the dataset's LSM in-memory component. AsterixDB implements a no-steal/no-force buffer management policy with write-ahead-logging (WAL) to ensure the durability and atomicity of ingested data. When the in-memory component is full and cannot accommodate new records, the \textit{LSM Tree Manager} (called the ``tree manager'' hereafter) schedules a \textit{flush operation}. Once the flush operation is triggered, the tree manager writes the in-memory component's records into a new LSM on-disk component on the partition's storage device, Figure~\ref{fig:lsm-flush}. On-disk components during their flush operation are considered \textit{INVALID} components. Once it is completed, the tree manager marks the flushed component as \textit{VALID} by setting a validity bit in the component's \textit{metadata page}. After this point, the tree manager can safely delete the logs for the flushed component. During crash recovery, any disk component with an unset validity bit is considered invalid and removed. The recovery manager can then replay the logs to restore the state of the in-memory component before the crash.

Once flushed, LSM on-disk components are immutable and, hence, updates and deletes are handled by inserting new entries. A delete operation adds an "anti-matter" entry~\cite{storage} to indicate that a record with a specified key has been deleted. An upsert is simply a delete followed by an insert with the same key. For example, in Figure~\ref{fig:lsm-flush}, we delete the record with $id = 0$. Since the target record is stored in $C_0$, we insert an "anti-matter" entry to indicate that the record with $id = 0$ is deleted. As on-disk components accumulate, the tree manager periodically merges them into larger components according to a merge policy \cite{storage, luo2019efficient} that determines when and what to merge. Deleted and updated records are garbage-collected during the merge operation. In Figure~\ref{fig:lsm-merge}, after merging $C_0$ and $C_1$ into $[C_0, C_1]$, we do not write the record with $id = 0$ as the record and the anti-matter entry annihilate each other. As in the flush operation, on-disk components created by a merge operation are considered \textit{INVALID} until their operation is completed. After completing the merge, older on-disk components ($C_0$ and $C_1$) can be safely deleted.

On-disk components in AsterixDB are identified by their \textit{component IDs}, where flushed components have monotonically increasing component IDs (e.g., $C_0$ and $C_1$) and merged components have components IDs that represent the range of component IDs that were merged (e.g., $[C_0, C_1]$). AsterixDB infers the recency ordering of components by inspecting the component ID, which can be useful for maintenance~\cite{luo2019efficient}. In this work, we explain how to use this property later in Section~\ref{sec:schema-struct}.

Datasets' records (of both \textit{open} and \textit{closed} types) in the LSM primary index are stored in a binary-encoded physical ADM format \cite{asterixdb-serialization}. Records of \textit{open} types that have undeclared fields are self-describing, i.e., the records contain additional information about the undeclared fields such as their types and their names. For our example in Figure~\ref{fig:lsm-lifecycle}, AsterixDB stores the information about the field \textbf{age} as it is not declared. For declared fields (\textit{id} and \textit{name} in this example), their type and name information are stored separately in the metadata node (NC0).

\subsection{Runtime Engine and Query Execution}
\label{sec:asterix_runtime}
To run a query, the user submits an SQL++ query to the CC, which optimizes and compiles it into a Hyracks job. Next, the CC distributes the compiled Hyracks job to the query executors in all partitions where each executor runs the submitted job in parallel \footnote{The default number of query executors is equal to the number of data partitions in AsterixDB.}. 

\begin{figure}[h]
    \centering
    \includegraphics[width=0.48\textwidth]{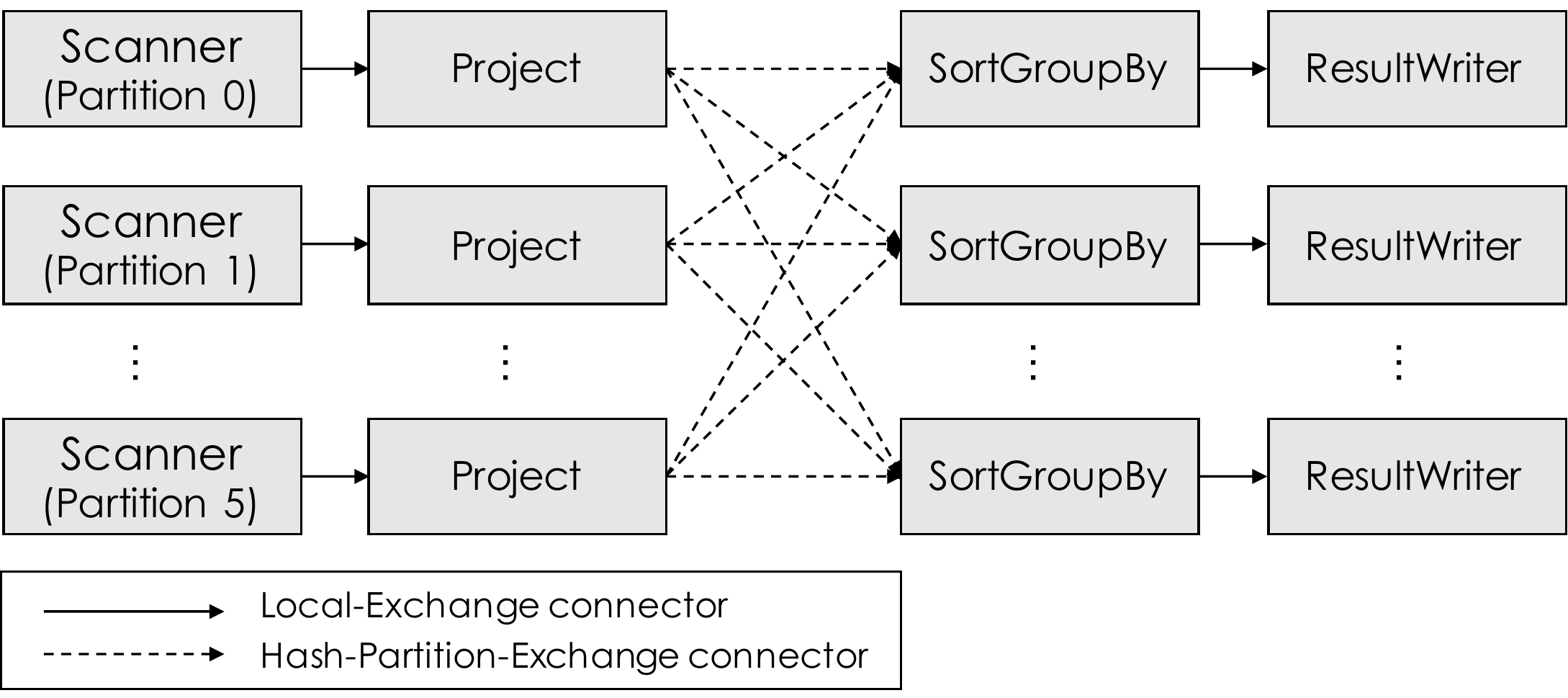}
    \caption{A compiled Hyracks job for the query in Figure~\ref{fig:sqlpp_example}}
    \label{fig:hyracks_job_example}
    \vspace{-0.5cm}
\end{figure}

Hyracks jobs consist of \textit{operators} and \textit{connectors}, where data flows between operators over connectors as a batch of records (or a frame of records in Hyracks terminology). Figure~\ref{fig:hyracks_job_example} depicts the compiled Hyracks job for the query in Figure~\ref{fig:sqlpp_example}. As shown in Figure~\ref{fig:hyracks_job_example}, records can flow within an executor's operators through \textit{Local-Exchange} connectors or they can be repartitioned or broadcast to other executors' operators through non-local exchange connectors such as the \textit{Hash-Partition-Exchange} connector in this example.

Operators in a Hyracks job process the ADM records in a received frame using AsterixDB-provided functions. For instance, a field access expression in a SQL++ query is translated into AsterixDB's internal function $getField()$. AsterixDB's compiler \textit{Algebricks}~\cite{algebricks} may rewrite the translated function when necessary. As an example, the field access expression \textit{e.name} in the query shown in Figure~\ref{fig:sqlpp_example} is first translated into a function call $getField(emp, ``name")$ where the argument $emp$ is a record and \textit{``name''} is the name of the requested field. Since \textit{name} is a declared field, Algebricks can rewrite the field access function to $getField(emp, 1)$ where the second argument \textbf{1} corresponds to the field's index in the schema provided by the Metadata Node.

\subsection{Page-level Compression}
\label{sec:compression}

As shown in Figure~\ref{fig:lsm-lifecycle}, the information of the undeclared field \textit{age} is stored within each record. This could incur an unnecessary higher storage overhead if all or most records of the Employee dataset have the same undeclared field, as records store redundant information. MongoDB and Couchbase have introduced compression to reduce the impact of storing redundant information in self-describing records. In AsterixDB, we introduce page-level compression, which compresses leaf pages of the B$^+$-tree of the primary index.

AsterixDB's new page-level compression is designed to operate at the buffer-cache level. On write, pages are compressed and then persisted to disk. On read, pages are decompressed to their original configured fixed-size and stored in memory in AsterixDB's buffer cache. Compressed pages can be of any arbitrary size. However, the AsterixDB storage engine was initially designed to work with fixed-size data pages where the size is a configurable parameter. Larger data pages can be stored as multiple fixed-size pages, but there is no mechanism to store smaller compressed pages. Any proposed solution to support variable-size pages must not change the current storage physical layout of AsterixDB.

 \begin{figure}[h]
    \centering
    \includegraphics[width=0.45\textwidth]{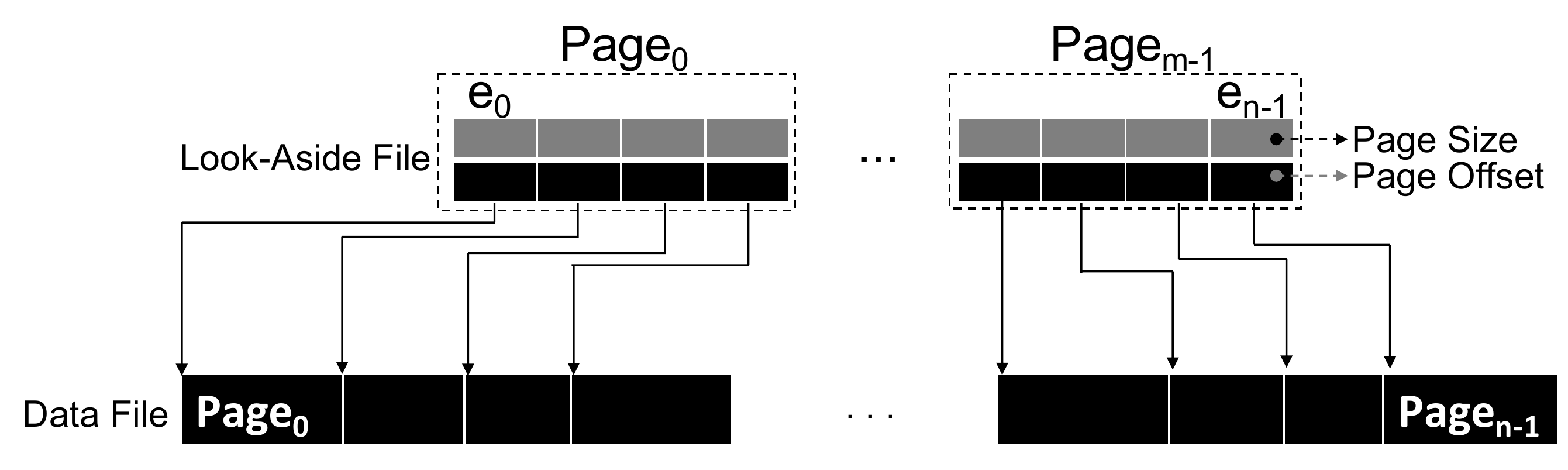}
    \caption{Compressed file with its Look-Aside File (LAF)}
    \label{fig:laf}
    \vspace{-.3cm}
\end{figure}

To address this issue, we use Look-aside Files (LAFs) to store offset-length entry pairs for the stored compressed data pages. When a page is compressed, we store both the page's offset and its length in the LAF before writing it to disk. Figure~\ref{fig:laf} shows a data file consists of $n$ compressed pages and its corresponding LAF. The number of entries in the LAF equals the number of pages in the data file, where each entry (e.g., $e_0$) stores the size and the offset of its corresponding compressed data page (e.g., $Page_0$). LAF entries can occupy more than one page, depending on the number of pages in the data file. Therefore, to access a data page, we need first to read the LAF page that contains the required data page's size and offset and then use them to access the compressed data page. This may require AsterixDB to perform an extra IO operation to read a data page. However, the number of LAF pages is usually small due to the fact that the entry size is small (12-bytes in our implementation). For instance, a 128KB LAF page can store up to 10,922 entries. Thus, LAF pages can be easily cached and read multiple times.

In addition to this ``syntactic'' approach based on compression, the next section introduces a ``semantic'' approach to reducing the storage overhead by inferring and stripping the schema out of self-describing records in AsterixDB. In Section~\ref{sec:exper}, we evaluate both approaches (syntactic and semantic) when they are applied separately and when they are combined and show their impact on storage size, data ingestion rate, and query performance.


\section{LSM-based Schema Inference and Tuple Compaction Framework}
\begin{figure}[h]
    \centering
    \begin{subfigure}[b]{0.22\textwidth}
        \includegraphics[width=\textwidth]{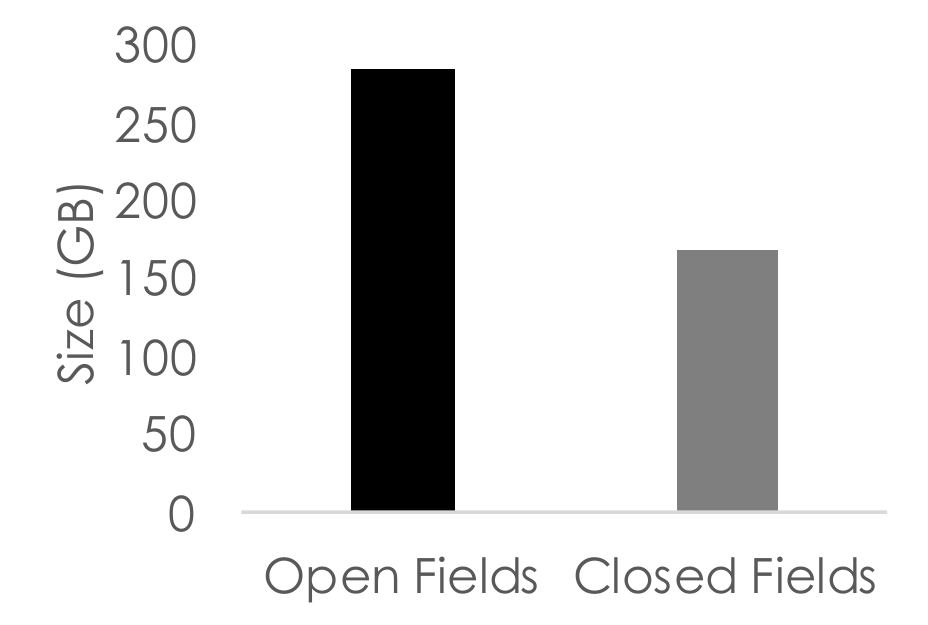}
        \caption{On-disk storage size}
        \label{fig:bigFun-storage}
    \end{subfigure}
    \begin{subfigure}[b]{0.2501\textwidth}
        \includegraphics[width=\textwidth]{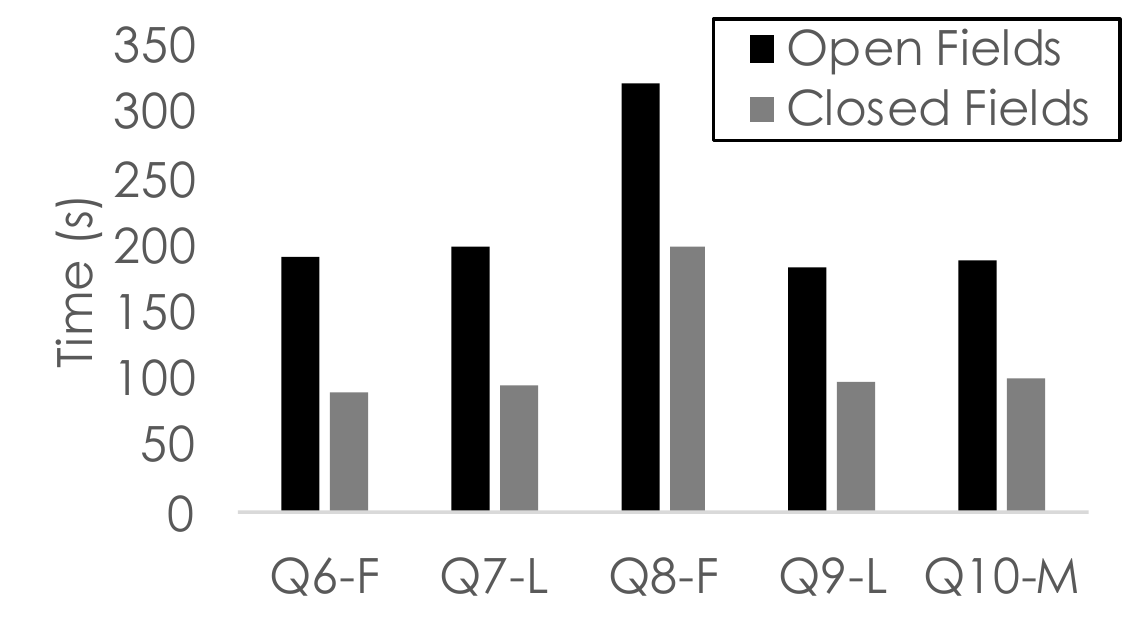}
            \vspace{-1.em}
        \caption{Query execution time}
        \label{fig:bigFun-query-time}
    \end{subfigure}
    \vspace{-1.em}
    \caption{Summary of the findings in~\protect\cite{bigFUN}}
    \label{fig:bigFUN}
        \vspace{-0.2cm}
\end{figure}

The flexibility of schema-less NoSQL systems attracts applications where the schema can change without declaring those changes. However, this flexibility is not free. In the context of AsterixDB, Pirzadeh et al.~\cite{bigFUN} explored query execution performance when all the fields are declared (\textit{closed type}) and when they are left undeclared (\textit{open type}). One conclusion from their findings, summarized in Figure~\ref{fig:bigFUN}, is that queries with non-selective predicates (using secondary indexes) and scan queries took twice as much time to execute against \textit{open} type records compared to \textit{closed} type records due to their storage overhead.

In this section, we present a tuple compactor framework (called the ``tuple compactor'' hereafter) that addresses the storage overhead of storing self-describing semi-structured records in the context of AsterixDB. The tuple compactor automatically infers the schema of such records and stores them in a compacted form without sacrificing the user experience of schema-less document stores. Throughout this section, we run an example of ingesting and querying data in the $Employee$ dataset declared as shown in Figure~\ref{fig:empddl_schemaless}. The \textit{Employee} dataset here is declared with a configuration parameter | \lstinline[language=AQLSchema,basicstyle=\noindent\small]|{"tuple-compactor-enabled": true}| | which enables the tuple compactor.

\begin{figure}[h]
        \vspace{-0.2cm}
    \begin{aqlschema}
  CREATE TYPE EmployeeType AS OPEN { id: int };
    
  CREATE DATASET Employee(EmployeeType)
  PRIMARY KEY id WITH {"tuple-compactor-enabled": true};
    \end{aqlschema}
    \vspace{-0.4cm}
    \caption{Enabling the tuple compactor for a dataset}
    \label{fig:empddl_schemaless}
        \vspace{-0.2cm}
\end{figure}

We present our implementation of the tuple compactor by first showing the workflow of inferring schema and compacting records during data ingestion and the implications of crash recovery in Section~\ref{sec:tuple-compactor}. In Section~\ref{sec:schema-struct}, we show the structure of an inferred schema and a way of maintaining it on update and delete operations. Then, in Section~\ref{sec:compacted-record}, we introduce a physical format for self-describing records that is optimized for the tuple compactor operations (schema inference and record compaction). Finally, in Section~\ref{sec:query_prcoessing}, we address the challenges of querying compacted records stored in distributed partitions of an AsterixDB cluster.

\subsection{Tuple Compactor Workflow}
\label{sec:tuple-compactor}
We first discuss the tuple compactor workflow during normal operation of data ingestion and during crash recovery.

\subsubsection{Data Ingestion}
\begin{figure}[h]
    \hspace{-0.55cm}
    \begin{subfigure}[b]{0.5\textwidth}
        \includegraphics[width=\textwidth]{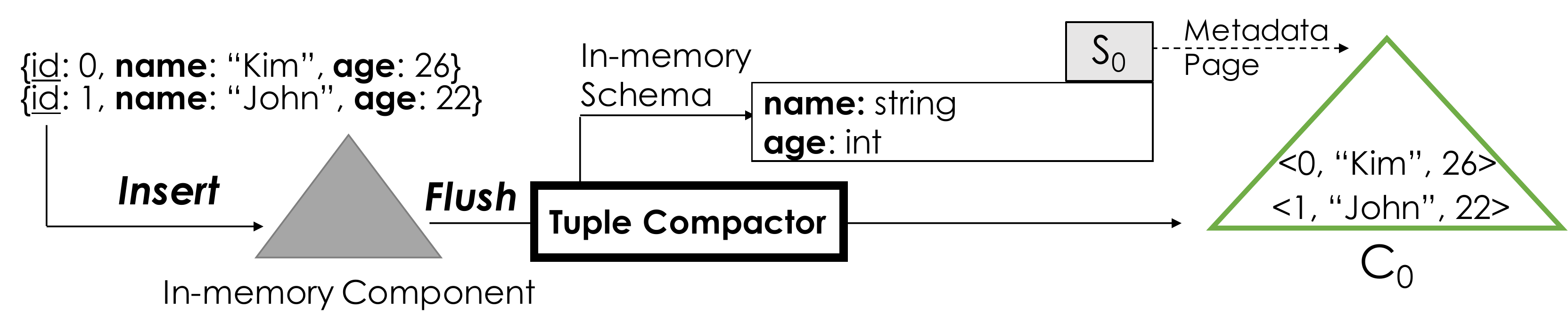}
        \vspace{-0.5cm}
        \caption{}
        \label{fig:schema-first}
    \end{subfigure}
    \hspace{-0.55cm}
    \begin{subfigure}[b]{0.5\textwidth}
        \includegraphics[width=\textwidth]{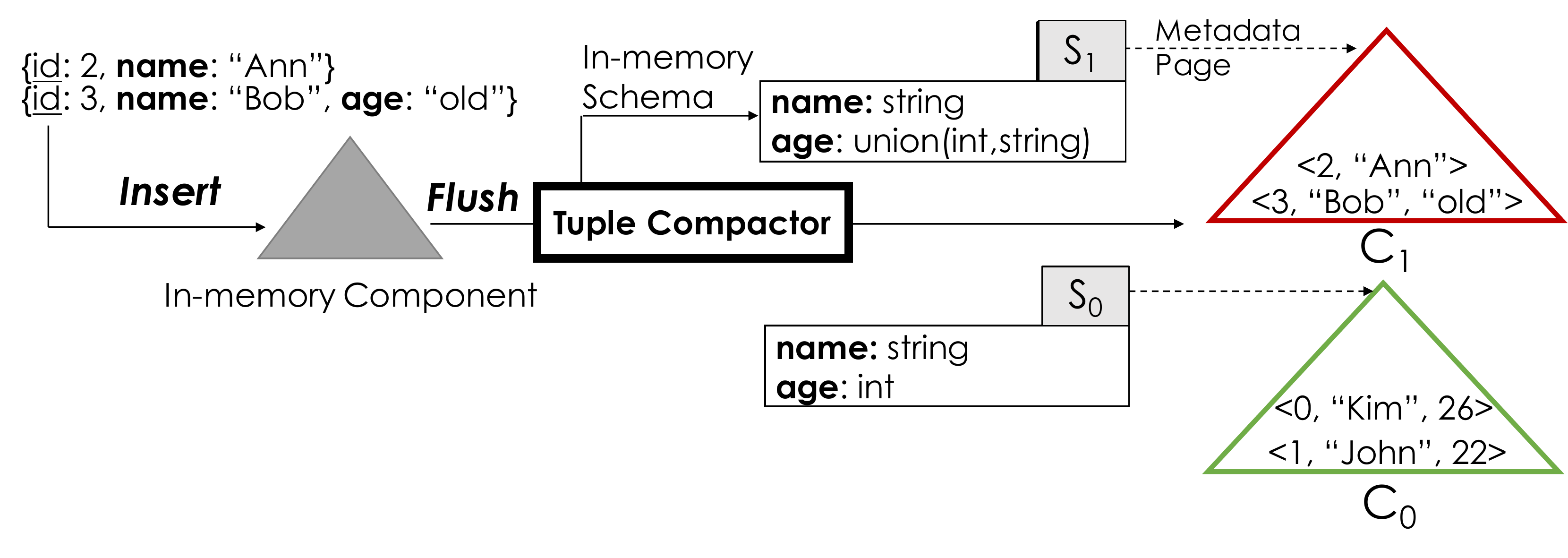}
        \vspace{-0.5cm}
        \caption{}
        \label{fig:schema-second}
    \end{subfigure}
    \hspace{-0.55cm}
    \begin{subfigure}[b]{0.5\textwidth}
        \includegraphics[width=\textwidth]{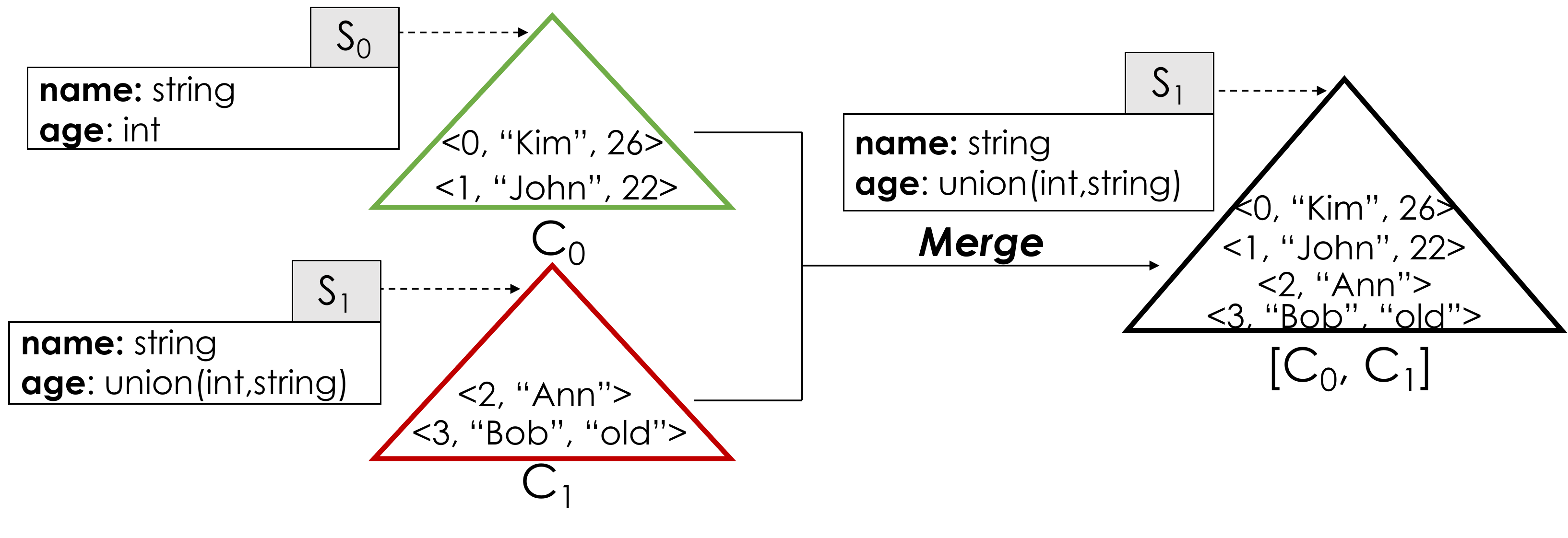}
                \vspace{-0.5cm}
        \caption{}
        \label{fig:schema-merge}
    \end{subfigure}
            \vspace{-0.5cm}
    \caption{\textbf{(a)} Flushing the first component $C_0$ \textbf{(b)} Flushing the second component $C_1$ \textbf{(c)} Merging the two components $C_0$ and $C_1$ into the new component $[C_0, C_1]$}
    \label{fig:schema-inference}
    \vspace{-.3cm}
\end{figure}

When creating the $Employee$ dataset (shown in Figure~\ref{fig:empddl_schemaless}) in the AsterixDB cluster illustrated in Figure~\ref{fig:asterix_arch}, each partition in every NC starts with an empty dataset and an empty schema. During data ingestion, newly incoming records are hash-partitioned on the primary keys ($id$ in our example) across all the configured partitions (Partition 0 to Partition 5 in our example). Each partition inserts the received records into the dataset's in-memory component until it cannot hold any new record. Then, the tree manager schedules a flush operation on the full in-memory component. During the flush operation, the tuple compactor, as shown in the example in Figure~\ref{fig:schema-first}, factors the schema information out of each record and builds a traversable in-memory structure that holds the schema (described in Section \ref{sec:schema-struct}). At the same time, the flushed records are written into the on-disk component $C_0$ in a compacted form where their schema information (such as field names) are stripped out and stored in the schema structure. After inserting the last record into the on-disk component $C_0$, the inferred schema $S_0$ in our example describes two fields $name$ and $age$ with their associated types denoted as $FieldName: Type$ pairs. Note that we do not store the schema information of any explicitly declared fields (field $id$ in this example) as they are stored in the Metadata Node (Section~\ref{sec:asterix_storage_ingestion}). At the end of the flush operation, the component's inferred in-memory schema is persisted in the component's Metadata Page before setting the component as $VALID$ (Section~\ref{sec:asterix_storage_ingestion}). Once persisted, on-disk schemas are immutable.

As more records are ingested by the system, new fields may appear or fields may change, and the newly inferred schema has to incorporate the new changes. The newly inferred schema will be a super-set (or union) of all the previously inferred schemas. To illustrate, during the second flush of the in-memory component to the on-disk component $C_1$ in Figure~\ref{fig:schema-second}, the records of the new in-memory component, with \textit{id} 2 and 3, have their \textit{age} values as \textit{missing} and \textit{string}, respectively. As a result, the tuple compactor changes the type of the inferred age field in the in-memory schema from $int$ to $union(int, string)$, which describes the records' fields for both components $C_0$ and $C_1$. Finally, $C_1$ persists the latest in-memory schema $S_1$ into its metadata~page.

Given that the newest schema is always a super-set of the previous schemas, during a merge operation, we only need to store the most recent schema of all the mergeable components as it covers the fields of all the previously flushed components. For instance, Figure~\ref{fig:schema-merge} shows that the resulting on-disk component $[C_0, C_1]$ of the merged components $C_0$ and $C_1$ needs only to store the schema $S_1$ as it is the most recent schema of $\{S_0, S_1\}$. Note that the merge operation does not need to access the in-memory schema and, hence, merge and flush operations can be performed concurrently without synchronization.

We chose to ignore compacting records of the in-memory component because (i) the in-memory component size is relatively small compared to the total size of the on-disk components, so any storage savings will be negligible, and (ii) maintaining the schema for in-memory component, which permits concurrent modifications (inserts, deletes and updates), would complicate the tuple compactor's workflow and slow down the ingestion rate.

\subsubsection{Crash Recovery Implications}
Apache AsterixDB's LSM-based engine guarantees that (i) records of in-memory components during the flush operation are immutable and (ii) flush operations of in-memory components are atomic (no half-flushed components are allowed). Those guarantees make the flush operation suitable for applying transformations on the flushed records before writing them to disk. The tuple compactor takes this opportunity to infer the schema during the flush operation as (i) guarantees that records cannot be modified. At the same time, the tuple compactor transforms the flushed records into a compacted form and ensures that the compaction process is guaranteed to be atomic as in (ii).

Now, let us consider the case where a system crash occurs during the second flush shown in Figure~\ref{fig:schema-second}. When the system restarts, the recovery manager will start by activating the dataset and then inspecting the validity of the on-disk components by checking their validity bits. The recovery manager will discover that $C_1$ is not valid and remove it. As $C_0$ is the ``newest'' valid flushed component, the recovery manager will read and load the schema $S_0$ into memory. Then, the recovery manager will replay the log records to restore the state of the in-memory component before the crash. Finally, the recovery manager will flush the restored in-memory component to disk as $C_1$, during which time the tuple compactor operates normally.

\subsection{Schema Structure}
\label{sec:schema-struct}
Previously, we showed the flow of inferring the schema and compacting the tuples during data ingestion. In this section, we focus on the inferred schema and present its structure. We also address the issue of maintaining the schema in case of delete and update operations, which may result in removing inferred fields or changing their types.

\subsubsection{Schema Structure Components}
\begin{figure*}[h]
    \centering
    \begin{subfigure}[b]{0.36\textwidth}
        \includegraphics[width=\textwidth]{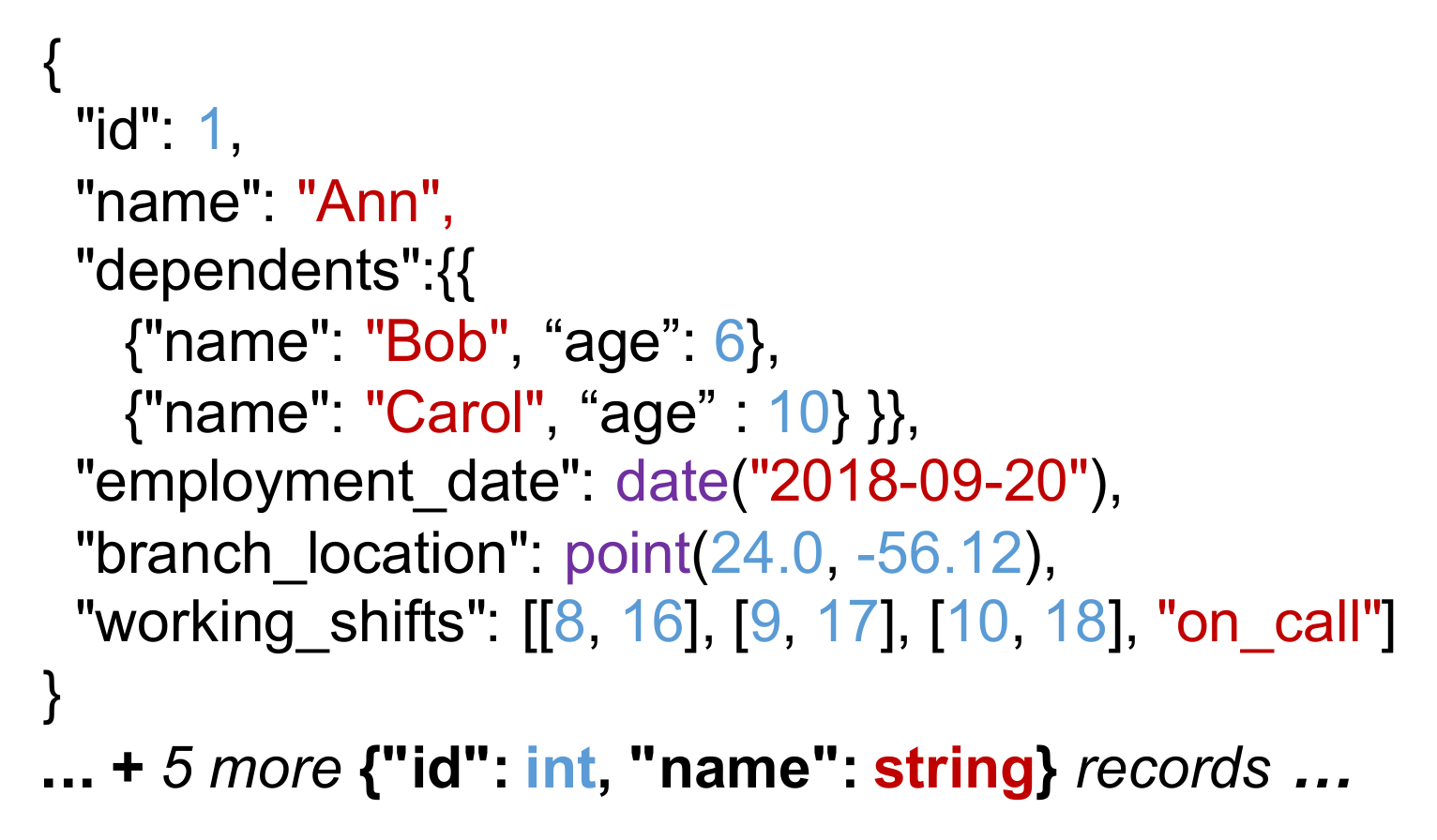}
        \caption{}
        \label{fig:schema-struct-1}
    \end{subfigure}
    \hspace{-0.2cm}
    \begin{subfigure}[b]{0.34\textwidth}
        \includegraphics[width=\textwidth]{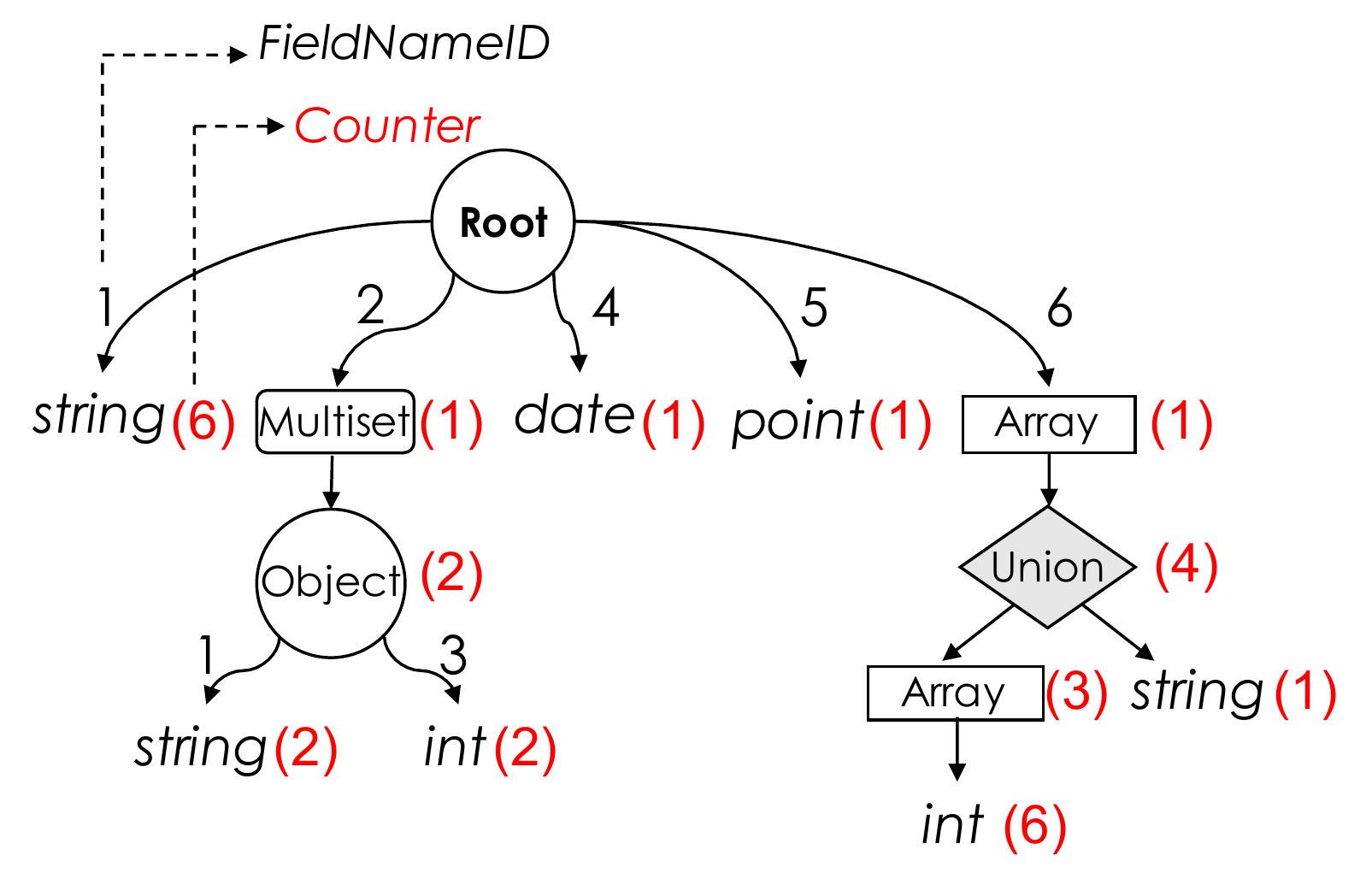}
        \caption{}
        \label{fig:schema-struct-2}
    \end{subfigure}
    \hspace{0.4cm}
    \begin{subfigure}[b]{0.20\textwidth}
        \includegraphics[width=\textwidth]{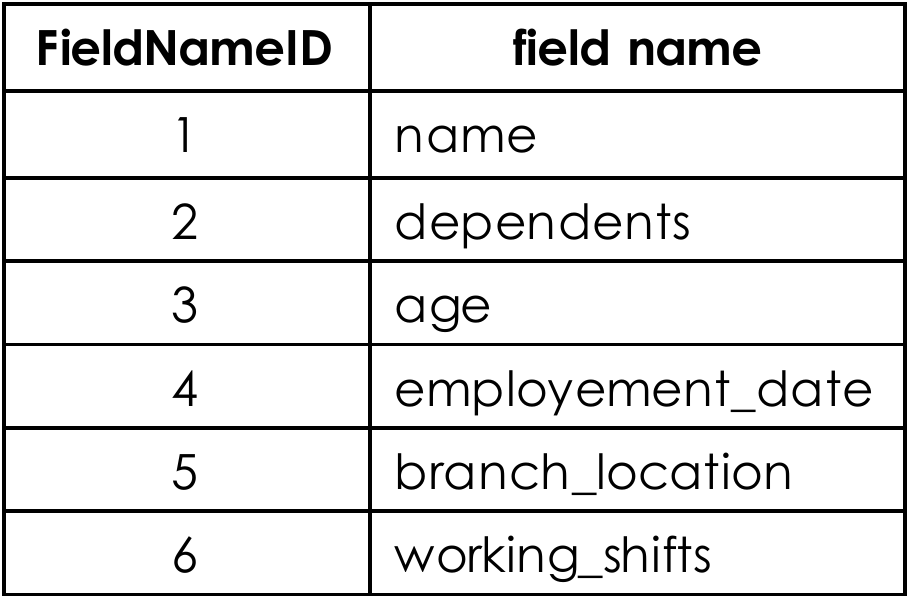}
        \caption{}
        \label{fig:schema-struct-3}
    \end{subfigure}
        \vspace{-0.2cm}
    \caption{\textbf{(a)} An ADM record \textbf{(b)} Inferred schema tree structure \textbf{(c)} Dictionary-encoded field names}
    \label{fig:schmea-strcut}
    \vspace{-0.4cm}
\end{figure*}

Semi-structured records in document store systems are represented as a tree where the inner nodes of the tree represent nested values (e.g., JSON objects or arrays) and the leaf nodes represent scalar values (e.g., strings). ADM records in AsterixDB also are represented similarly. Let us consider the example where the tuple compactor first receives the ADM record shown in Figure~\ref{fig:schema-struct-1} during a flush operation followed by five other records that have the structure \lstinline[language=AQLSchema,basicstyle=\noindent\small]|{"id": int, "name": string}|. The tuple compactor traverses the six records and constructs: (i) a tree-structure that summarizes the records structure, shown in Figure~\ref{fig:schema-struct-2}, and (ii) a dictionary that encodes the inferred field names strings into \textit{FieldNameID}s, as shown in Figure~\ref{fig:schema-struct-3}. The \textit{Counter} in the schema tree-structure represents the number of occurrences of a value, which we further explain in Section~\ref{sec:schema_maintainance}.

The schema tree structure starts with the root object node which has the fields at the first level of the record (\textit{name}, \textit{dependents}, \textit{employment\_date}, \textit{branch\_location}, and \textit{working\_shifts}). We do not store any information here about the dataset's declared field $id$ as explained in previously in Section~\ref{sec:tuple-compactor}. Each inner node (e.g., \textit{dependents}) represents a nested value (object, array, or multiset) and the leaf nodes (e.g., \textit{name}) represent the scalar (or primitive) values. Union nodes are for object fields or collection (array and multiset) items if their values can be of different types. In this example, the tuple compactor infers the array item type of the field \textit{working\_shifts} as a union type of an array of integers and a string. 

The edges between the nodes in the schema tree structure represent the nested structure of an ADM record. Each inner node of a nested value in the schema tree structure can have one or more children depending on the type of the inner node. Children of object nodes (e.g., fields of the \textit{Root} object) are accessed by \textit{FieldNameID}s (shown as integers on the edges of object nodes in Figure~\ref{fig:schema-struct-2}) that reference the stored field names in the dictionary shown in Figure~\ref{fig:schema-struct-3}. Each field name (or $FieldNameID$ in the schema tree structure) of an object is unique, i.e., no two children of an object node share the same field name. However, children of different object nodes can share the same field name. Therefore, storing field names in a dictionary allows us to canonicalize repeated field names such as the field name \textit{name}, which has appeared twice in the ADM record shown in Figure~\ref{fig:schema-struct-1}. A collection node (e.g., \textit{dependents}) have only one child, which represents the items' type. An object field or a collection item can be of heterogeneous value types. So, their types may be inferred as a union of different value types. In a schema tree structure, the number of children a union node can have depends on the number of supported value types in the system. For instance, AsterixDB has 27 different value types~\cite{asterix-doc}. Hence, a union node could have up to 27 children.

\subsubsection{Schema Structure Maintenance}
\label{sec:schema_maintainance}
In Section~\ref{sec:tuple-compactor} we described the flow involved in inferring the schema of newly ingested records, where we ``add'' more information to the schema structure. However, when deleting or updating records, the schema structure might need to be changed by ``removing'' information. For example, the record with \textit{id} 3 shown in Figure~\ref{fig:schema-inference} is the only record that has an $age$ field of type $string$. Therefore, deleting this record should result in changing the type of the field $age$ from $union(int, string)$ to $int$ as the dataset no longer has the field $age$ as a string. From this example, we see that on delete operations, we need to (i) know the number of appearances of each value, and (ii) acquire the old schema of a deleted or updated record.

\begin{figure}[h]
    \centering
    \includegraphics[width=0.30\textwidth]{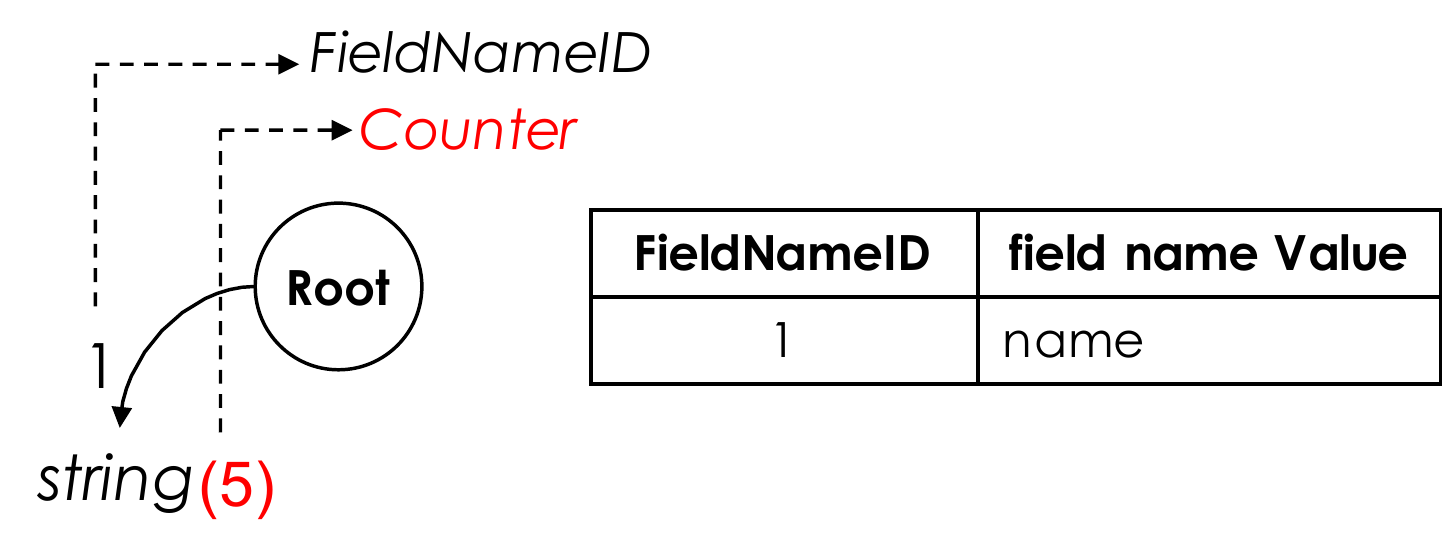}
    \caption{After deleting the record shown in Figure\ref{fig:schema-struct-1}}
    \label{fig:schema_after_delete}
    \vspace{-0.5cm}
\end{figure}

During the schema inference process, the tuple compactor counts the number of appearances of each value and stores it in the schema tree structure's nodes. In Figure~\ref{fig:schema-struct-2}, each node has a $Counter$ value that represents the number of times the tuple compactor has seen this node during the schema inference process. From the schema tree structure in Figure~\ref{fig:schema-struct-2}, we can see that there are six records that have the field \textit{name}, including the record shown in Figure~\ref{fig:schema-struct-1}. Also, we can infer from the schema structure that all fields other than \textit{name} belong to the record shown in Figure~\ref{fig:schema-struct-1}. Therefore, after deleting this record, the schema structure should only have the field $name$ as shown in Figure~\ref{fig:schema_after_delete}.

On delete, AsterixDB performs a point lookup to get the old record from which the tuple compactor extracts its schema (we call the schema of a deleted record the ``\textit{anti-schema}''). Then, it constructs an anti-matter entry that includes the primary key of the deleted record and its \textit{anti-schema} and then inserts it into the in-memory component. During the flush operation, the tuple compactor processes the anti-schema by traversing it and decrementing the $Counter$ of each node of the schema tree structure. When the counter's value of a node in the schema tree structure reaches zero, we know that there is no record that still has this value (whether it is nested or a primitive value). Then, the tuple compactor can safely delete the node from the schema structure. As shown in Figure~\ref{fig:schema_after_delete}, after deleting the record in Figure~\ref{fig:schema-struct-1}, the counter value corresponding to the field $name$ is decremented from 6 to 5 whereas the other nodes of the schema structure (shown in Figure~\ref{fig:schema-struct-1}) have been deleted as they were unique to the deleted record. After processing the anti-schema, the tuple compactors discard it before writing the anti-matter entry to disk. Upserts can be performed as deletes followed by inserts.

It is important to note that performing point lookups for maintenance purposes is not unique to the schema structure. For instance, AsterixDB performs point lookups to maintain secondary indexes \cite{storage} and LSM filters~\cite{lsm-filter}. Luo et al. \cite{luo2018lsm, luo2019efficient} showed that performing point lookups for every upsert operation can degrade data ingestion performance. More specifically, checking for key-existence for every upserted record is expensive, especially in cases where the keys are mostly new. As a solution, a primary key index, which stores primary keys only, can be used to check for key-existence instead of using the larger primary index. In the context of retrieving the anti-schema on upsert, one can first check if a key exists by performing a point lookup using the primary key index. Only if the key exists, an additional point lookup is performed on the primary index to get the anti-schema of the upserted record. If the key does not yet exist (new key), the record can be inserted as a new record. In Section~\ref{sec:exper}, we evaluate the data ingestion performance of our tuple compactor under heavy updates using the suggested primary key index.


\subsection{Compacted Record Format}
\label{sec:compacted-record}
Since the tuple compactor operates during data ingestion, the process of inferring the schema and compacting the records needs to be efficient and should not degrade the ingestion rate. As the schema can change significantly over time, previously ingested records must not be affected or require updates. Additionally, sparse records should not need to store additional information about missing values such as null bitmaps in RDBMSs' records. For example, storing the record \verb|{"id": 5, "name": "Will"}| with the schema shown in Figure~\ref{fig:schema-struct-2} should not include any information about other fields (e.g., dependents). \rev{Moreover, uncompacted records (in-memory components) and compacted records (on-disk components) should be evaluated and processed using the same evaluation functions to avoid any complexities when generating a query plan}. To address those issues, we introduce a compaction-friendly physical record data format into AsterixDB, called the vector-based~format.


\subsubsection{Vector-based Physical Data Format}
\label{sec:vector-based-format}
\begin{figure}[h]
	\centering
	\includegraphics[width=0.47\textwidth]{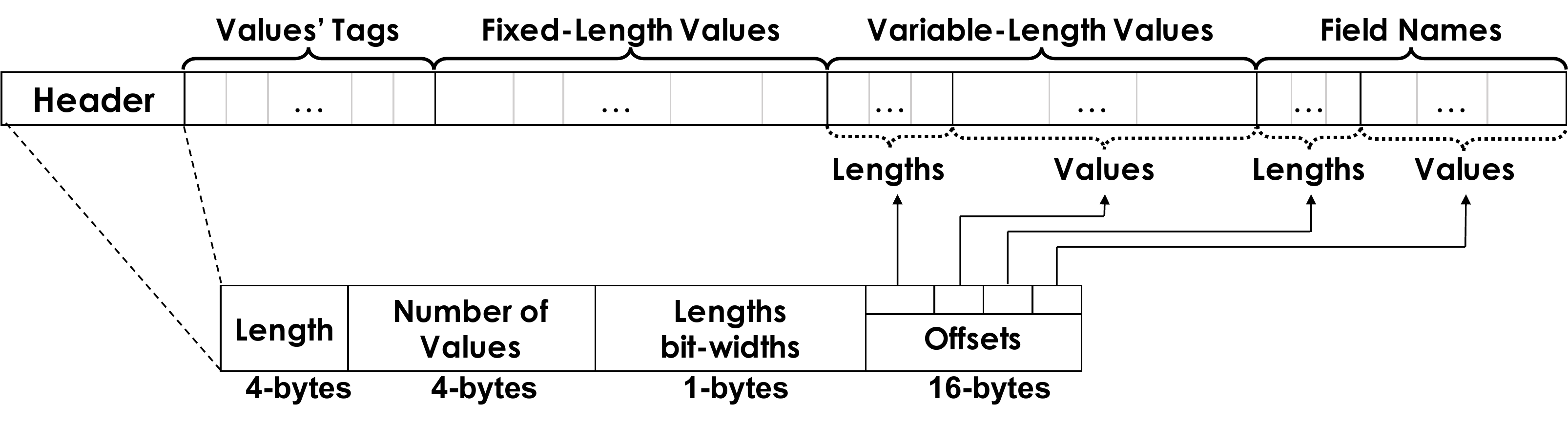}
    \caption{The structure of the vector-based format.}
    \label{fig:vector_based_format_structure}
        \vspace{-0.5cm}
\end{figure}

The main idea of the vector-based format is that it separates the metadata and values of self-describing records into vectors that allow us to manipulate the record's metadata efficiently during the schema inference and record compaction processes. \rev{To not be confused with a columnar format, the vectors are stored within each record and the records are stored contiguously in the primary index (Section~\ref{sec:asterix_storage_ingestion})}. Figure~\ref{fig:vector_based_format_structure} depicts the structure of a record in the vector-based format. First comes the record's header, which contains information about the record such as its length. Next comes the values' tags vector, which enumerates the types of the stored primitive and nested values. Fixed-length primitive (or scalar) values such as integers are stored in the fixed-length values vector. The next vector is split into two sub-vectors, where the first stores lengths and the second stores the actual values of variable-length values. Lastly, the field names sub-vectors (lengths and values) store field name information for all objects' fields in the record.


Figure \ref{fig:vector_based_format_example} shows an example of a record in the vector-based format (See Appendix~\ref{appen:example2} for an additional example). The record has four fields: \textit{id}, \textit{name}, \textit{salaries}, and \textit{age} with the types integer, string, array of integers and integer, respectively. Starting with the header, we see that the record's total size is 73-bytes and there are nine tags in the values' type tags vector. Lengths for variable-length values and field names are stored using the minimum amount of bytes. In our example, the maximum lengths of the variable-length values and field names are 3 (Ann) and 8 (salaries), respectively. Thus, we need at most 3-bits and 5-bits to store the length of each variable-length value or field name, respectively. We only actually need 4-bits for field name lengths; however, the extra bit is used to distinguish inferred fields (e.g.,~name) from declared ones (e.g.,~id) as we explain~next.

\begin{figure}[h]
	\centering
	\includegraphics[width=0.47\textwidth]{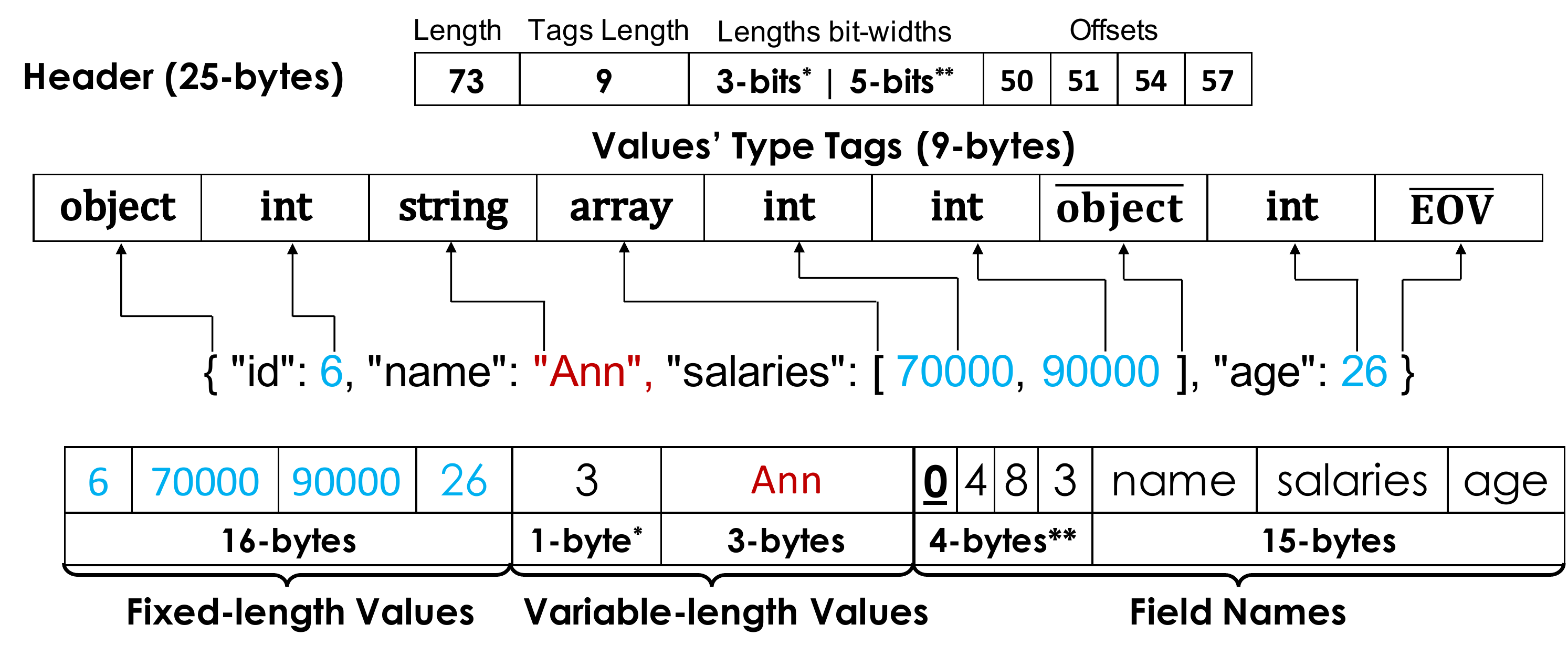}
    \caption{An example record in the vector-based format}
    \label{fig:vector_based_format_example}

\end{figure}

After the header, we store the values' type tags. The values' type tags encode the tree structure of the record in a \textit{Depth-First-Search} order. In this example, the record starts with an object type to indicate the root's type. The first value of the root object is of type integer, and it is stored in the first four bytes of the fixed-length values. Since an object tag precedes the integer tag, this value is a child of that object (root) and, hence, the first field name corresponds to it. Since the field \textit{id} is a declared field, we only store its index (as provided by the metadata node) in the lengths sub-vector. We distinguish index values from length values by inspecting the first bit. If set, we know the length value is an index value of a declared field. The next value in the example record is of type string, which is the first variable-length value in the record. The string value is stored in the variable-length values' vector with its length. Similar to the previous integer value, this string value is also a child of the root and its field name (\textit{name}) is next in the field names' vector. As the field \textit{name} is not declared, the record stores both the name of the field and its length. After the string value, we have the array tag of the field \textit{salaries}. The subsequent integers' tags indicate the array items' types. The array items do not correspond to any field name, and their integer values are stored in the fixed-length values' vector. After the last item of the array, we store a control tag $\overline{object}$ to indicate the end of the array as the current nesting type and a return to the parent nesting type (object type in this example). Hence, the subsequent integer value (age) is again a child of the root object type. At the end of the value's tags, we store a control tag $\overline{EOV}$ to mark the end of the values of the record.

As can be inferred from the previous example, the complexity of accessing a value in the vector-based format is linear in the number of tags, which is inferior to the logarithmic time provided by some traditional formats~\cite{asterixdb-serialization,bson}. We address this issue in more detail in Section~\ref{sec:process-vector-based}.

\subsubsection{Schema Inference and Tuple Compaction}
\label{sec:schema_infer_and_compact_vb}
Records in vector-based format separate values from metadata. The example shown in Figure~\ref{fig:vector_based_format_example} illustrates how the fixed-length and variable-length values are separated from the record's nested structure (values' types tags) and field names. When inferring the schema, the tuple compactor needs only to scan the values' type tags and the field names' vectors to build the schema structure. 

\begin{figure}[h]
	\centering
	\includegraphics[width=0.47\textwidth]{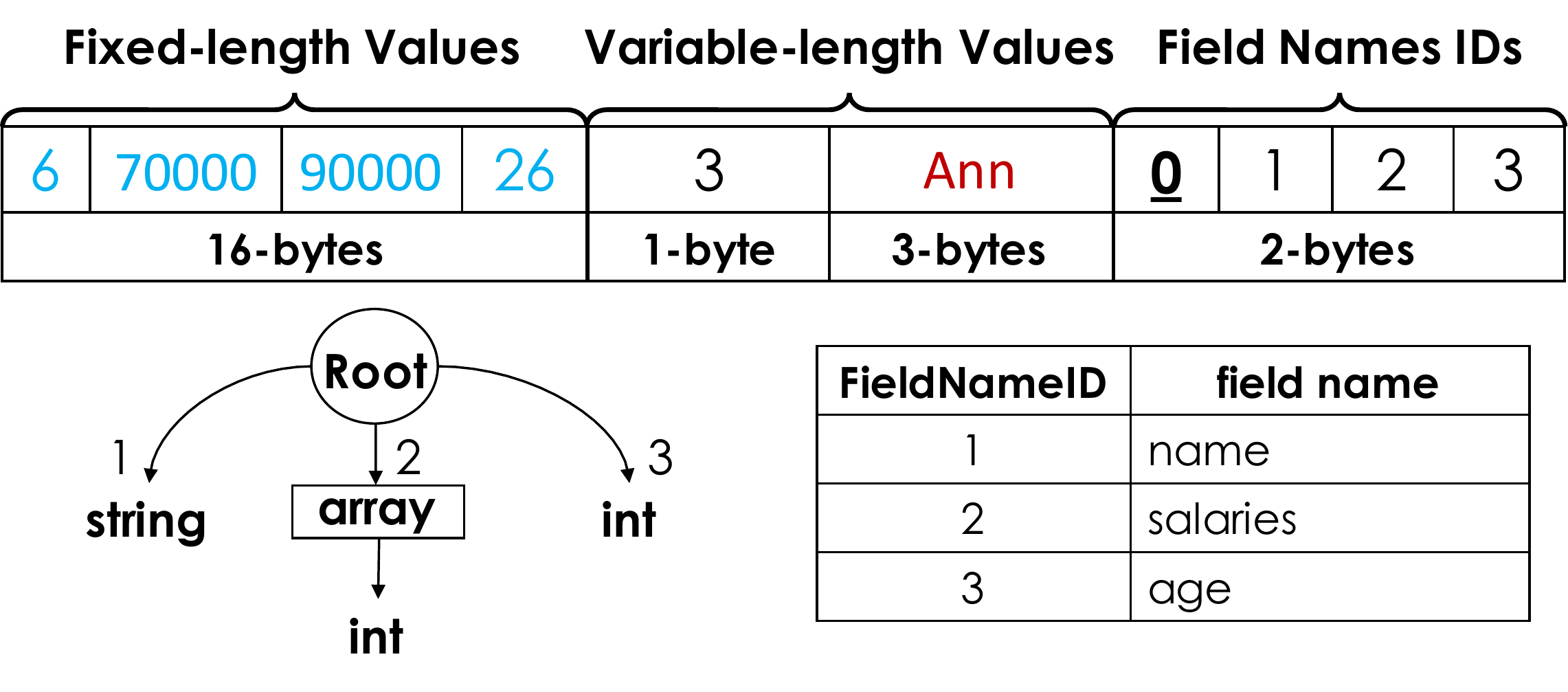}
	    \vspace{-0.3cm}
    \caption{The record in Figure \ref{fig:vector_based_format_example} after compaction}
    \label{fig:tuple_compaion_example}
    \vspace{-0.4cm}
\end{figure}

Compacting vector-based records is a straightforward process. Figure \ref{fig:tuple_compaion_example} shows the compacted structure of the record in Figure \ref{fig:vector_based_format_example} along with its schema structure after the compaction process. The compaction process simply replaces the field names string values with their corresponding FieldNameIDs after inferring the schema. It then sets the fourth offset to the field names' values sub-vector in the header (Figure \ref{fig:vector_based_format_structure}) to zero to indicate that field names were removed and stored in the schema structure. As shown in the example in Figure \ref{fig:tuple_compaion_example}, the record after the compaction needs just two bytes to store the field names' information, where each FieldNameID takes three bits (one bit for distinguishing declared fields and two for field name IDs), as compared to the 19 (4+15) bytes in the uncompacted form in Figure~\ref{fig:vector_based_format_example}.  

\subsection{Query Processing}
\label{sec:query_prcoessing}
In this section, we explain our approach of querying compacted records in the vector-based format. We, first, show the challenges of having distributed schemas in different partitions and propose a solution that addresses this issue. Next, we zoom in into each query executor and show the optimizations needed to process compacted records.
 
\subsubsection{Handling Heterogeneous Schemas}
\label{sec:hetro-schema}
\begin{figure}[h]
	\centering
	\includegraphics[width=0.49\textwidth]{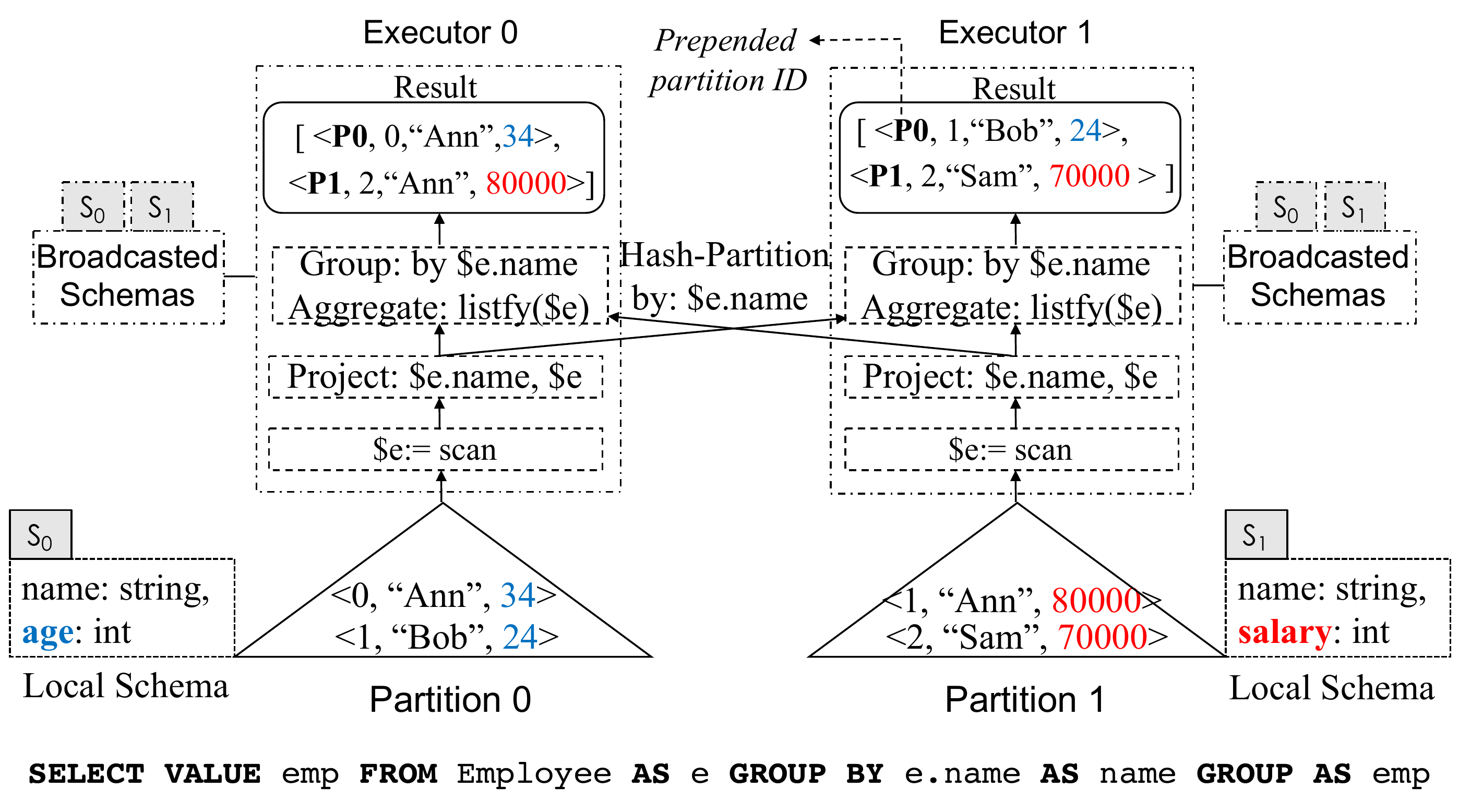}
    \caption{Two partitions with two different schemas}
    \label{fig:schema_distribution}
\end{figure}

As a scalability requirement, the tuple compaction framework operates in each partition without any coordination with other partitions. Therefore, the schema in each partition can be different from other schemas in other partitions. When a query is submitted, each distributed partition executes the same job. Having different schemas becomes an issue when the requested query needs to repartition the data to perform a join or group-by. To illustrate, suppose we have two partitions for the same dataset but with two different inferred schemas, as shown in Figure~\ref{fig:schema_distribution}. We see that the schemas in both partitions have the field \textit{name} of type string. However, the second field is \textit{age} in partition 0 and \textit{salary} in partition 1. After hash-partitioning the records by the \textit{name} value, the resulting records are shuffled between the two query executors and the last field can be either \textit{age} or \textit{salary}. Recall that partitions can be in different machines within the AsterixDB cluster and have no runtime access to the schema information of other partitions. Consequently, query executors cannot readily determine whether the last field corresponds to \textit{age} or \textit{salary}.


To solve the schema heterogeneity issue, we added functionality to broadcast the schema information of each partition to all nodes in the cluster at the beginning of a query's execution. Each node receives each partition's schema information along with its partition ID and serves the schemas to each executor in the same node. Then, \rev{we prepend each record resulting from the scan operator with the source partition ID}. When an operator accesses a field, the operator uses both the prepended partition ID of the record and the distributed schema to perform the field access. Broadcasting the partitions' schemas can be expensive, especially in clusters with a large number of nodes. Therefore, we only broadcast the schemas when the query plan contains a non-local exchange operator such as the hash-partition-exchange in our example in Figure~\ref{fig:schema_distribution}. \rev{When comparing the schema broadcasting mechanism to handling self-describing records, a broadcasted schema represents a \textit{batch of records}, whereas the redundant schemas embedded in self-describing records are carried through the operators on a \textit{record-by-record basis}. Thus, transmitting the schema once per partition instead of once per record is more efficient.} 

\subsubsection{Processing Compacted Records}
\label{sec:process-vector-based}
One notable difference between the vector-based format and the ADM physical format is the time complexity of accessing a value (as discussed in Section~\ref{sec:vector-based-format}). 
The AsterixDB query optimizer can move field access expressions within the plan when doing so is advantageous. For instance, AsterixDB's query optimizer inlines field access expressions with WHERE clause conjunct expressions as in:

\vspace{1mm}
\centerline {\noindent $emp.age > 25~AND~emp.name = ``Ann"$}
\vspace{1mm}

The inlined field access expression $emp.name$ is evaluated only if the expression $emp.age > 25$ is true. However, in the vector-based format, each field access requires a linear scan on the record's vectors, which could be expensive. To minimize the cost of scanning the record's vectors, we added one rewrite rule to the AsterixDB query optimizer to consolidate field access expressions into a single function expression. Therefore, the two field access expressions in our example will be written as follows:


\vspace{1mm}
\centerline {\noindent $[\$age, \$name] \gets getValues(emp, ``age", ``name")$}
\vspace{1mm}

\noindent The function $getValues()$ takes a record and path expressions as inputs and outputs the requested values of the provided path expressions. The two output values are assigned to two variables $\$age$ and $\$name$ and the final conjunct expression of our WHERE clause example is transformed as: 

\vspace{1mm}
\centerline {\noindent$\$age > 25 \ AND \ \$name = ``Ann"$}
\vspace{1mm}

The function $getValues()$ is also used for accessing array items by providing the item's index. For example, the expression $emp.dependents[0].name$ is translated as follows:

\vspace{1mm}
\centerline {\noindent $[\$d\_name] \gets getValues(emp, ``dependents", 0, ``name")$}
\vspace{1mm}

\noindent Additionally, we allow ``wildcard'' index to access nested values of all items of an array. For instance, the output of the expression $emp.dependents[*].name$ is an array of all $name$s' values in the array of objects $dependents$.

\section{Experiments}
\label{sec:exper}
In this section, we experimentally evaluate the implementation of our tuple compactor in AsterixDB. In our experiments, we compare our compacted record approach with AsterixDB's current \textit{closed} and \textit{open} records in terms of (i) on-disk storage size after data ingestion, (ii) data ingestion rate, and (iii) the performance of analytical queries.

\vspace{2.mm}
\noindent We also conduct additional experiments to evaluate:
\vspace{-.27cm}
\begin{enumerate}
	\item The performance accessing values in records in vector-based format (Section~\ref{sec:vector-based-format}) with and without the optimization techniques explained in Section~\ref{sec:process-vector-based}.
	\vspace{-.27cm}
	\item The impact of our approach on query performance using secondary indexes.
	\vspace{-.27cm}
	\item The scalability of our framework using computing clusters with different number Amazon EC2 instances.
\end{enumerate}

\vspace{1.mm}
\noindent\textbf{Experiment Setup}
We conducted our initial experiments using a single machine with an 8-core (Intel i9-9900K) processor and 32GB of main memory. The machine is equipped with two storage drive technologies SATA SSD and NVMe SSD, both of which have 1TB of capacity. The SATA SSD drive can deliver up to 550 MB/s for sequential read and 520 MB/s for sequential write, and the NVMe SSD drive can deliver up to 3400 MB/s for sequential read and 2500 MB/s for sequential write. Section~\ref{sec:exper-scale-out} details the setup for our additional scale-out experiments.

We used AsterixDB v9.5.0 after extending it with our tuple compaction framework. We configured AsterixDB with 15GB of total memory, where we allocated 10GB for the buffer cache and 2GB for the in-memory component budget. The remaining 3GB is allocated as temporary buffers for operations such as sort and join.

In Section~\ref{sec:compression}, we introduced our implementation of the page-level compression in AsterixDB. Throughout our experiments, we also evaluate the impact of compression (using Snappy~\cite{snappy} compression scheme) on the storage size, data ingestion rate, and query performance.

 \vspace{2.mm}
\noindent\textbf{Schema Configuration.}
In our experiments, we evaluated the storage size, data ingestion rate, and query performance when defining a dataset as \textbf{(i) open, (ii) closed, and (iii) inferred} using our tuple compactor. For the open and inferred datasets, we only declare the primary key field, whereas in closed datasets, we pre-declare all the fields. The records of open and closed datasets are stored using the ADM physical format, whereas the inferred datasets are using the new vector-based~format. Note that the AsterixDB open case is similar to what schema-less NoSQL systems, like MongoDB and Couchbase, do for storage.

\subsection{Datasets}
In our experiments, we used three datasets (summarized in Table~\ref{tab:datasets-summary}) which have different characteristics in terms of their record's structure, size, and value types. 

Using the first dataset, we want to evaluate ingesting and querying social network data. We obtained a sample of tweets using the Twitter API~\cite{twitter-api}. Due to the daily limit of the number of tweets that one can collect from the Twitter API, we replicated the collected tweets ten times to have 200GB worth of tweets in total. \rev{Replicating the data would not affect the experiment results as (i) the tuple compactor's scope is the records' metadata (not the values) and (ii) the original data is larger than the compressible page size.}

The second dataset we used is the Web of Science (WoS)~\cite{cadre} publication dataset\footnote{We obtained the dataset from Thomson Reuters. Currently, Clarivate Analytics maintains it \cite{clarivate-wos}.}. The WoS dataset encompasses meta information about scientific publications (such as authors, fundings and abstracts) from 1980 to 2016 with a total dataset size of 253GB. We transformed the dataset's record format from its XML original structure to a JSON one using an existing XML-to-JSON converter~\cite{xml-to-json}. The resulting JSON documents contain some fields with heterogeneous types, specifically a union of object and array of objects. The reason behind using such a converter is to mimic the challenges a data scientist can experience when resorting to existing solutions. (Due to a lack of support for declared union types in AsterixDB, we could only pre-declare the fields with homogeneous types in the closed schema case.)
 
To evaluate more numeric Internet of Things (IoT)-like workloads, we generated a third synthetic dataset that mimics data generated by sensors. Each record in the sensors' dataset contains captured readings and their timestamps along with other information that monitors the health status of the sensor. The sensor data contains mostly numerical values and has a larger field-name-size to value-size ratio. The total size of the raw Sensors data is 122GB.

\begin{center}
\begin{table*}[!ht]  
    \vspace{-.3cm}
  \centering
  \begin{tabular}{|l|c|c|c|}
    \cline{2-4}
    \multicolumn{1}{c|}{} & Twitter & WoS & Sensors \\ \hline
    Source & Scaled   & Real-world  & Synthetic  \\ \hline
    Total Size & 200GB & 253GB & 122GB \\ \hline
    \# of Records & 77.6M & 39.4M & 25M \\ \hline
    Record Size & $\sim$2.7KB & $\sim$6.2KB & 5.1KB  \\ \hline
    \# of Scalar val. (min, max, avg) & 53, 208, 88 & 71, 193, 1430 & 248, 248, 248 \\ \hline
    Max. Depth & 8  & 7  & 3 \\ \hline
    Dominant Type & String  & String  & Double \\ \hline
    Union Type? & No  & \textbf{Yes}  & No \\ \hline
  \end{tabular}
    \vspace{-.3cm}
  \caption{Datasets summary}
  \label{tab:datasets-summary}

\end{table*}
\end{center}
\vspace{-0.9cm}
\subsection{Storage Size}
\label{sec:storage-size}
In this experiment, we evaluate the on-disk storage size after ingesting the Twitter, WoS and Sensors datasets into AsterixDB using the three formats (open, closed and inferred) and we compare it with MongoDB's storage size. Our goal of comparing with MongoDB's size is simply to show that the compressed open case is comparable to what other NoSQL systems take for storage using the same compression scheme (Snappy). (It is not the focus of this paper to compare both systems' data ingestion and query performance.)

\begin{figure}[h]
	\centering
    \begin{subfigure}[b]{0.3\textwidth}
        \includegraphics[width=\textwidth]{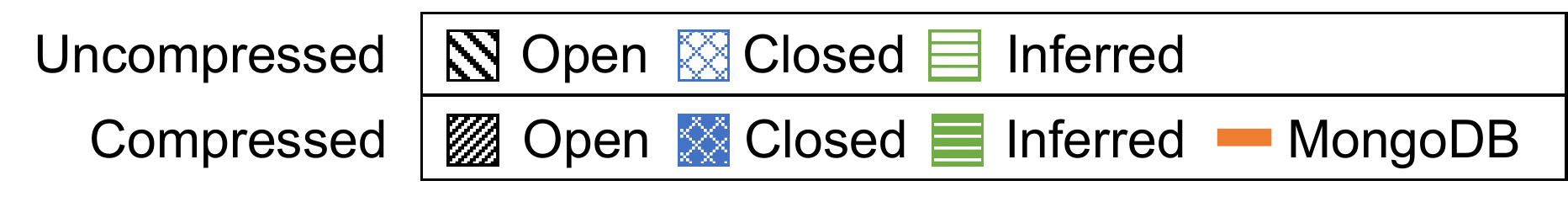}
    \end{subfigure}
    
    \begin{subfigure}[b]{0.159\textwidth}
        \includegraphics[width=\textwidth]{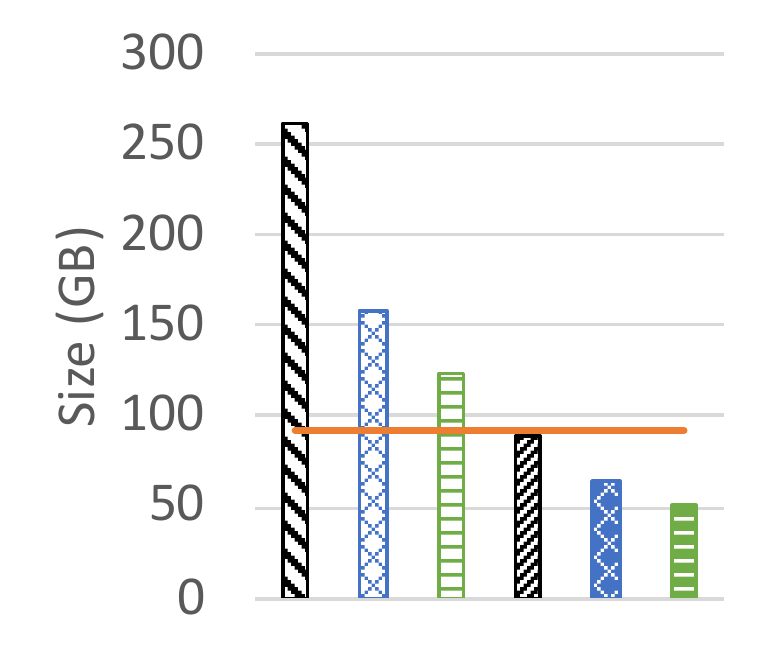}
        \caption{Twitter dataset}
        \label{fig:twitter-storage}
    \end{subfigure}
    \begin{subfigure}[b]{0.153\textwidth}
        \includegraphics[width=\textwidth]{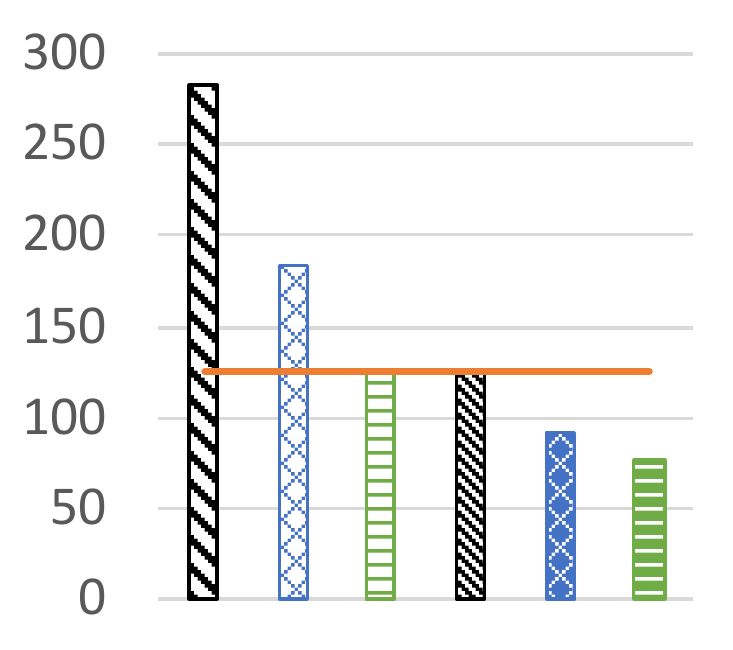}
        \caption{WoS dataset}
        \label{fig:wos-storage}
    \end{subfigure}
    \begin{subfigure}[b]{0.153\textwidth}
        \includegraphics[width=\textwidth]{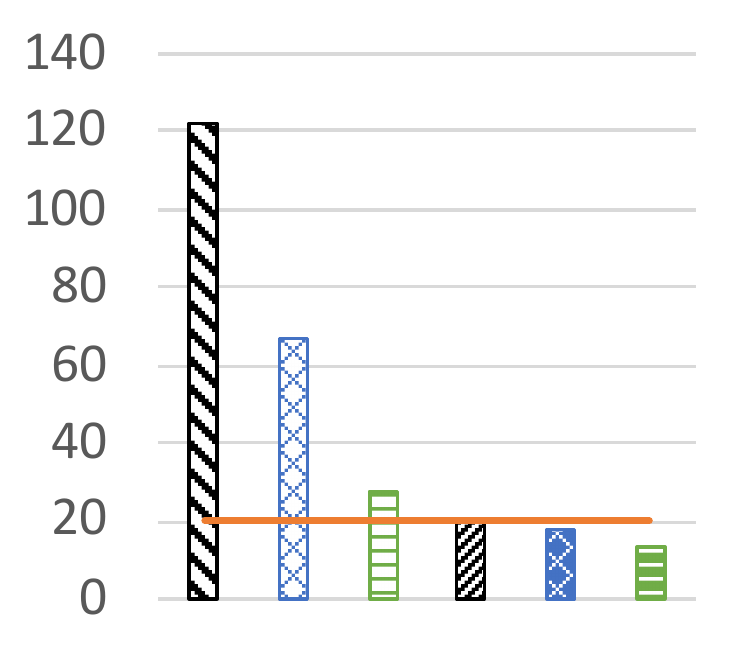}
        \caption{Sensors dataset}
        \label{fig:iot-storage}
    \end{subfigure}

    \caption{On-disk sizes}
    \label{fig:storage-all}
    \vspace{-0.6cm}
\end{figure}

We first evaluate the total on-disk sizes after ingesting the data into the open, closed and inferred datasets. We begin with the Twitter dataset. Figure~\ref{fig:twitter-storage} shows its total on-disk sizes. We see that the inferred and closed schema datasets have lower storage footprints compared to the open schema dataset, as both avoid storing field names in each record. When compression is enabled, both formats still have smaller size compared to the open format and to MongoDB's compressed collection size. The size of the inferred dataset is slightly smaller than the closed schema dataset since the vector-based format does not store offsets for every nested value (as opposed to the ADM physical format in the closed schema dataset).

For the WoS dataset, Figure~\ref{fig:wos-storage} shows that the inferred dataset again has the lowest storage overhead. Even after compression, the open dataset (and the compressed MongoDB collection) had about the same size as the uncompressed inferred dataset. The reason is that WoS dataset structure has more nested values compared with the Twitter dataset. The vector-based format has less overhead for such data, as it does not store the 4-byte offsets for each nested value.

The Sensors dataset contains only numerical values that describe the sensors' status along with their captured readings, so this dataset's field name size to value size ratio is higher compared to the previous datasets. Figure~\ref{fig:iot-storage} shows that, in the uncompressed dataset, the closed and inferred datasets have about 2x and 4.3x less storage overhead, respectively, than the open dataset. The additional savings for the inferred dataset results from eliminating the offsets for readings objects, which contain reading values along with their timestamps | \lstinline[language=AQLSchema,basicstyle=\noindent\small]|{"value": double, "timestamp": bigint}|. Compression reduced the sizes of the open and closed datasets by a factor of 6.2 and 3.8, respectively, as compared to their uncompressed counterparts. For the inferred dataset, compression reduced its size only by a factor of 2.1. This indicates that both the open and closed dataset records incurred higher storage overhead from storing redundant offsets for nested fixed-length values (readings objects). As in the Twitter and WoS datasets, the sizes of both the compressed open dataset in AsterixDB and the compressed collection in MongoDB were comparable in the Sensors dataset.

To summarize our size findings, both the syntactic (page-level compression) and semantic (tuple compactor) approaches alleviated the storage overhead as shown in Figure~\ref{fig:storage-all}. The syntactic approach was more effective than the semantic approach for the Twitter dataset and the two were comparable for the WoS dataset. For the Sensors dataset, the semantic approach (with our vector-based format) was more effective for the reasons explained earlier. When combined, the approaches were able to reduce the overall storage sizes by 5x, 3.7x and 9.8x for the Twitter, WoS and Sensors datasets, respectively, compared to the open schema case in AsterixDB.

\subsection{Ingestion Performance}
We evaluated the performance of continuous data ingestion for the different formats using AsterixDB's data feeds for the Twitter dataset. We first evaluate the insert-only ingestion performance, without updates. In the second experiment, we evaluate the ingestion performance for an update-intensive workload, where previously ingested records are updated by either adding or removing fields or changing the types of existing data values. The latter experiment measures the overhead caused by performing point lookups to get the anti-schemas of previously ingested records. The Sensor dataset was also ingested through a data feed and showed similar behavior to the Twitter dataset; we omit these results here. 

Continuous data ingestion from a data feed is sensitive to LSM configurations such as the merge-policy and the memory budget. For instance, when cutting the memory budget by half, the size of flushed components would become 50\% smaller. AsterixDB's default "prefix-merge policy" \cite{storage} could then suffer from higher write-amplification by repeatedly merging smaller on-disk components until their combined size reaches a certain threshold. To eliminate those factors, we also evaluated the performance of bulk-loading, which builds a single on-disk component for the loaded dataset. (We evaluated the performance of bulk-loading into open, closed and inferred datasets using the WoS dataset.)
 

\begin{figure}[h]
    \centering
    \begin{subfigure}[b]{0.3\textwidth}
        \includegraphics[width=\textwidth]{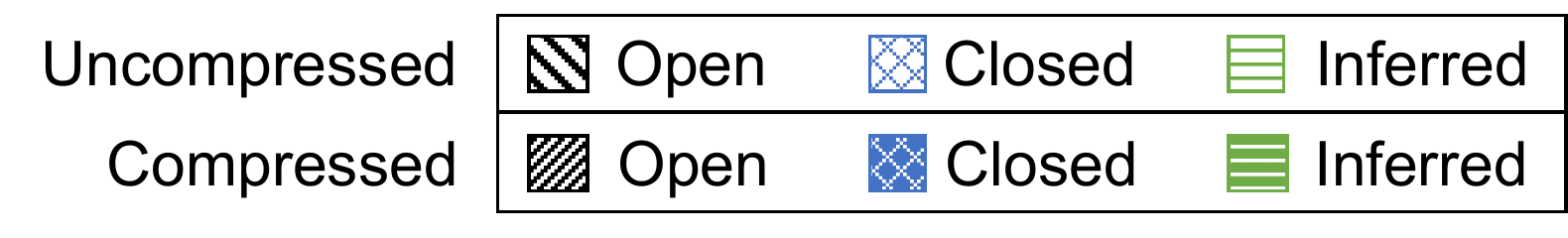}
    \end{subfigure}
    
    \begin{subfigure}[b]{0.225\textwidth}
        \includegraphics[width=\textwidth]{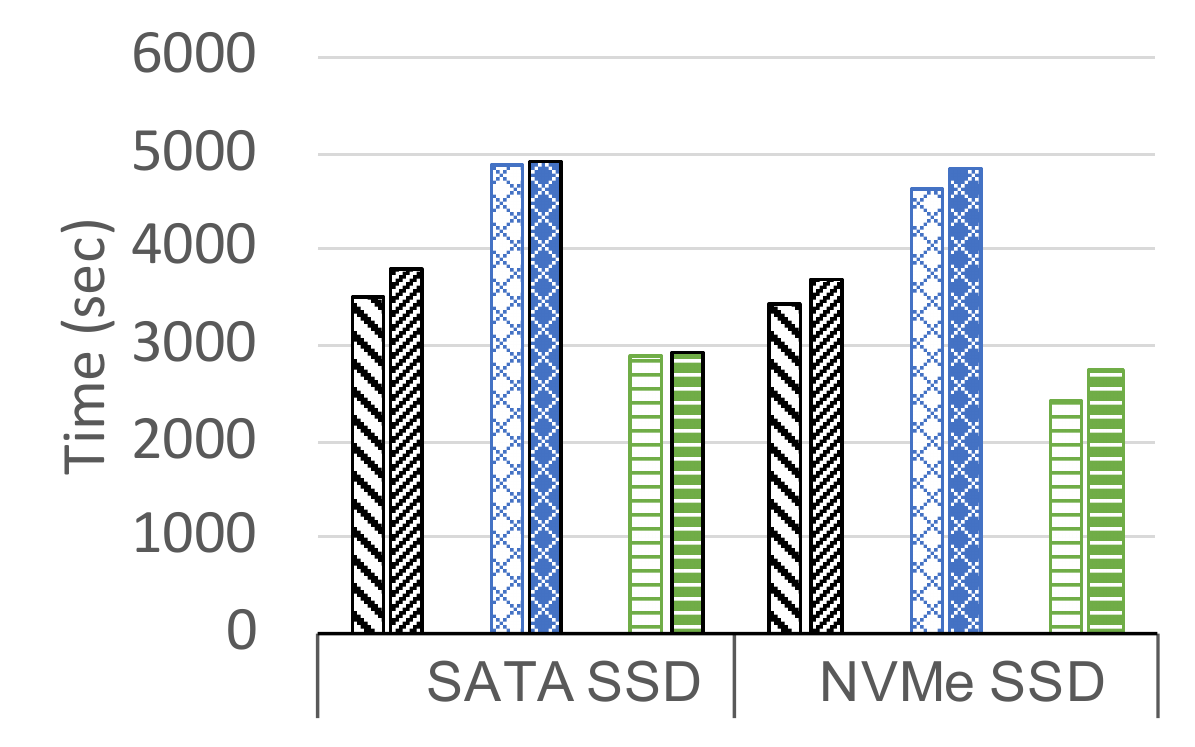}
        \caption{Twitter dataset | feed}
        \label{fig:twitter-feed}
    \end{subfigure}
    \begin{subfigure}[b]{0.225\textwidth}
        \vspace{0.2cm}
        \includegraphics[width=\textwidth]{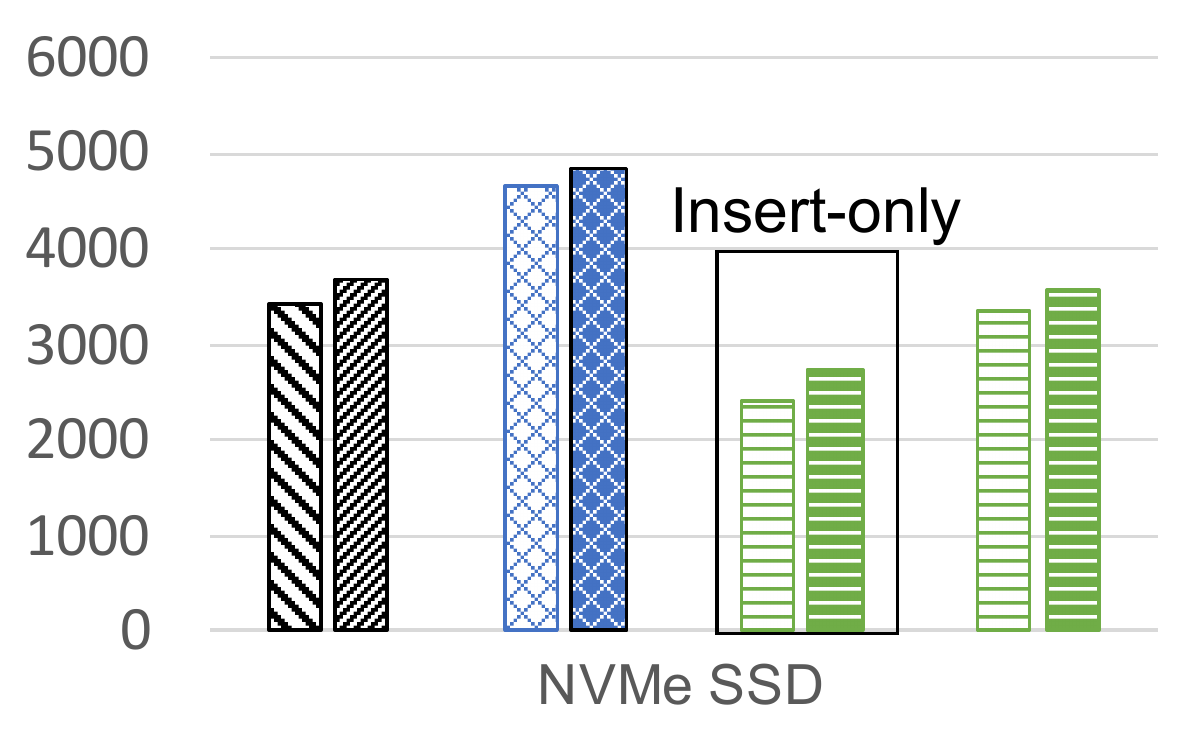}
        \caption{Twitter with 50\% updates}
        \label{fig:ingestion-update}
    \end{subfigure}
    \begin{subfigure}[b]{0.245\textwidth}
        \includegraphics[width=\textwidth]{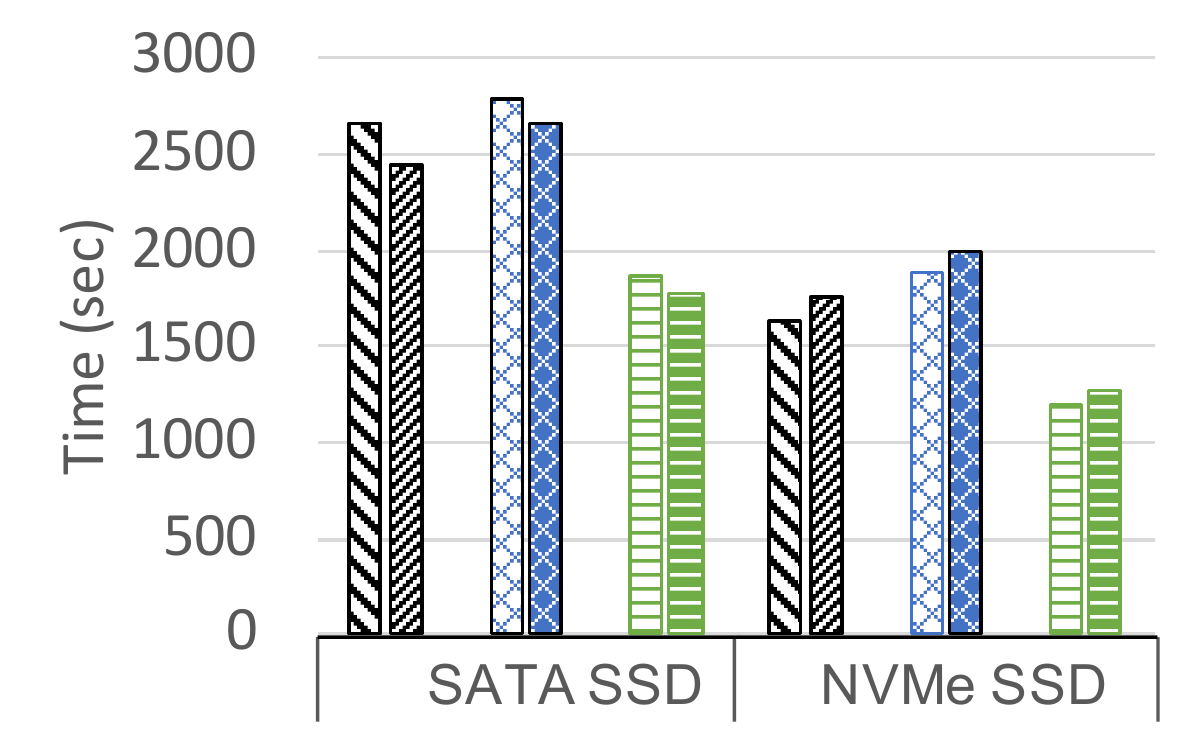}
        \caption{WoS dataset | bulkload}
        \label{fig:wos-bulkload}
    \end{subfigure}
        \vspace{-0.2cm}
    \caption{Data ingestion time}
    \label{fig:ingestion}
    \vspace{-0.5cm}
\end{figure}

\vspace{2.mm}
\noindent\textbf{Data Feed (Insert-only).} To evaluate the performance of continuous data ingestion, we measured the time to ingest the Twitter dataset using a data-feed to emulate Twitter's firehose. We set the maximum mergeable component size to 1GB and the maximum tolerable number of components to 5, after which the tree manager triggers a merge operation.

Figure~\ref{fig:twitter-feed} shows the time needed to complete the data ingestion for the 200GB Twitter dataset. Ingesting records into the inferred dataset took less time than ingesting into the open and closed datasets. Two factors played a role in the data ingestion rate. First, we observed that the record construction cost of the system's current ADM physical format was higher than the vector-based format by $\sim$40\%. Due to its recursive nature, the ADM physical format requires copying the values of the child to the parent from the leaf to the root of the record, which means multiple memory copy operations for the same value. Closed records took even more time to enforce the integrity constraints such as the presence and types of none-nullable fields. The second factor was the IO cost of the flush operation. We noticed that the inferred dataset's flushed on-disk components are $\sim$50\% and $\sim$25\% smaller than the open and closed datasets, respectively. This is due to the fact that compacted records in the vector-based format were smaller in size than the closed and open records in ADM format (see Figure~\ref{fig:twitter-storage}). Thus, the cost of writing larger LSM components of both open and closed datasets was higher.

The ingestion rate for the SATA SSD and the NVMe SSD were comparable, as both were actually bottlenecked by flushing transaction log records to the disk. Enabling compression had a slight negative impact on the ingestion rate for each format due to the additional CPU cost.

\vspace{2.mm}
\noindent\textbf{Data Feed (50\% Updates)} As explained in Section~\ref{sec:schema_maintainance}, updates require point lookups to maintain the schema, which can negatively impact the data ingestion rate. We evaluated the ingestion performance for update-intensive workload when the tuple compactor is enabled. In this experiment, we randomly updated 50\% of the previously ingested records by either adding or removing fields or changing existing value types. The updates followed a uniform distribution, where all records are updated equally. We created a primary key index, as suggested in \cite{luo2018lsm,luo2019efficient}, to reduce the cost of point lookups of non-existent (new) keys. Figure~\ref{fig:ingestion-update} shows the ingestion time of Twitter dataset, using the NVMe SSD drive, for the open, closed and inferred dataset with updates. The ingestion times for both open and closed datasets were about the same as t with no updates (Figure~\ref{fig:twitter-feed}). For the inferred dataset, the ingestion time with updates took $\sim$27\% and $\sim$23\% more time for the uncompressed and compressed datasets, respectively, compared to one with no updates. The ingestion times of the inferred and open datasets were comparable and both took less time than the closed dataset.

\vspace{2.mm}
\noindent\textbf{Bulk-load.} As mention earlier, continuous data ingestion is sensitive to LSM configurations such as the allocated memory budget for in-memory components. Additionally, the sizes of flushed on-disk components are smaller for the inferred and closed datasets as they have smaller storage overhead than the open dataset (as we saw while ingesting the Twitter dataset). Smaller on-disk components may trigger more merge operations to reach the maximum mergeable component size. To eliminate those factors, we also evaluated the time it takes AsterixDB to bulk-load the WoS dataset. When loading a dataset, AsterixDB sorts the records and then builds a single on-disk component of the B$^+$-tree in a bottom-up fashion. The tuple compactor infers the schema and compacts the records during this process. When loading finishes, the single on-disk component will have a single inferred schema for the entire set of records.

Figure~\ref{fig:wos-bulkload} shows the time needed to load the WoS dataset into open, closed, and inferred datasets. As for continuous data ingestion,  
the lower per-record construction cost of the vector-based format was the main contributor to the performance gain for the inferred dataset. We observed that the cost of the sort was relatively the same for the three schema datasets. However, the cost of building the B$^+$-tree was higher for both the open and closed schema datasets due to their higher storage overheads (Figure~\ref{fig:wos-storage}).

As loading a dataset in AsterixDB does not involve maintaining transaction logs, the higher throughput of the NVMe SSD was noticeable here compared to continuous data ingestion. When compression is enabled, the SATA SSD slightly benefited from the lower IO cost; however, the faster NVMe SSD was negatively impacted by the compression due to its CPU cost.



\subsection{Query Performance}
We next evaluated the impact of our work on query performance by running analytical queries against the ingested Twitter, WoS, and Sensor datasets. The objective of our experiments is to evaluate the IO cost of querying against open, closed, and inferred datasets. Each executed query was repeated six times and we report the average execution time of the last five.

We ran four queries (listed in Appendix \ref{appen:twitter}) against the Twitter dataset which retrieve:
\vspace{-.2cm}
\begin{enumerate}
	\item[Q1.] The number of records in the dataset | \lstinline[language=AQLSchema,basicstyle=\noindent\small]|COUNT(*)|.
	\vspace{-.2cm}
	\item[Q2.] The top ten users whose tweets' average length are the largest | \lstinline[language=AQLSchema,basicstyle=\noindent\small]|GROUP BY/ORDER BY|.
	\vspace{-.2cm}
	\item[Q3.] The top ten users who have the largest number of tweets that contain a popular hashtag | \lstinline[language=AQLSchema,basicstyle=\noindent\small]|EXISTS/GROUP BY|
	\lstinline[language=AQLSchema,basicstyle=\noindent\small]|ORDER BY|.
	\vspace{-.2cm}
	\item[Q4.] All records of the dataset ordered by the tweets' posting timestamps | \lstinline[language=AQLSchema,basicstyle=\noindent\small]| SELECT */ORDER BY|\footnote{In Q4, we report only the time for executing the query, excluding the time for actually retrieving the final formatted query result.}.
\end{enumerate}

\begin{figure}[h]
    \centering
    \hspace{-0.55cm}
    \includegraphics[width=0.5\textwidth]{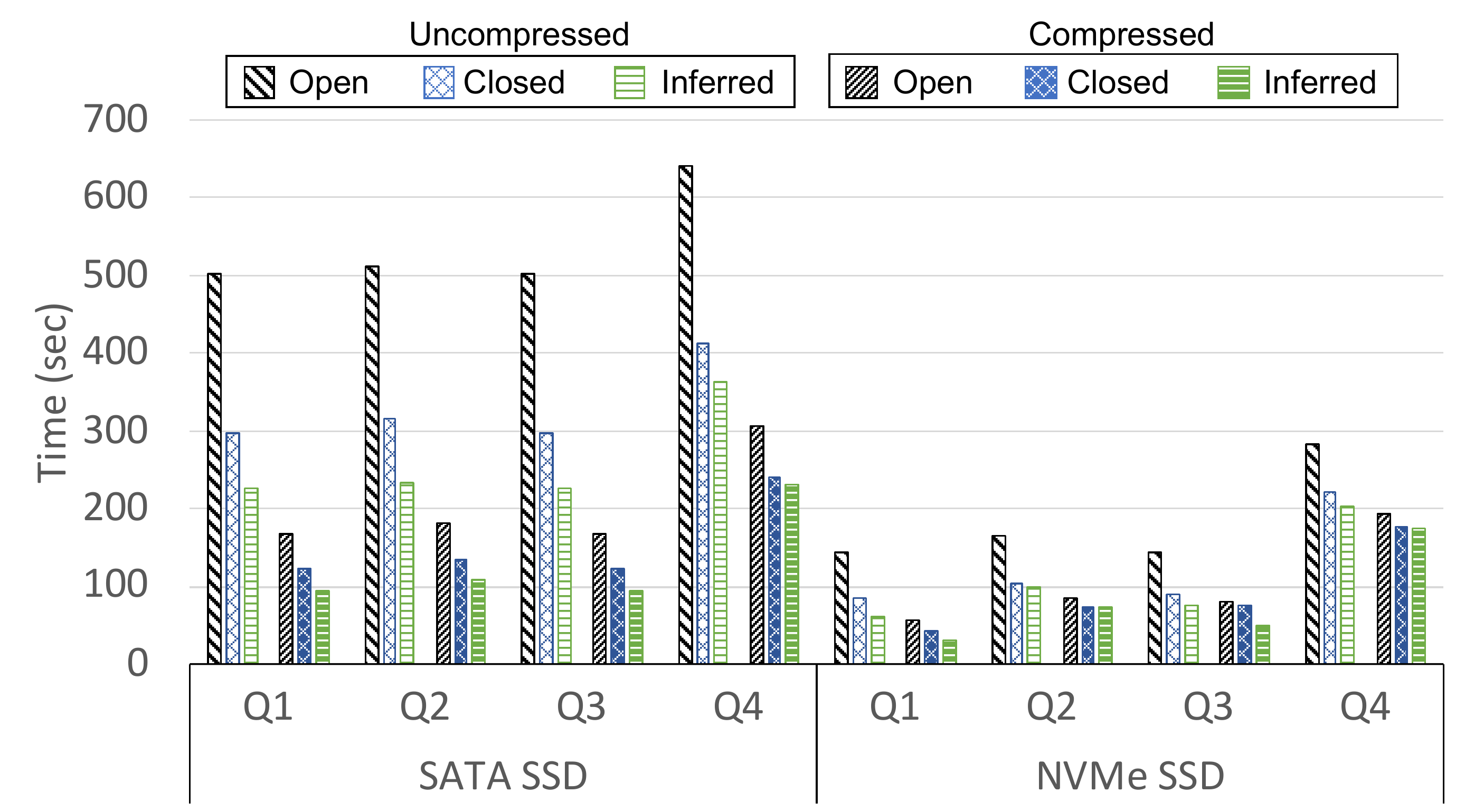}
    \caption{Query execution the Twitter dataset}
    \label{fig:twitter-exper}
    \vspace{-0.5cm}
\end{figure}

Figure~\ref{fig:twitter-exper} shows the execution time for the four queries in the three datasets (open, closed, and inferred) when the data is on the SATA SSD drive and the NVMe SSD drive. On the SATA SSD, the execution times of the four queries, with and without compression, correlated with their on-disk sizes from Figure~\ref{fig:twitter-storage}. This correlation indicates that the IO cost dominates the execution time. However, on the NVMe SSD drive, the CPU cost becomes more evident, especially when page-level compression is enabled. For Q2 and Q4, the $\sim$2X reduction in storage after compression reduced their execution times in the SATA case in all three datasets. However, the execution times for Q2 and Q4 in the closed and inferred datasets did not improve as much after compression in the NVMe case, as the CPU became the bottleneck here. Q3, which filters out all records that do not contain the required hashtag, took less time to execute in the inferred dataset. This is due to the way that nested values of records in the vector-based format are accessed. In the Twitter dataset, hashtags are modeled as an array of objects; each object contains the hashtag text and its position in the tweet's text. We consolidate field access expressions for records in the vector-based format (as discussed in Section~\ref{sec:process-vector-based}), and the query optimizer was able to push the consolidated field access through the unnest operation and extract only the hashtag text instead of the hashtag objects. Consequently, Q3's intermediate result size was smaller in the inferred dataset compared to the other two datasets, and executing Q3 against the inferred dataset was faster. This experiment shows that our schema inference and tuple compaction approach can match (or even improve in some cases) the performance of querying datasets with fully declared schemas | without a need for pre-declaration.
  
\subsubsection{Twitter Dataset}
\label{sec:twitter-exper}
  

\subsubsection{WoS Dataset}

We also ran four queries (listed in Appendix~\ref{appen:wos}) against the WoS dataset: 
\vspace{-.2cm}
\begin{enumerate}
	\item[Q1.] The number of records in the dataset | \lstinline[language=AQLSchema,basicstyle=\noindent\small]|COUNT(*)|.
	\vspace{-.2cm}
	\item[Q2.] The top ten scientific fields with the highest number of publications | \lstinline[language=AQLSchema,basicstyle=\noindent\small]|GROUP BY/ORDER BY|.
	\vspace{-.2cm}
	\item[Q3.] The top ten countries that co-published the most with US-based institutes| \lstinline[language=AQLSchema,basicstyle=\noindent\small]|UNNEST/EXISTS/|
	\lstinline[language=AQLSchema,basicstyle=\noindent\small]|GROUP BY/ORDER BY|.
	\vspace{-.5cm}
	\item[Q4.] The top ten pairs of countries with the largest number of co-published articles | \lstinline[language=AQLSchema,basicstyle=\noindent\small]|UNNEST/GROUP BY/ORDER BY|
\end{enumerate}

\begin{figure}[h]
    \centering
    \hspace{-0.55cm}
    \includegraphics[width=0.5\textwidth]{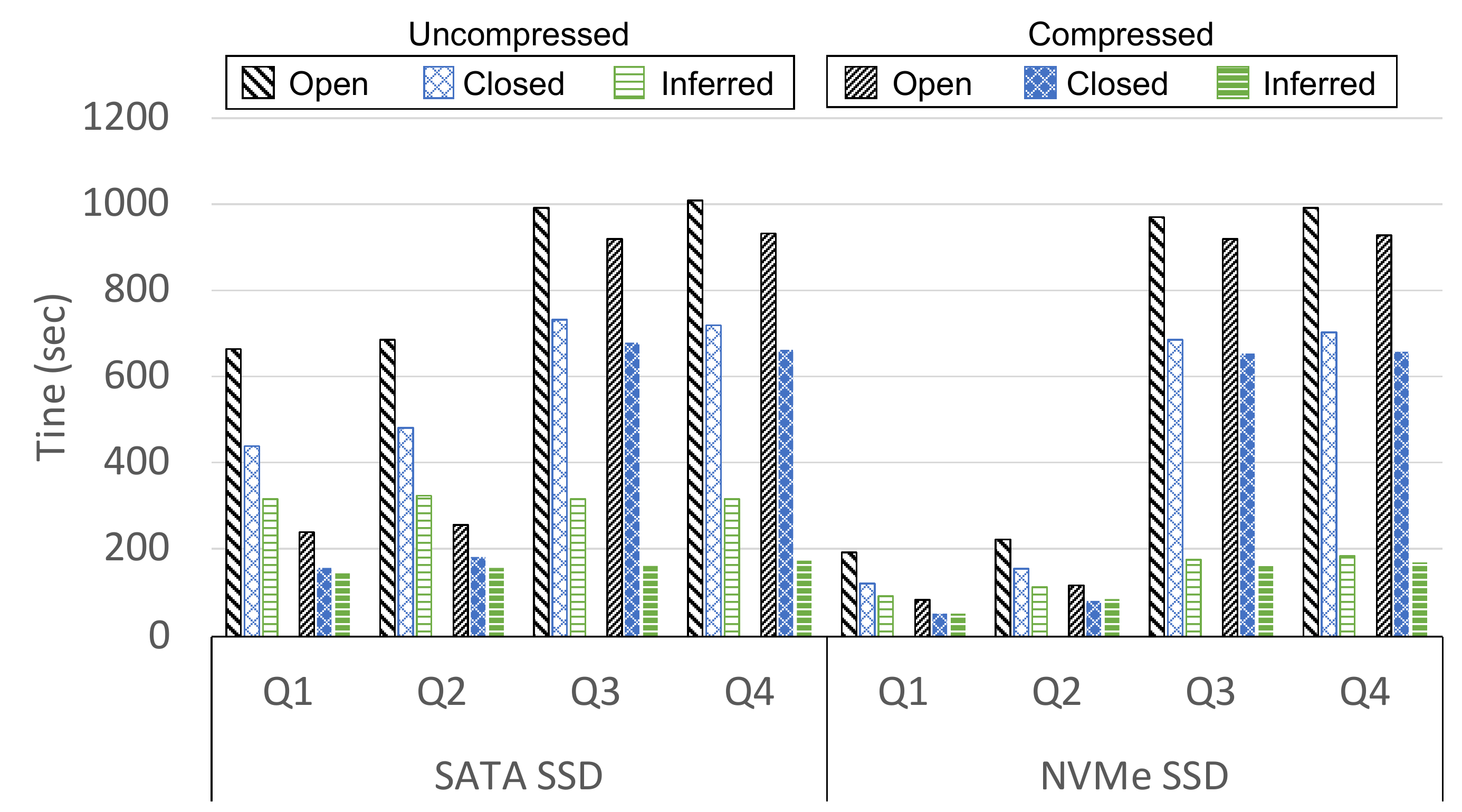}
    \caption{Query execution time for the WoS dataset}
    \label{fig:wos-exper}
    \vspace{-0.5cm}
\end{figure}

As Figure~\ref{fig:wos-exper} illustrates, the execution times for Q1 and Q2 are correlated with the storage sizes of the three datasets (Figure~\ref{fig:wos-storage}). For Q3 and Q4, the execution times were substantially higher in the open and closed datasets as compared to the inferred dataset. Similar to Q3 in the Twitter dataset, field access expression consolidation and pushdown were beneficial. Even after enabling compression, the open and closed schema execution times for Q3 and Q4 remained about the same despite the storage savings. (We will evaluate that behavior in more detail in Section~\ref{sec:eval-pushdown}).

%

\begin{figure}[h]
    \centering
    \hspace{-0.55cm}
    \includegraphics[width=0.5\textwidth]{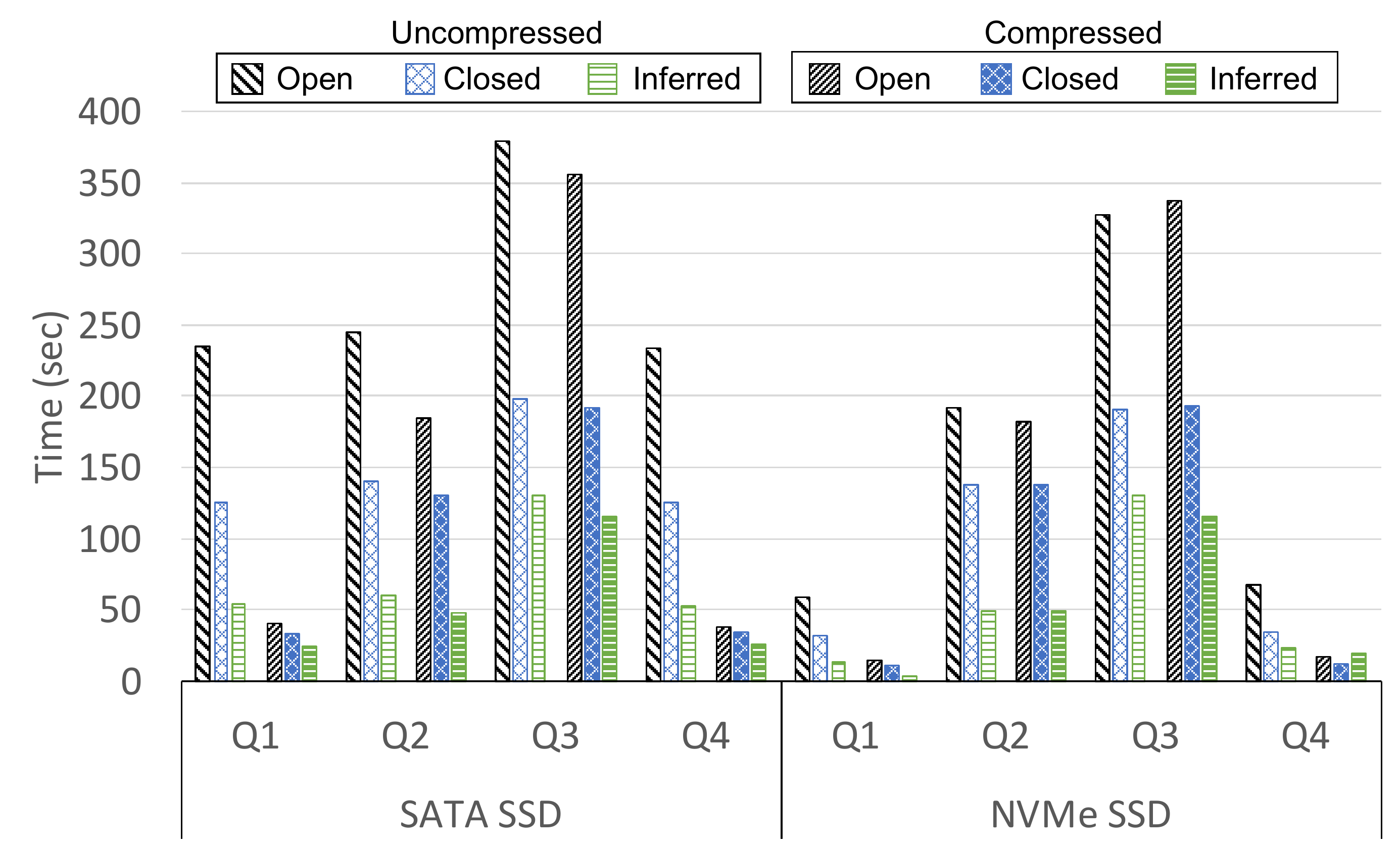}
    \caption{Query execution time for the Sensors dataset.}
    \label{fig:iot-exper}
    \vspace{-0.3cm}
\end{figure}

\subsubsection{Sensors Dataset} 

We again ran four queries (listed in Appendix~\ref{appen:sensors}):
\vspace{-.2cm}
\begin{enumerate}
	\item[Q1.] The number of records in the dataset | \lstinline[language=AQLSchema,basicstyle=\noindent\small]|COUNT(*)|.
	\vspace{-.2cm}
	\item[Q2.] The minimum and maximum reading values that were ever recorded across all sensors | \lstinline[language=AQLSchema,basicstyle=\noindent\small]|UNNEST/GROUP BY|.
	\vspace{-.2cm}
	\item[Q3.] The IDs of the top ten sensors that have recorded the highest average reading value | \lstinline[language=AQLSchema,basicstyle=\noindent\small]|UNNEST/GROUP BY/ORDER BY|	
	\vspace{-.5cm}
	\item[Q4.] Similar to Q4, but look for the recorded readings in a given day | \lstinline[language=AQLSchema,basicstyle=\noindent\small]|WHERE/UNNEST/GROUP BY/ORDER BY|
\end{enumerate}

The execution times are shown in Figure~\ref{fig:iot-exper}. The execution times for Q1 on both uncompressed and compressed datasets correlate with the storage sizes of the datasets from Figure~\ref{fig:iot-storage}. Q2 and Q3 exhibit the effect of consolidating and pushing down field value accesses of vector-based format, where both queries took significantly less time to execute in the inferred dataset. However, pushing the field access down is not always advantageous. When compression is enabled, the execution time of Q4 for the inferred dataset using NVMe SSD was the slowest. This is because the consolidated field accesses (of sensor ID, reading and reporting timestamp) are evaluated before filtering using a highly selective predicate (0.001\%). In the open and closed datasets, delaying the evaluation of field accesses until after the filter for Q4 was beneficial. However, the execution times for the inferred dataset was comparable to the open~case.

\subsubsection{Impact of the Vector-based Optimizations}
\label{sec:eval-pushdown}

\noindent\textbf{Breakdown of the storage savings.} \rev{As we showed in our experiments, the time it takes for ingesting and querying records in the vector-based format (inferred) was smaller even when the schema is fully declared for the ADM format (closed). This is due to fact that the vector-baed format encodes nested values more efficiently using only the type tags (as in Section~\ref{sec:vector-based-format}). To measure the impact of the newly proposed format, we reevaluate the storage size of the vector-based without inferring the schema or compacting the records (i.e., a schema-less version using the vector-based format), which we refer to as \textit{SL-VB}.}

\begin{figure}[h]
\vspace{-0.1cm}
    \centering
    \begin{subfigure}[b]{0.3\textwidth}
        \includegraphics[width=\textwidth]{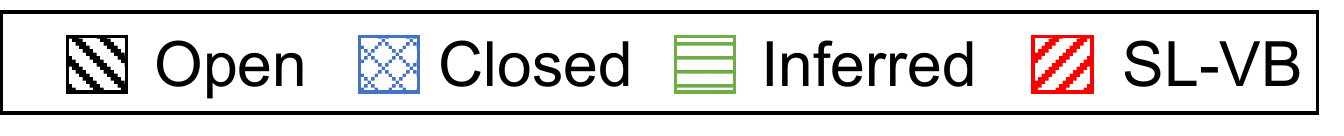}
        \vspace{-2.em}
    \end{subfigure}
    \begin{subfigure}[b]{0.2\textwidth}
        \includegraphics[width=\textwidth]{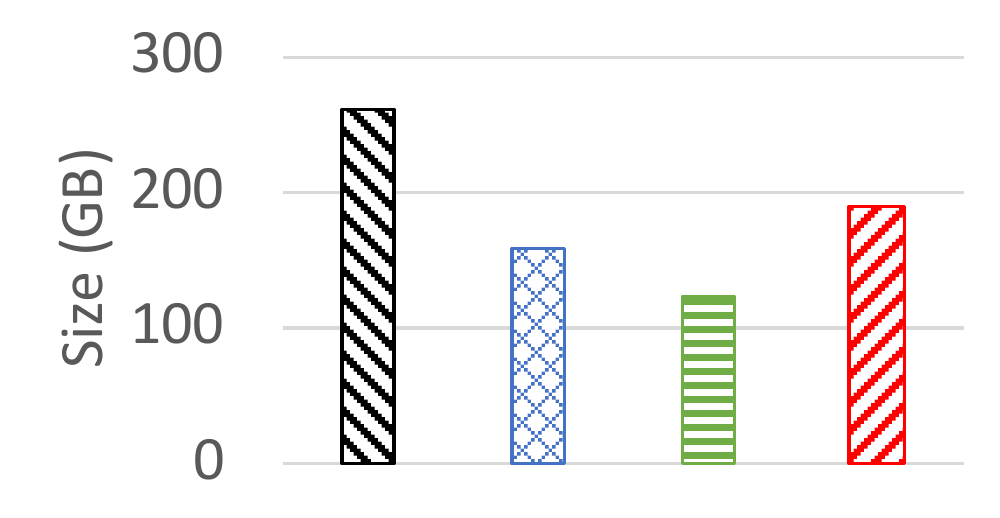}
        \caption{Twitter}
        \label{fig:twitter-storage2}
    \end{subfigure}
    \begin{subfigure}[b]{0.2\textwidth}
        \vspace{0.2cm}
        \includegraphics[width=\textwidth]{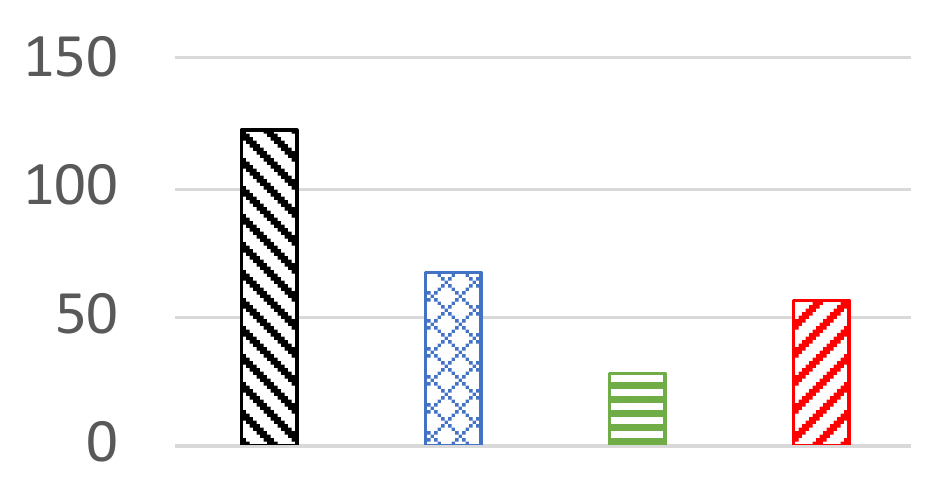}
        \caption{Sensors}
        \label{fig:sensors-storage2}
    \end{subfigure}
    \vspace{-0.3cm}
    \caption{Impact of the vector-based format on storage}
    \label{fig:vector-based-storage-expr}
\end{figure}

\rev{In Figure~\ref{fig:twitter-storage2}, we see the total sizes of the four datasets \textit{open, closed, inferred}, and \textit{SL-VB} after ingesting the Twitter dataset. We see that the SL-VB dataset is smaller than the open dataset but slightly larger than the closed one. More specifically, about half of the storage savings in the inferred dataset (compared to the open dataset) is from the more efficient encoding of nested values in the vector-based format, and the other half is from compacting the record. For the Sensors dataset, Figure~\ref{fig:sensors-storage2} shows a similar pattern; however, the SL-VB Sensors dataset is smaller than the closed dataset for the reasons explained in Section~\ref{sec:storage-size}.}

\begin{figure}[h]
\vspace{-0.1cm}
    \begin{subfigure}[b]{0.45\textwidth}
	    \includegraphics[width=\textwidth]{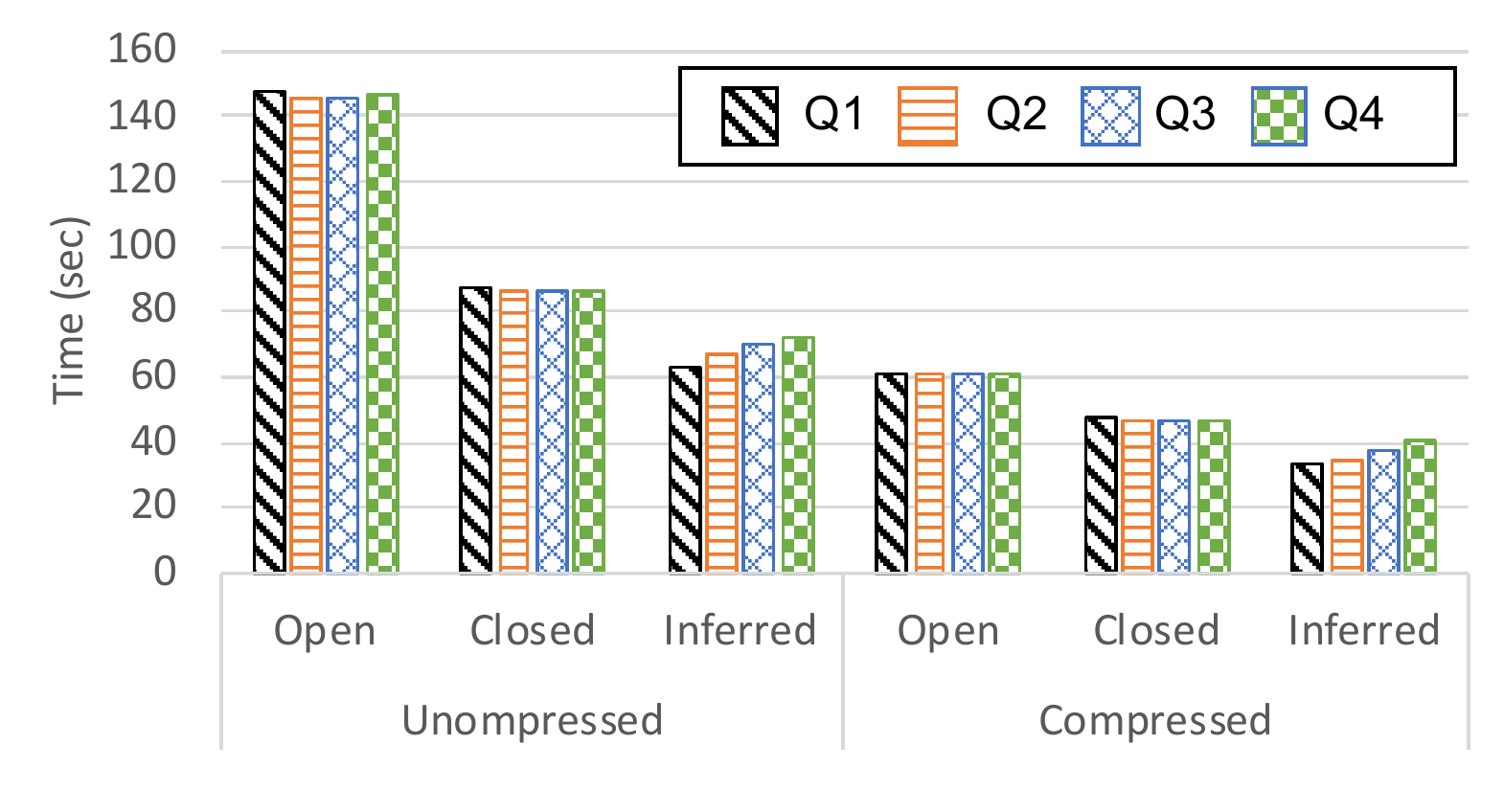}
  	    \vspace{-0.5cm}
	    \caption{Large: 200GB}
	    \label{fig:sensitivity}
    \end{subfigure}
    \begin{subfigure}[b]{0.45\textwidth}
	    \includegraphics[width=\textwidth]{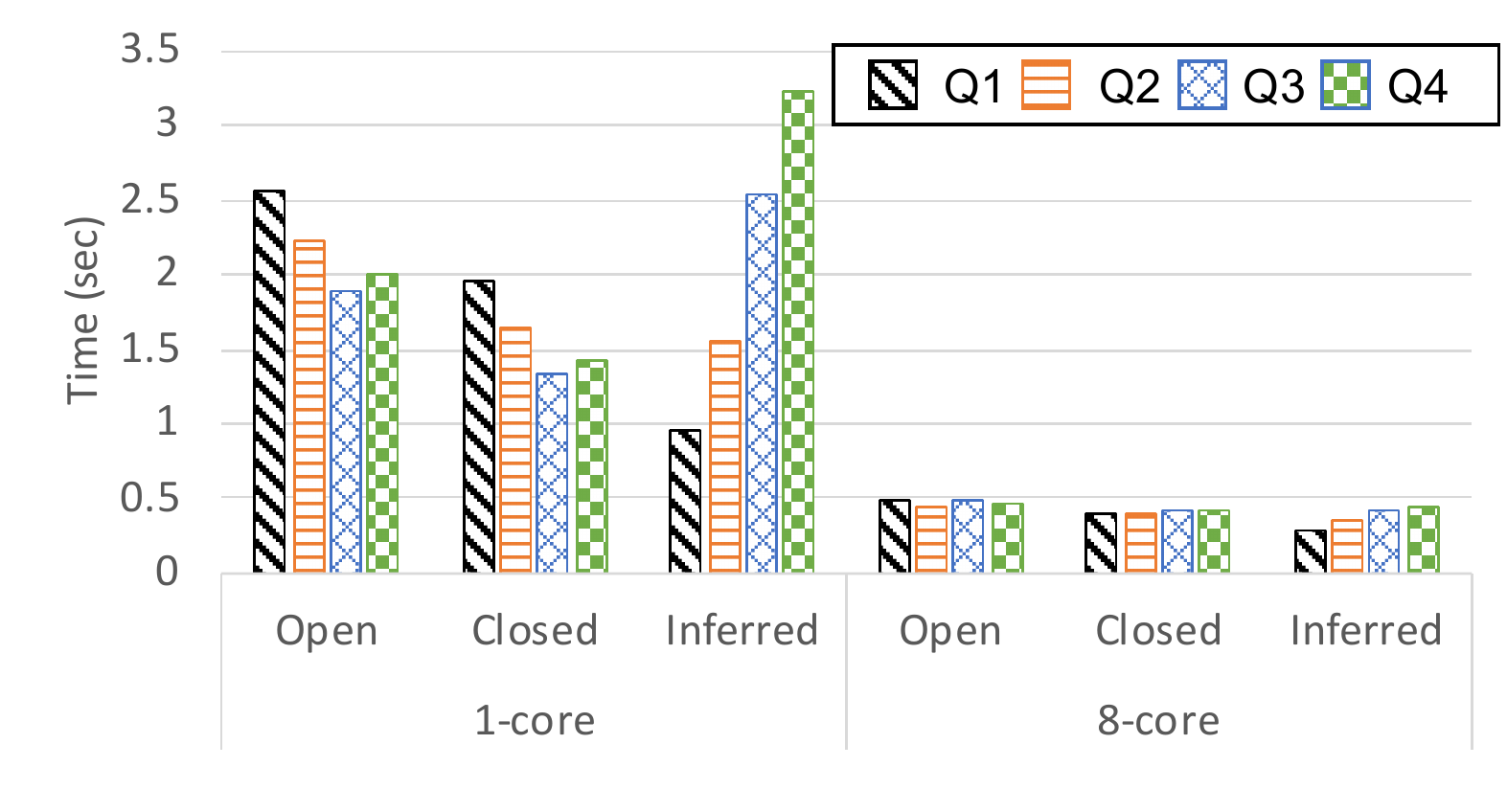}
	    \vspace{-0.5cm}
	    \caption{Small: 5GB}
	    \label{fig:sensitivity-small}
    \end{subfigure}
    \vspace{-0.2cm}
    \caption{Impact of the vector-based format on storage}
    \label{fig:sensitivity-large-samll}
\end{figure}

\noindent\textbf{Linear-time field access.} Accessing values in the vector-based format is sensitive to the position of the requested value. For instance, accessing a value that appears first in a record is faster than accessing a value that resides at the end. To measure the impact of linear access in the vector-based format, we ran four queries against the Twitter dataset (using the NVMe SSD drive) where each counts the number of appearances of a value. The positions (or indexes) of those values in the vector-based format are 1, 34, 68, and 136 for Q1, Q2, Q3 and Q4, respectively, where position 1 means the first value in the record and the position 136 is the last. Figure~\ref{fig:sensitivity} shows the time needed to execute the queries. In the inferred datasets, the position of the requested value affected the query times, where Q1 was the fastest and Q4 was the slowest. For the open and closed datasets, the execution times for all queries were about the same. However, all four queries took less time to execute in the inferred cases, due to the storage savings. \rev{When all the data fits in-memory, the CPU cost becomes more apparent as shown in Figure~\ref{fig:sensitivity-small}. In the case of a single core, the vector-based format was the slowest to execute Q3 and Q4. When using all 8-cores, the execution time for all queries were about the same for the three datasets.}

\noindent\textbf{Field-access consolidation and pushdown.} Also in our experiments, we showed that our optimizations of consolidating and pushing down field access expressions can tremendously improve query execution time. To isolate the factors that contributed to the performance gains, we reevaluated the execution times for Q2-Q4 of the Sensors dataset with and without these optimizations. 

\begin{figure}[h]
    \centering
    \hspace{-0.55cm}
    \includegraphics[width=0.5\textwidth]{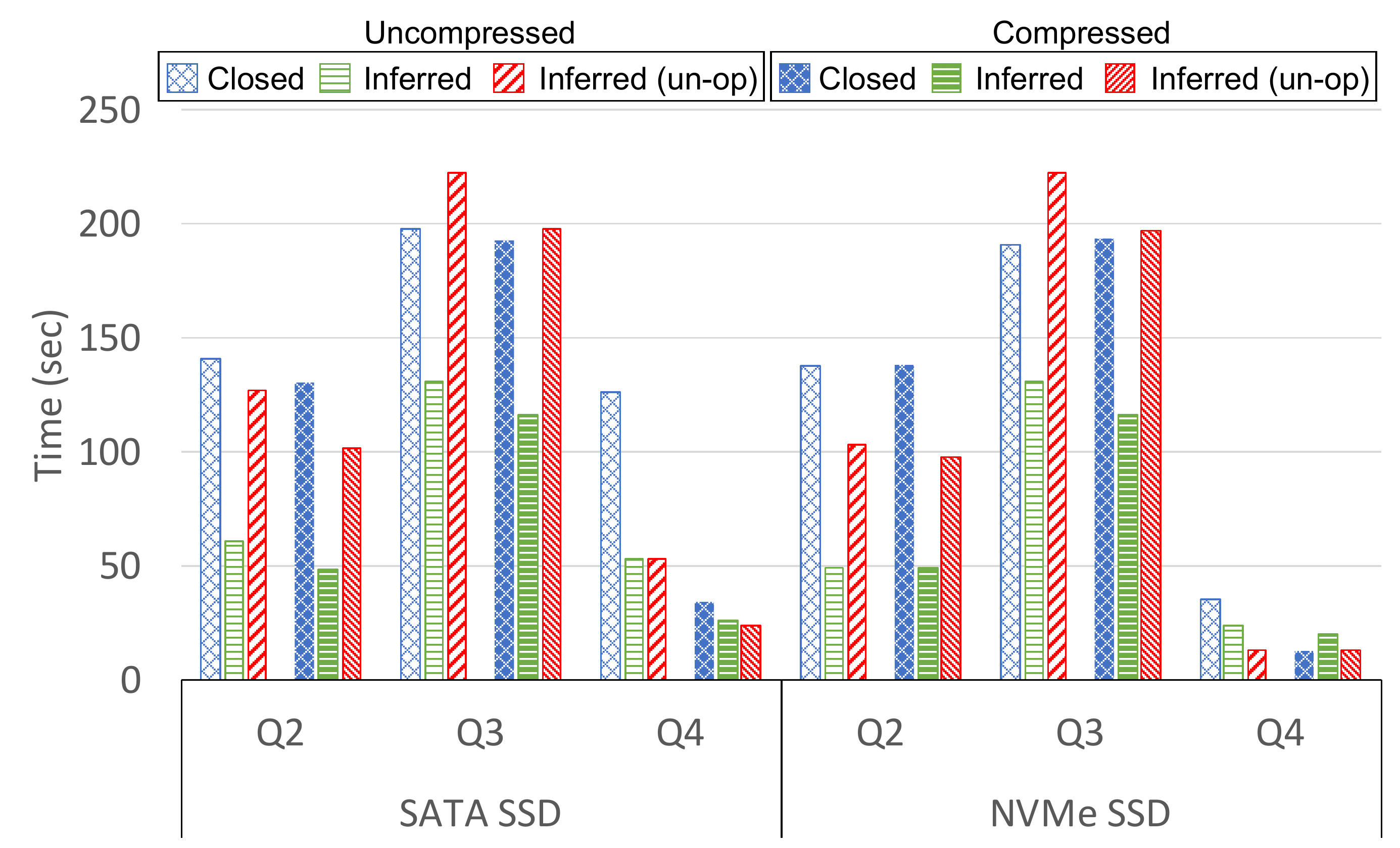}
    \caption{Impact of consolidating and pushing down field access expressions}
    \label{fig:iot-exper-unop}
      \vspace{-1.5em}
\end{figure}

The execution times of the queries are shown in Figure~\ref{fig:iot-exper-unop}. We refer to \textit{Inferred (un-op)} as querying the inferred dataset without our optimization of consolidating and pushing down field access expressions. When we disable our optimizations, the linear-time field accesses of the vector-based format are performed as many times as there are field access expressions in the query. For instance, Q3 has three field access expressions to get the (i) sensor ID, (ii) readings array, and (iii) reporting timestamp. Each field access requires scanning the record's vectors, which is expensive. Additionally, the size of the intermediate results of Q2 and Q3 were then larger (array of objects vs. array of doubles). As a result, Q2 and Q3 took twice as much time to finish for Inferred~(un-op). Q2 is still faster to execute in the Inferred~(un-op) case than in the closed case whereas Q3 took slightly more time to execute. Finally, disabling our optimizations improved the execution time for Q4 on the NVMe SSD, as delaying the evaluation of field accesses can be beneficial for queries with highly selective predicates.

\noindent\textbf{Vector-based format vs. others.} \rev{Other formats, such as Apache Avro \cite{avro}, \sloppy{Apache} Thrift \cite{thrift}, and Google Protocol Buffers~\cite{protobuf}, also exploit schemas to store semi-structured data more efficiently. In fact, providing a schema is not optional for writing records in such formats | as opposed to the vector-based format, where the schema is optional. Nonetheless, we compared the vector-based format to Apache Avro, Apache Thrift using both Binary Protocol (BP) and Compact Protocol (CP), and Protocol Buffers to evaluate 1) the storage size and 2) the time needed to construct the records in each format using 52MB of the Twitter dataset. Table~\ref{tab:serder_exper} summarizes the result of our experiment. We see that the storage sizes of the different formats were mostly comparable. In terms of the time needed to construct the records, Apache Thrift (for both protocols) took the least construction time followed by the vector-based format. Apache Avro and Protocol Buffers took 1.9x and 2.9x more time to construct the records compared to the vector-based format, respectively.}

\vspace{-1.2em}
\begin{center}
\begin{table}[h]  
\vspace{-.8em}
\small
  \centering
  \begin{tabular}{|l|c|c|}
    \cline{2-3}
    \multicolumn{1}{c|}{} & Space (MB) & Time (msec) \\ \hline
   Avro         & 27.49 & 954.90  \\ \hline
   Thrift (BP)  & 34.30 & 341.05  \\ \hline
   Thrift (CP)  & 25.87 & 370.93  \\ \hline
   ProtoBuf     & 27.16 & 1409.13 \\ \hline
   Vector-based & 29.49 & 485.48  \\ \hline
  \end{tabular}
  \caption{Writing 52MB of Tweets in different formats }
  \label{tab:serder_exper}
  \vspace{-2.5em}
\end{table}
\end{center}

\subsubsection{Secondary Index Query Performance}
Pirzadeh et al.~\cite{bigFUN} previously showed that predeclaring the schema in AsterixDB did not improve (notably) the performance of range-queries with highly selective predicates in the presence of a secondary index. In this experiment, we evaluated the impact of having a secondary index using the Twitter dataset.

\begin{figure}[h]
    \centering
    \begin{subfigure}[b]{0.4\textwidth}
        \includegraphics[width=\textwidth]{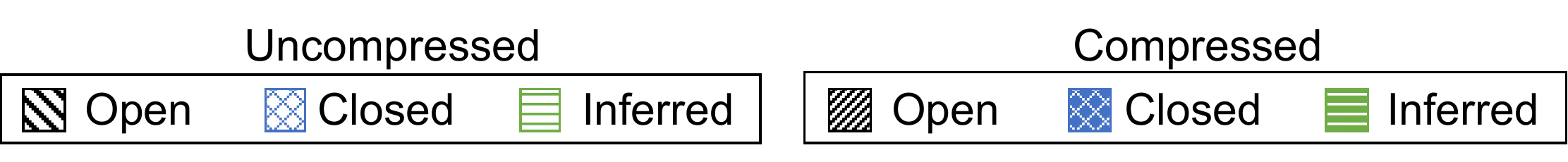}
    \end{subfigure}
    
    \begin{subfigure}[b]{0.24\textwidth}
        \includegraphics[width=\textwidth]{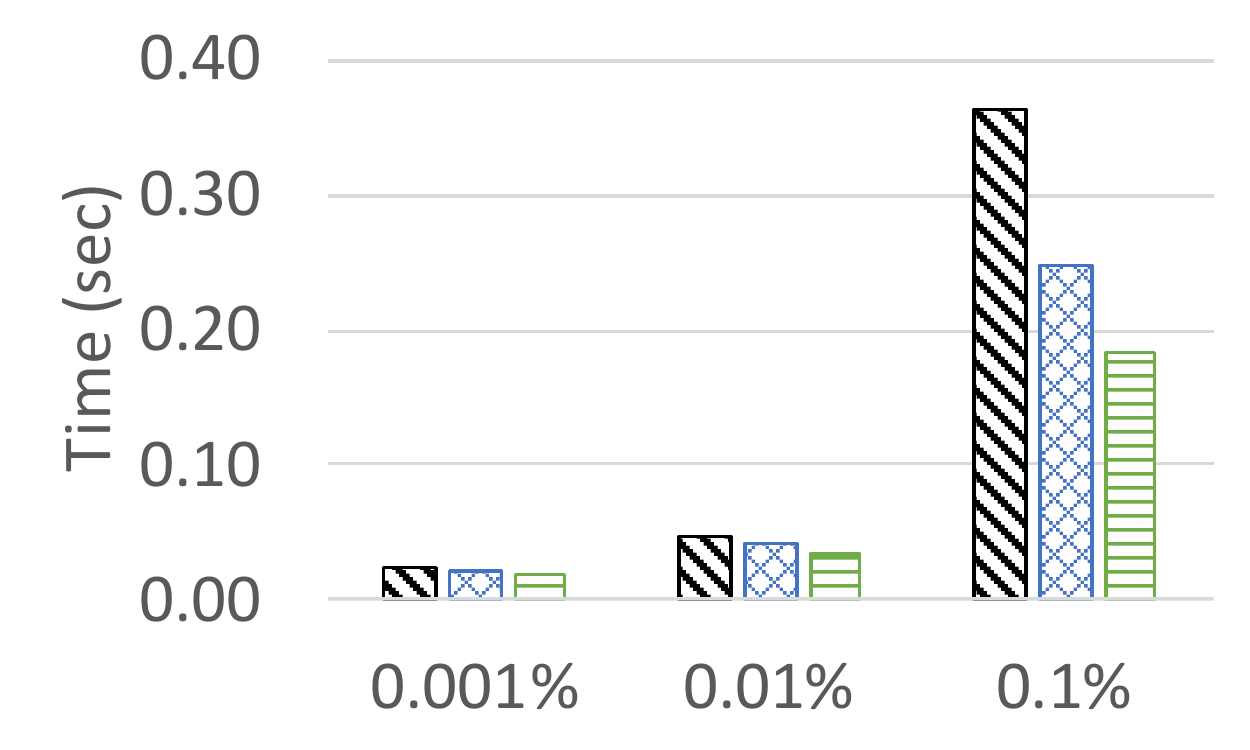}
        \caption{Low selectivity}
        \label{fig:low_uncompressed}
    \end{subfigure}
    \begin{subfigure}[b]{0.23\textwidth}
        \includegraphics[width=\textwidth]{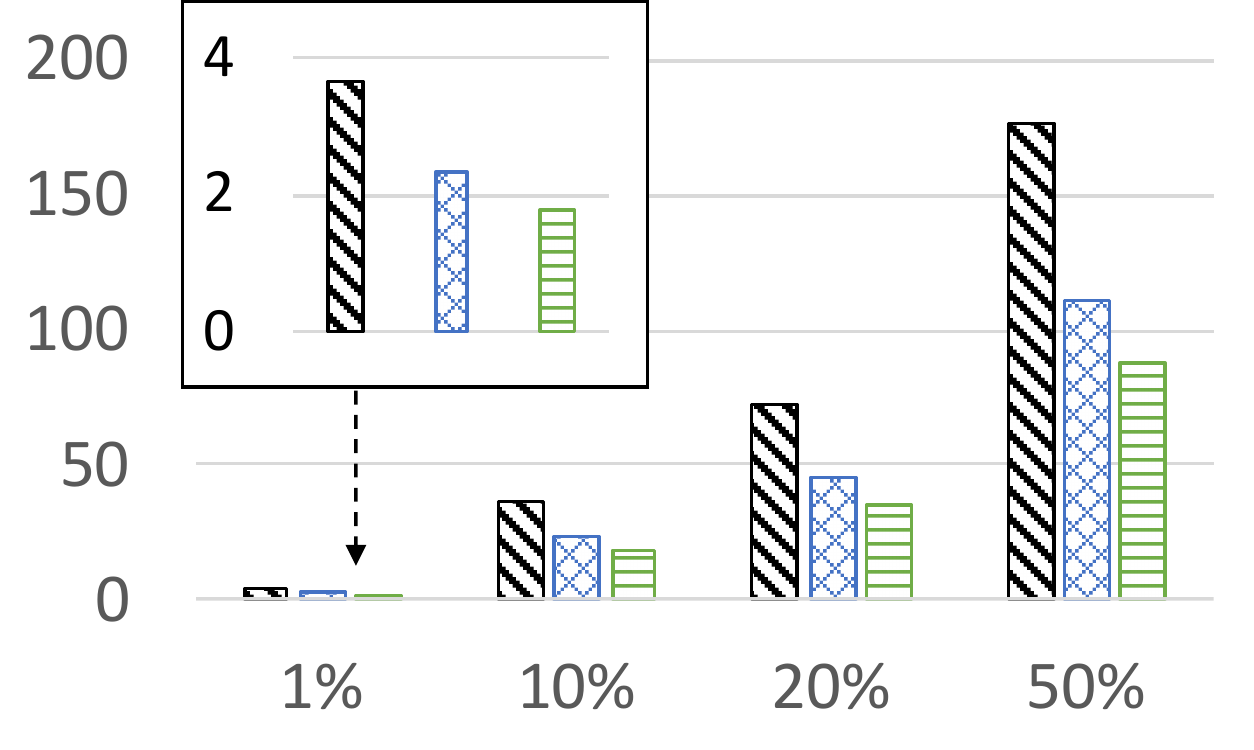}
        \caption{High selectivity}
        \label{fig:high_uncompressed}
    \end{subfigure}
    
    \begin{subfigure}[b]{0.24\textwidth}
        \includegraphics[width=\textwidth]{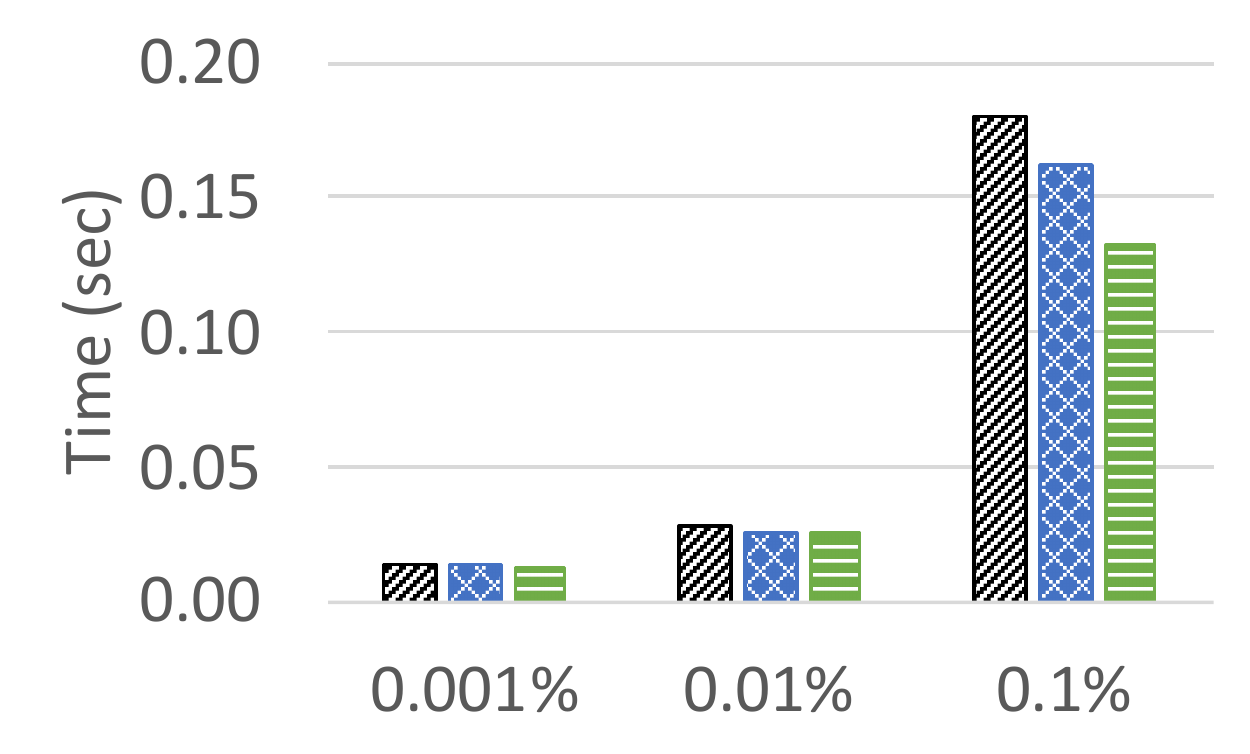}
        \caption{Low selectivity}
        \label{fig:low_compressed}
    \end{subfigure}
    \begin{subfigure}[b]{0.23\textwidth}
        \includegraphics[width=\textwidth]{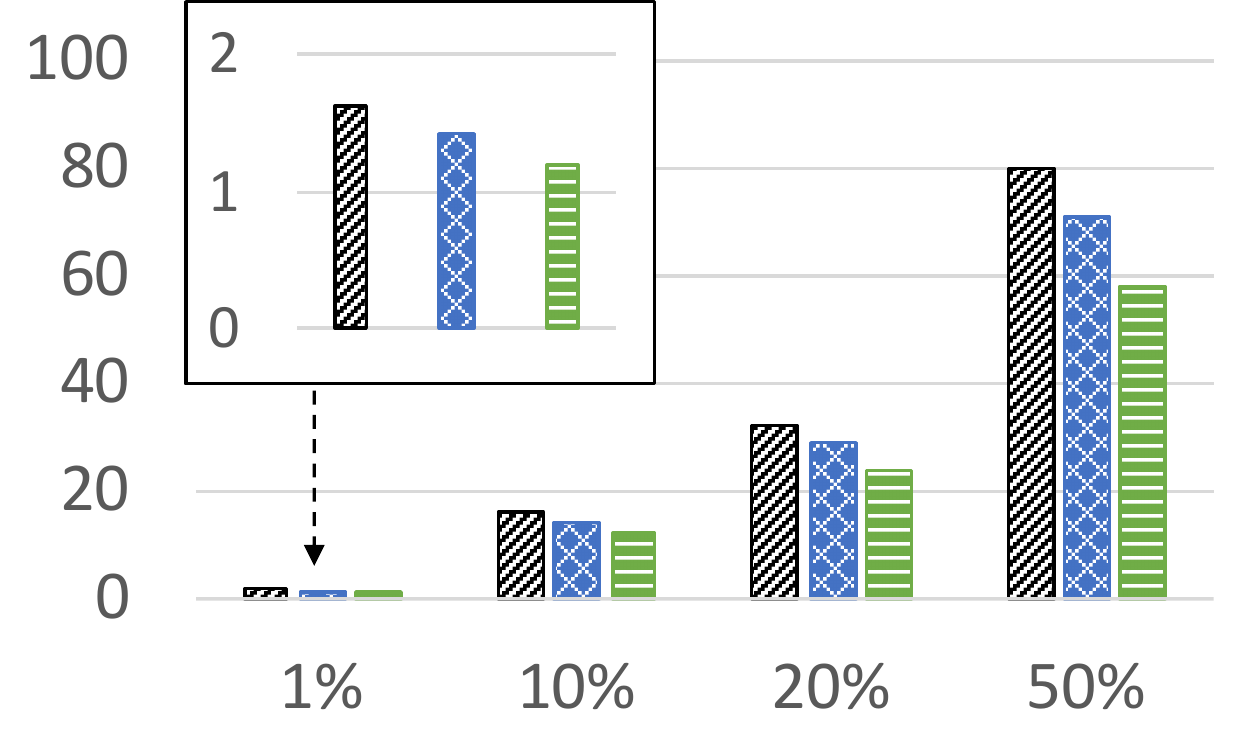}
        \caption{High selectivity}
        \label{fig:high_compressed}
    \end{subfigure}
    
    \caption{Query with secondary index (NVMe)}
    \label{fig:secondary}
    \vspace{-0.5cm}
\end{figure}

We modified the scaled Twitter dataset by generating monotonically increasing values for the attribute \textit{timestamp} to mimic the time at which users post their tweets. We created a secondary index on this generated timestamp attribute and ran multiple range-queries with different selectivities. For each query selectivity, we executed queries with different range predicates to warm up the system's cache and report the average stable execution time. Figures~\ref{fig:low_uncompressed}~and~\ref{fig:high_uncompressed} show the execution times for queries, with both low and high selectivity predicates, using the NVMe SSD, for uncompressed datasets. The execution times for all queries correlated with the storage sizes (Figure~\ref{fig:twitter-storage}), where the closed and inferred datasets have lower storage overhead compared to the open dataset. The execution times for compressed datasets (Figures~\ref{fig:low_compressed}~and~\ref{fig:high_compressed}) showed similar relative behavior.

\subsection{Scale-out Experiment}
\label{sec:exper-scale-out}
Finally, to evaluate the scalability our approach, we conducted a scale-out experiment using a cluster of Amazon EC2 instances of type \texttt{c5d.2xlarge} (each with 16GB of memory and 8 virtual cores). We evaluate the ingestion and query performance of the Twitter dataset using clusters with 4, 8, 16 and 32 nodes. We configure each node with 10GB of total memory, with 6GB for the buffer cache and 1GB for the in-memory component budget. The remaining 3GB is allocated for working buffers. We used the instance \textit{ephemeral} storage to store the ingested data. Due to the lack of storage space in a \texttt{c5d.2xlarge} instance (200GB), we only evaluate the performance on compressed datasets.

\begin{figure}[h]
    \centering
    \begin{subfigure}[b]{0.25\textwidth}
        \includegraphics[width=\textwidth]{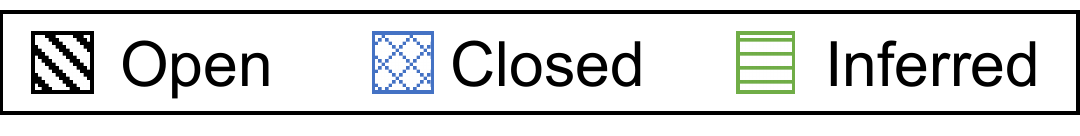}
    \end{subfigure}
    
    \begin{subfigure}[b]{0.235\textwidth}
        \includegraphics[width=\textwidth]{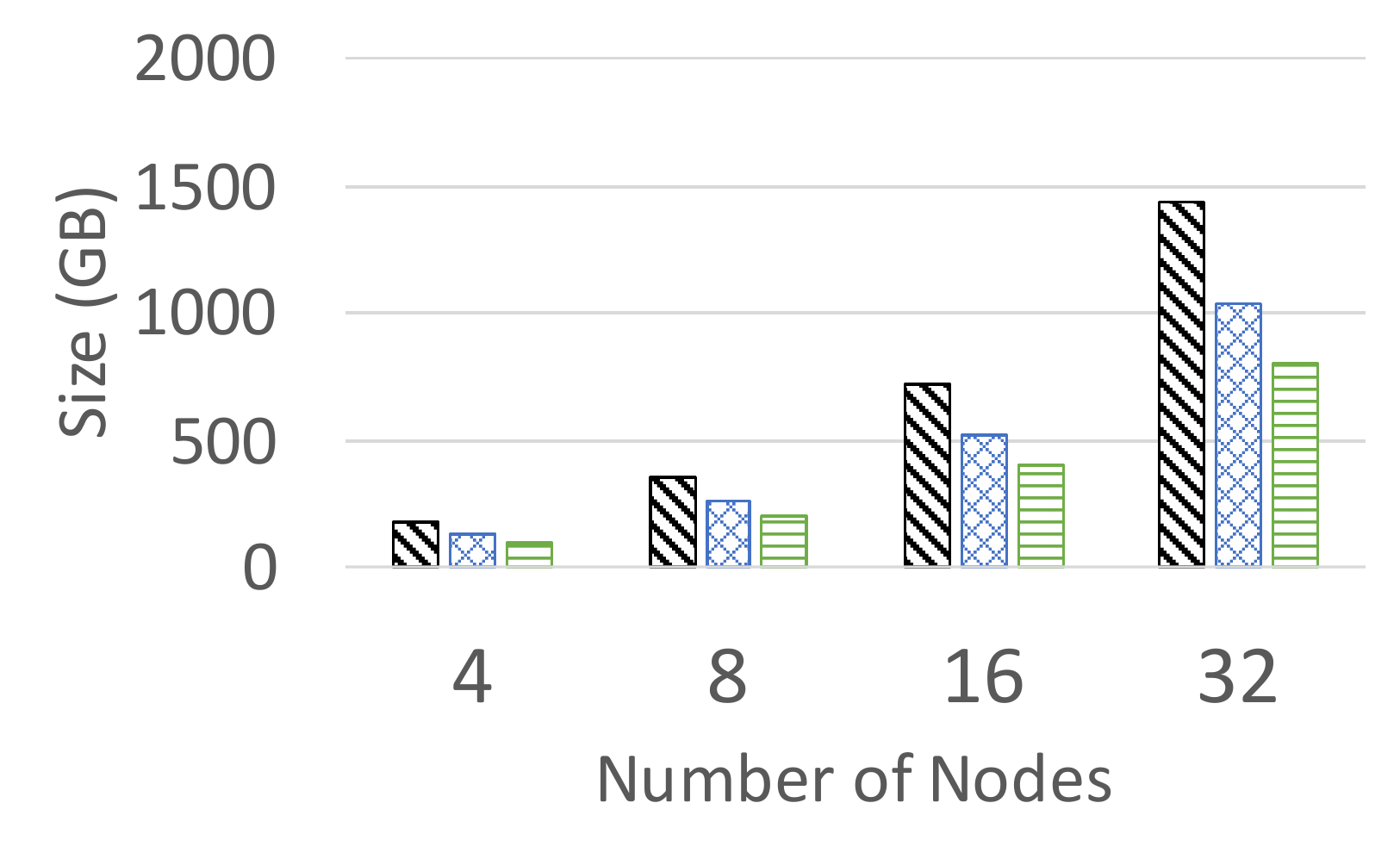}
        \caption{On-disk size}
        \label{fig:scale-out-storage}
    \end{subfigure}
    \begin{subfigure}[b]{0.235\textwidth}
        \includegraphics[width=\textwidth]{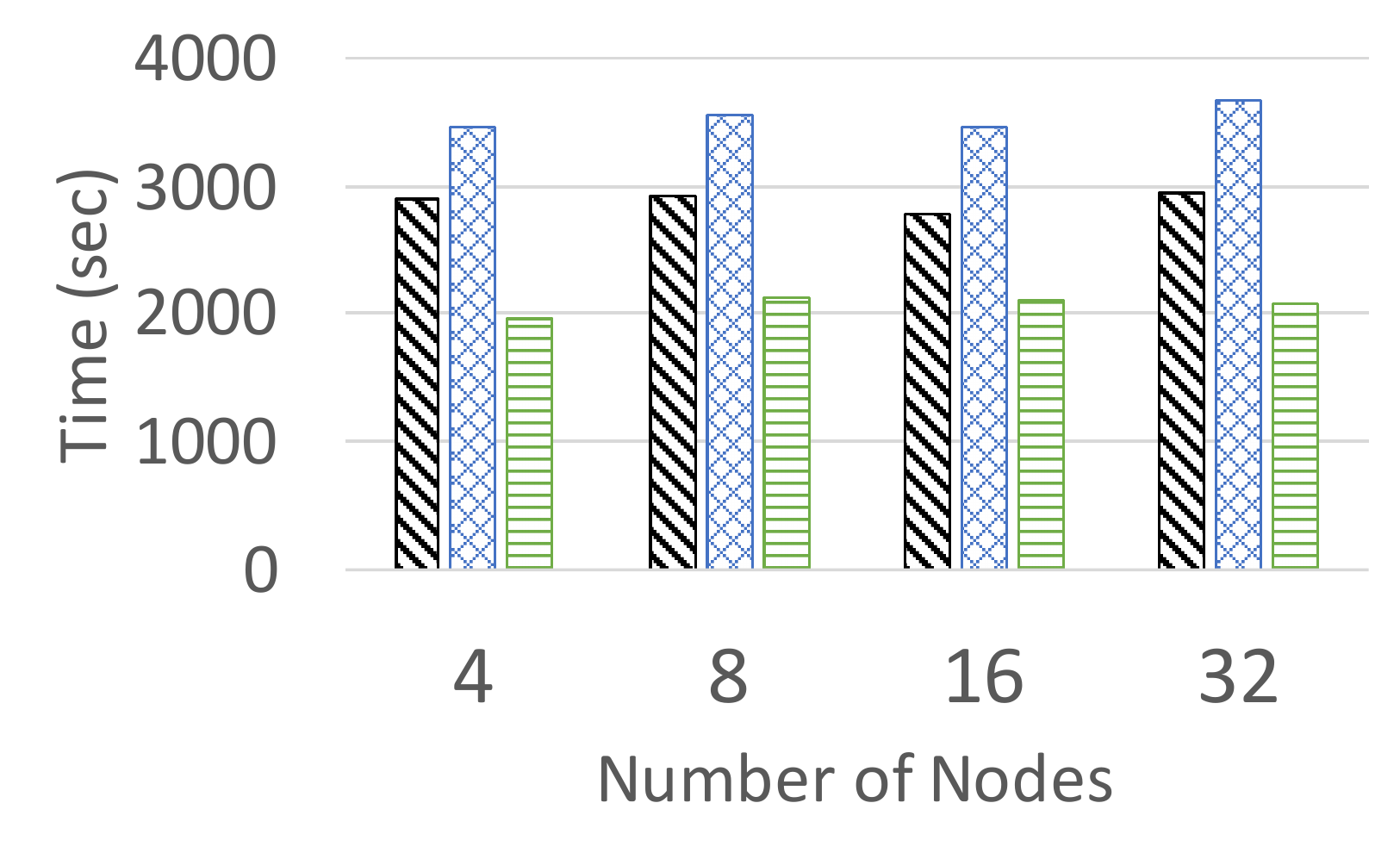}
        \caption{Ingestion time}
        \label{fig:scale-out-ingestion}
    \end{subfigure}
	    \vspace{-0.5cm}
    \caption{Storage and ingestion performance (scale-out)}
    \label{fig:scale-out-storage_and_ingestion}
\end{figure}

Figure~\ref{fig:scale-out-storage} shows the total on-disk size after ingesting the Twitter data into the open, closed and inferred datasets. The raw sizes of the ingested data were 400, 800, 1600 and 3200 GB for the 4, 8, 16 and 32 node clusters, respectively. Figure~\ref{fig:scale-out-ingestion} shows the time taken to ingest the Twitter data into the three datasets. As expected, we observe the same trends seen for the single node cluster (see Figure~\ref{fig:twitter-storage} and Figure~\ref{fig:twitter-feed}), where the inferred dataset has the lowest storage overhead with the highest data ingestion rate.
 
\begin{figure}[h]
    \centering
    \hspace{-0.55cm}
    \includegraphics[width=0.5\textwidth]{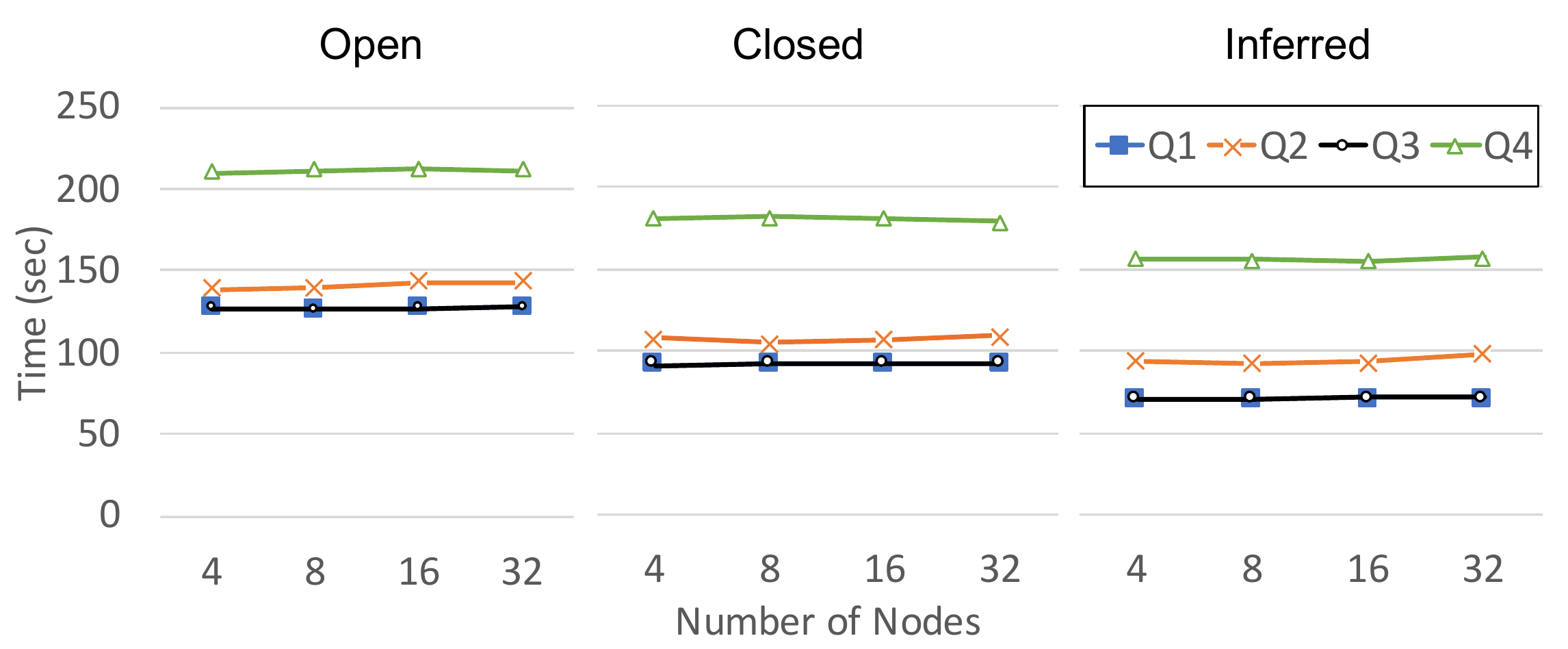}
    \vspace{-0.5cm}
    \caption{Query performance (scale-out)}
    \label{fig:scale-out-exper}

\end{figure}

 To evaluate query performance, we ran the same four Twitter queries as in Section~\ref{sec:twitter-exper}. Figure~\ref{fig:scale-out-exper} shows the execution times for the queries against the open, closed and inferred datasets. All four queries scaled linearly, as expected, and all four queries were faster in the inferred dataset. Since the data is shuffled in Q2 and Q3 to perform the parallel aggregation, each partition broadcasts its schema to the other nodes in the cluster (Section~\ref{sec:hetro-schema}) at the start of a query. However, the performance of the queries was essentially unaffected and was still faster to execute in the inferred~dataset.

\vspace{-0.3cm}
\section{Related Work}
\label{sec:related-work}
\textbf{Schema inference} for self-describing, semi-structured data has appeared in early work for Object Exchange Model (OEM) and later for XML and JSON documents. For OEM (and later for XML\footnote{Adopted later for XML as Lore project changed from OEM to the XML data-model.}), \cite{dataguides} presented the concept of a dataguide, which is a summary structure for schema-less semi-structured documents. A dataguide could be accompanied with values' summaries and samples (annotations) about the data, which we also use in our schema structure to keep the number of occurrences in each value. In \cite{wang2015schema}, Wang et al. present an efficient framework for extracting, managing and querying schema-view of JSON datasets. Their work targeted data exploration, where showing a frequently appearing structure can be good enough. However, in our work, the purpose of inferring the schema is to use it for compacting and querying the records, so, we infer the exact schema of the ingested dataset. In another work~\cite{json-schema-relational}, the authors detail an approach for automatically inferring and generating a normalized (flat) schema for JSON-like datasets, which then can be utilized in an RDBMS to store the data. Our work here is orthogonal; we target document store systems with LSM-based storage engines.

Creating \textbf{secondary indexes} is related to declaring attributes in a schema-less document store. Systems such as Azure DocumentDB~\cite{azure_documentdb} and MongoDB support indexing all fields at once without declaring the indexed fields explicitly. For instance, MongoDB allows users to create an index on all fields using a \textit{wildcard index} that was initially introduced in v4.2. Doing so requires the system to ``infer'' the fields of a collection a priori. Despite the similarities, our objective is different. In our work, we infer the schema to reduce storage overhead by compacting self-describing records residing in the primary index. 

\textbf{Semantically compacting} self-describing, semi-structured records using schemas appears in popular big data systems such as Apache Spark~\cite{spark} and Apache Drill~\cite{drill}. For instance, Apache Drill uses schemas of JSON datasets (provided by the user or inferred by scanning the data) to transform records into a compacted in-memory columnar format (Apache Arrow~\cite{arrow}). File formats such as Apache Parquet~\cite{parquet} (or Google Dremel~\cite{dremel}) use the provided schema to store nested data in a columnar format to achieve higher compressibility. An earlier effort to semantically compact and compress XML data is presented in \cite{xml-compressed, xmill}. Our work is different in targeting more ``row''-oriented document stores~with LSM-based storage engines. Also, we support data values with heterogeneous types, in contrast to Spark and Parquet.

\textbf{Exploiting LSM lifecycle events} to piggyback other operations to improve the query execution time is not new by itself and has been proposed in several contexts~\cite{lsm-stats, lsm-filter, c-store}. LSM-backed operations can be categorized as either non-transformative operations, such as computing information about the ingested data, or transformative operations, e.g., in which the records are transformed into a read-optimized format. An example of a non-transformative operation is \cite{lsm-filter}, which shows how to utilize LSM flush and merge operations to compute range-filters that can accelerate time-correlated queries by skipping on-disk components that do not satisfy the filter predicate. \cite{lsm-stats} proposes a lightweight statistics collection framework that utilizes LSM lifecycle events to compute statistical summaries of ingested data that the query optimizer can use for cardinality estimation. An example of a transformative operation is \cite{c-store}, which utilizes LSM-like operations to transform records in the writeable-store into a read-optimized format for the~readable-store.

\section{Conclusion and Future work}
\label{sec:conclusion}
In this paper, we introduced a tuple compaction framework that addresses the overhead of storing self-describing records in LSM-based document store systems. Our framework utilizes the flush operations of LSM-based engines to infer the schema and compact the ingested records without sacrificing the flexibility of schema-less document store systems. We also addressed the complexities of adopting such a framework in a distributed setting, where multiple nodes run independently, without requiring synchronization. We further introduced the vector-based record format, a compaction-friendly format for semi-structured data. Experiments showed that our tuple compactor is able to reduce the storage overhead significantly and improve the query performance of AsterixDB. Moreover, it achieves this without impacting data ingestion performance. In fact, the tuple compactor and vector-based record format can actually improve data ingestion performance of insert-heavy workloads. In addition to the semantic approach, we also added the support for a syntactic approach using page-level compression to AsterixDB. With both approaches combined, we were able to reduce the total storage size by up to 9.8x and improve query performance by the same factor.

Our tuple compactor framework targets LSM-backed row-oriented document-store systems. We plan to extend this work to introduce a schema-adaptive columnar-oriented document store. First, we plan to explore the viability of adopting the PAX~\cite{pax} page format, which could potentially eliminate the CPU cost of the linear access time of the vector-based format. In a second direction, we want to explore ideas from popular static columnar file formats (such as Apache Parquet and Apache CarbonData~\cite{apache-carbo-data}) to build an LSM-ified version of columnar indexes for self-describing, semi-structured data. 

\noindent\textbf{\textit{Acknowledgements}}
This work was supported by a graduate fellowship from KACST. It was also supported by NSF awards IIS-1838248 and CNS-1925610, industrial support from Amazon, Google, Microsoft and Couchbase, and the Donald Bren Foundation (via a Bren Chair).

\vspace{0.1cm}
\bibliographystyle{abbrv}
\Urlmuskip=0mu plus 1mu
\balance
\bibliography{main} 
\clearpage
\appendix
\section{Queries}
In this section, we show the queries we ran in our experiments against the Twitter, WoS and Sensors datasets.
\subsection{Twitter Dataset's Queries}
\label{appen:twitter}

\begin{sqlpp}
SELECT VALUE count(*) 
FROM Tweets
\end{sqlpp}
\centerline{Q1}
\vspace{-0.2cm}
\begin{sqlpp}
SELECT VALUE uname,a
FROM Tweets t
GROUP BY t.users.name AS uname 
WITH a AS avg(length(t.text))
ORDER BY a DESC
LIMIT 10
\end{sqlpp}
\centerline{Q2}
\vspace{-0.2cm}
\begin{sqlpp}
SELECT uname, count(*) as c
FROM Tweets t
WHERE (
  SOME ht IN t.entities.hashtags 
  SATISFIES lowercase(ht.text) = "jobs"
)
GROUP BY user.name as uname
ORDER BY c DESC
LIMIT 10
\end{sqlpp}
\centerline{Q3}
\vspace{-0.2cm}
\begin{sqlpp}
SELECT *
FROM Tweets
ORDER BY timestamp_ms
\end{sqlpp}
\centerline{Q4}

\subsection{WoS Dataset's Queries}
\label{appen:wos}

\begin{sqlpp}

SELECT VALUE count(*)
FROM Publications as t
\end{sqlpp}
\centerline{Q1}
\vspace{-0.2cm}
\begin{sqlpp}
SELECT v, count(*) as cnt
FROM Publications as t, 
   t.static_data.fullrecord_metadata
   .category_info.subjects.subject 
   	   AS subject
WHERE subject.ascatype = "extended"
GROUP BY subject.`value` as v
ORDER BY cnt DESC	
\end{sqlpp}
\centerline{Q2}
\vspace{-0.2cm}
\begin{sqlpp}
SELECT country, count(*) as cnt
FROM (
  SELECT value distinct_countries
  FROM Publications as t
  LET address = t.static_data
                .fullrecord_metadata
                .addresses.address_name,
  countries = (
      SELECT DISTINCT VALUE 
         a.address_spec.country
      FROM address as a
  )
  WHERE is_array(address)
  AND array_count(countries) > 1
  AND array_contains(countries, "USA")
) as collaborators
UNNEST collaborators as country
WHERE country != "USA"
GROUP BY country
ORDER BY cnt DESC
LIMIT 10
\end{sqlpp}
\centerline{Q3}
\vspace{-0.2cm}
\begin{sqlpp}
SELECT pair, count(*) as cnt
FROM (
    SELECT value country_pairs
    FROM Publications as t
    LET address = t.static_data
                 .fullrecord_metadata
                 .addresses.address_name,
    countries = (
        SELECT DISTINCT VALUE 
            a.address_spec.country
        FROM address as a
        ORDER by a.address_spec.country
    ),
    country_pairs = (
        SELECT VALUE [countries[x], countries[y]]
        FROM range(0, 
                     array_count(countries) - 1) as x, 
             range(x + 1, 
                     array_count(countries) - 1) as y
	)
	WHERE is_array(address)
	AND array_count(countries) > 1
) as country_pairs
UNNEST country_pairs as pair
GROUP BY pair
ORDER BY cnt DESC
LIMIT 10
\end{sqlpp}
\centerline{Q4}

\subsection{Sensors Dataset's Queries}
\label{appen:sensors}

\begin{sqlpp}
SELECT count(*)
FROM Sensors s, s.readings r
\end{sqlpp}
\centerline{Q1}
\vspace{-0.2cm}
\begin{sqlpp}
Q2:
SELECT max(r.temp), min(r.temp)
FROM Sensors s, s.readings r
\end{sqlpp}
\centerline{Q2}
\vspace{-0.2cm}
\begin{sqlpp}
Q3:
SELECT sid, avg_temp
FROM Sensors s, s.readings as r
GROUP BY s.sensor_id as sid 
WITH avg_temp as AVG(r.temp)
ORDER BY t DESC
LIMIT 10
\end{sqlpp}
\centerline{Q3}
\vspace{-0.2cm}
\begin{sqlpp}
Q4:
SELECT sid, avg_temp
FROM Sensors s, s.readings as r
WHERE s.report_time > 1556496000000 
AND s.report_time < 1556496000000 
                    + 24 * 60 * 60 * 1000
GROUP BY s.sensor_id as sid 
WITH avg_temp as AVG(r.temp)
ORDER BY avg_temp DESC
LIMIT 10
\end{sqlpp}
\begin{figure*}[t!]
    \begin{subfigure}[b]{1.\textwidth}
        \includegraphics[width=\textwidth]{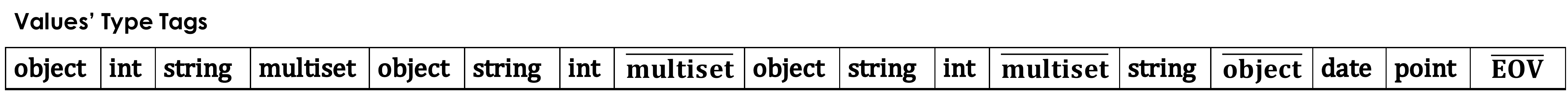}
        \caption{Type tag vector}
        \label{fig:example2-tags}
    \end{subfigure}
    \begin{subfigure}[b]{.5\textwidth}
        \includegraphics[width=\textwidth]{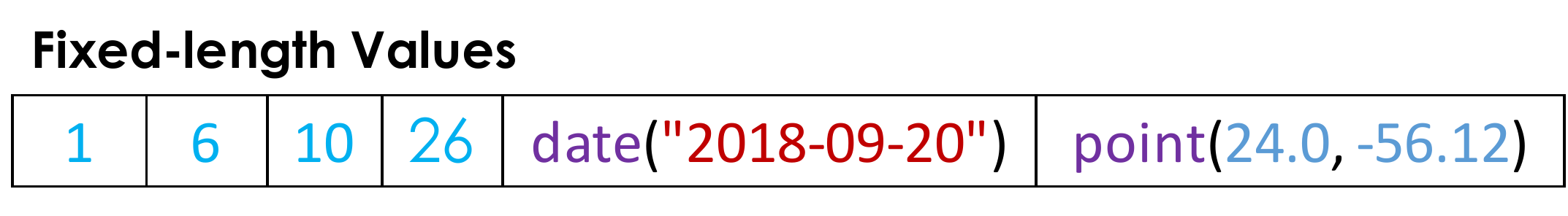}
        \vspace{.6em}
        \caption{Fixed-length values' vector}
        \label{fig:example2-fixed}
    \end{subfigure}
    \begin{subfigure}[b]{0.5\textwidth}
        \includegraphics[width=\textwidth, height=0.2\textwidth]{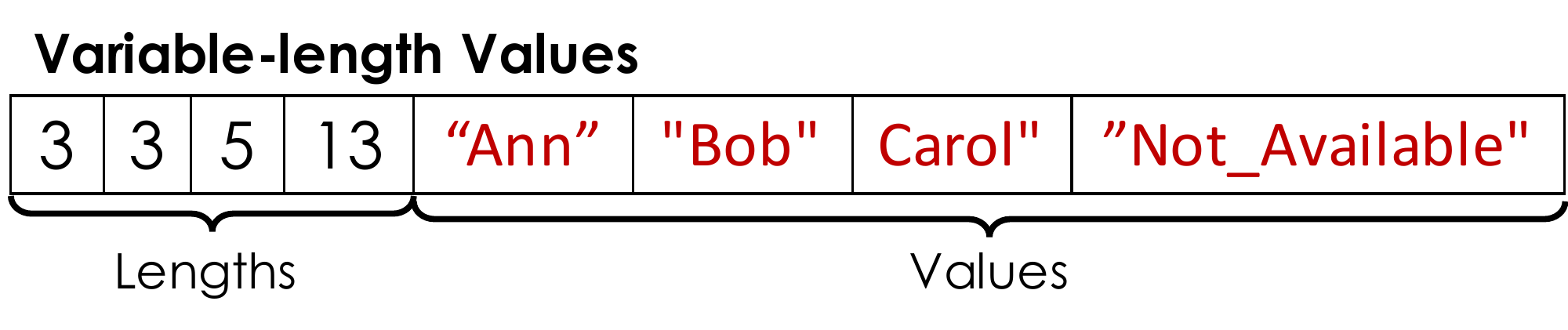}
        \caption{Variable-length values' vector}
        \label{fig:example2-var}
    \end{subfigure}
    \begin{subfigure}[b]{1.\textwidth}
        \includegraphics[width=\textwidth, height=0.1\textwidth]{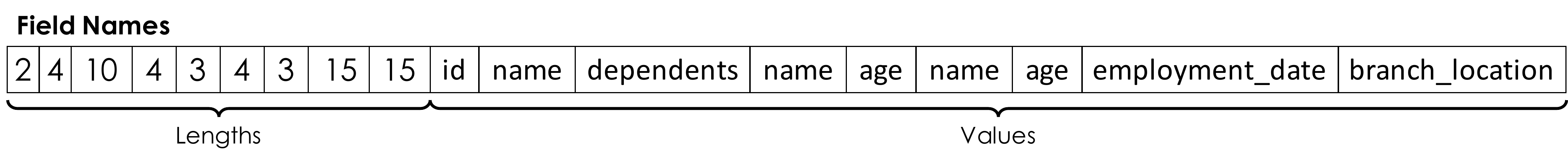}
        \caption{Field names' vector}
        \label{fig:example2-field}
    \end{subfigure}
    \caption{A record in a vector-based format (another example)}
    \label{fig:example2}
\end{figure*}
\section{Vector-based Format: Additional Example}
\label{appen:example2}
In Section~\ref{sec:vector-based-format}, we showed an example of a record in the vector-based format. However, the example may not clearly illustrate the structure of a record with complex nested values. In this section, we walk through another example of how to interpret a record in a vector-based format with more nested values.
\begin{figure}[H]
	\centering
	\includegraphics[width=0.3\textwidth]{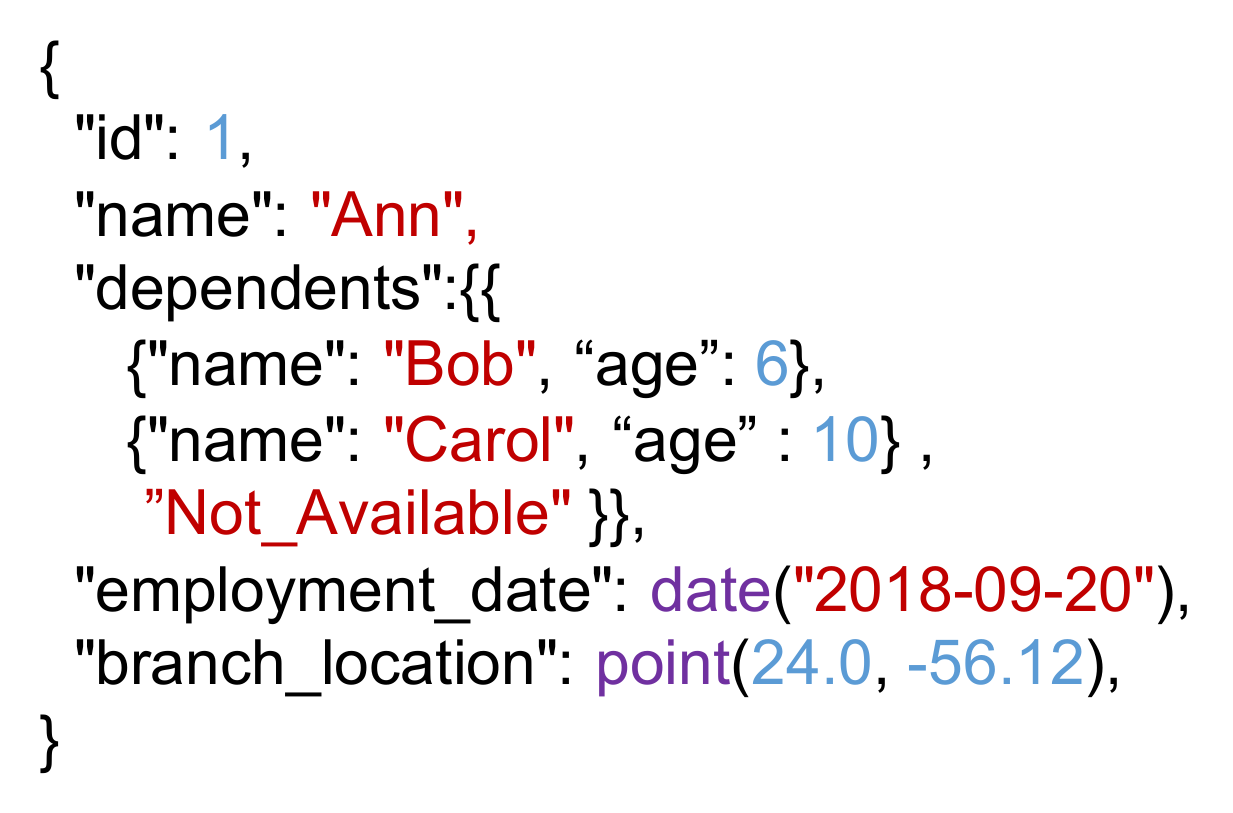}
	\caption{JSON document with more nested values}
	\label{fig:example2-json}
\end{figure}
Figure~\ref{fig:example2} shows the structure of the JSON document, shown in Figure~\ref{fig:example2-json}, in the vector-based format. Starting with the header (not show), we determine the four vectors, namely \textbf{(i)} the type tag vector (Figure~\ref{fig:example2-tags}), \textbf{(ii)} the fixed-lengths values' vector  (Figure~\ref{fig:example2-fixed}), \textbf{(iii)} the variable-length values' vector (Figure~\ref{fig:example2-var}), and finally \textbf{(v)} the field names' vector (Figure~\ref{fig:example2-field}). 

After processing the header, we start by reading the first tag (\textit{object}), which determine the root type. As explained in Section~\ref{sec:vector-based-format}, the tags of nested values (i.e., \textit{object, array}, and \textit{multi}) and control tags (i.e., $\overline{object}$, $\overline{array}$, $\overline{multiset}$, and $\overline{EOV}$) are neither fixed or variable length values. We continue to the next tag which is (\textit{int}). Since \textit{int} is a fixed-length value, we know it is stored in the first four bytes (assuming \textit{int} is a 4-byte int). Also, because it is preceded by the nested tag (\textit{object}), we know it is a field of this object, and thus its field name corresponds to the first field name \textbf{\textit{id}} in the field names' vector. Next, we see the tag (\textit{string}). As it is the first variable-length value, the length (i.e., \textbf{3}) and the value \texttt{"Ann"} in the variable-length vector belong to this string value. 

Followed by the string, we get the tag (\textit{multiset}), which is a nested value and a child of the root object type. Therefore, the third field name (length: 10, value: dependents) corresponds to the multiset value. We see in Figure~\ref{fig:example2-json} that the the field \textit{dependents} is a multiset of three elements of types: (\textit{object}), (\textit{object}) and (\textit{string}). Thus, the following tag (\textit{object}) corresponds to the first element of the multiset. As it is a child of a multiset, our object value does not have a field name. The next two tags are of type (\textit{string}) and (\textit{int}), which correspond to the field names \textbf{\textit{name}} and \textbf{\textit{age}}, respectively, as they are children of the preceded (\textit{object}) tag. The following tag ($\overline{multiset}$) marks the end of the current nesting type (i.e., object) and we are back to the nesting type (i.e., mutliset). The following tag (\textit{object}) marks the beginning of the second element of the mutliset (See Figure~\ref{fig:example2-json}) and we process it as we did for the first element. After the second control tag ($\overline{multiset}$), we get the tag (\textit{string}), which is the type of the third and the last element of our multiset. The following control tag ($\overline{object}$) tells it is the end of the multiset and we are going back to the root object type.

The next two tags (\textit{date}) and (\textit{point}) are the last two children of the root object type and they have the last two field names \textit{\textbf{employment\_date}} and \textit{\textbf{branch\_location}}, respectively. Finally, the control tag ($\overline{EOV}$) marks the end of the record.

\end{document}